\pdfoutput=1
\documentclass[twocolumn]{aastex62}

\usepackage[autostyle]{csquotes}

\usepackage{subfiles}
\usepackage{soul}

\graphicspath{{./}{figures/}}

\usepackage{amsmath}
\usepackage{hyperref}
\makeatletter
\newcommand{\mathrmnum}[1]{\romannumeral #1}
\makeatother

\begin{document}
\shortauthors{Phillips et al.}
\author{Caprice L. Phillips}
\affil{Department of Astronomy, The Ohio State University, Columbus, OH, 43210,USA}
\affil{LSSTC DSFP Fellow}

\author{Brendan P. Bowler}
\affil{Department of Astronomy, The University of Texas at Austin, Austin, TX, 78712, USA}
\shorttitle{A Young Companion at the Substellar Boundary}

\author{Gregory Mace}
\affil{McDonald Observatory $\&$ Department of Astronomy, University of Texas at Austin, 2515 Speedway, Stop C1400, Austin, TX 78712-1205, USA}

\author{Michael C. Liu}
\affil{Institute for Astronomy, University of Hawai`i at M\={a}noa, 2680 Woodlawn Drive, Honolulu, HI 96822, USA}

\author{Kimberly Sokal}
\affil{McDonald Observatory $\&$ Department of Astronomy, University of Texas at Austin, 2515 Speedway, Stop C1400, Austin, TX 78712-1205, USA}

\accepted{ by ApJ 7th April 2020}
\title{2MASS J04435686+3723033 B:

A Young Companion at the Substellar Boundary with Potential Membership in the $\beta$ Pictoris Moving Group}

\begin{abstract}

We present a detailed characterization of 2MASS
J04435750+3723031, a low-mass companion orbiting the young M2 star
''
2MASS J04435686+3723033 at 7$\farcs$6 (550 AU) with potential
membership in the 23 Myr $\beta$ Pictoris moving group
($\beta$PMG). Using near-infrared spectroscopy of the companion from
IRTF/SpeX we have found a spectral type of M6 $\pm$ 1 and indications
of youth through age-sensitive absorption lines and a low surface
gravity index (VL-G). A young age is supported by H$\alpha$
emission and lithium absorption in the host. We re-evaluate
the membership of this system and find that it is a marginally
consistent kinematic match to the $\beta$PMG using $Gaia$ parallaxes
and new radial velocities for the host and companion. If this system
does belong to the $\beta$PMG, it would be a kinematic outlier and the
companion would be over-luminous compared to other similar ultracool
objects like PZ Tel B; this would suggest 2M0443+3723 B could be a
close brown dwarf binary ($\approx$52+52 M$_\mathrm{Jup}$ if equal-flux,
compared with 99 $\pm$ 5 M$_\mathrm{Jup}$ if single), and would make it
the sixth substellar companion in this group.  To test this
hypothesis, we acquired NIR AO images with Keck II/NIRC2, but they
do not resolve the companion to be a binary down to the
diffraction limit of $\sim$3 AU. If 2M0443+3723 AB does not belong to
any moving group then its age is more uncertain.  In this case it is
still young ($\lesssim$30 Myr), and the implied mass of the companion
would be between $\sim$30--110 M$_\mathrm{Jup}$.

\end{abstract}


\section{Introduction} \label{sec:intro}
 
The study and characterization of the lowest-mass stars and brown dwarfs (BDs) is a relatively new field, with the first brown dwarfs, Gliese 229 B (\citealt{nakajima1995}; \citealt{oppenheimer1995}), Teide 1 \citep{rebolo1995}, and PPL 15 \citep{basri1996}  having been discovered less than a quarter century ago. BDs ($\lesssim$ 75 M$_\mathrm{Jup}$) are not massive enough to stably burn hydrogen in their cores and  thus represent the transition region between gas giant planets and low--mass stars.  Substellar objects  that fall below the hydrogen burning limit gradually cool and grow dimmer as they age  and thus follow a degenerate mass-age-luminosity relationship (\citealt{burrows2001}; \citealt{saumon&marley2008}). This is contrary to low--mass stars which have stable luminosities  over the main course of their lifetime. These objects are generally defined  based on the limit for the onset of hydrogen and deuterium burning, not necessarily their formation: stars have masses greater than 75 M$_\mathrm{Jup}$, BDs span 13--75 M$_\mathrm{Jup}$, and objects between 0.2  M$_\mathrm{Jup}$ and 13 M$_\mathrm{Jup}$ are considered gas giant planets if they are companions to stars \citep{burrows2001}.
\par

\par
Benchmark systems  are objects that have two or more measured fundamental quantities such as  luminosity and age; for brown dwarfs these parameters can be used to infer other properties like mass, temperature, and radius using substellar evolutionary models \citep{burrows2001}. For example, BDs that are companions to stars and members of young moving groups have ages that can be determined from the age of the host star or  from age-dating the moving group. Benchmark brown dwarfs provide valuable tests for substellar atmospheric and evolutionary models by anchoring parameters to provide assessments for mutual consistency or  to serve as direct comparisons with predictions from ultra-cool atmospheric and evolutionary models (e.g. \citealt{bowler2009}; \citealt{dupuy2009}; \citealt{crepp2012}; \citealt{brandt2018}).

\par

Similarly, young low-mass stars with well constrained ages and metallicities offer important tests of pre-main sequence evolutionary models (e.g., \citealt{montet2015}; \citealt{nielsen2016}).  There is growing evidence for discrepancies between the measured properties of young low-mass stars and predictions from evolutionary tracks  (e.g., \citealt{kraus2015}; \citealt{david2015}; \citealt{rizzuto2017}), which may point to a systematic error in the modeling of convective stars.  For example, \cite{feiden2016} found that including magnetic fields in the evolution of low-mass stars inhibits their convection and slows their contraction along the Hayashi track.  This puts the inferred isochronal ages for K and M dwarfs more in line with those for earlier-type stars of the same cluster and appears to largely reconcile the inferred masses determined from HR diagrams with dynamical masses (\citealt{simon2019}).  Identifying the lowest-mass stellar and substellar members of young clusters is therefore important to continue these tests over a variety of cluster ages, sizes, and environments.

Over the past few years, new high-contrast adaptive optics (AO) imaging instruments have paved the way for direct imaging studies of exoplanets and brown dwarf companions \citep{bowler2016}. The Spectro-Polarimetric High-contrast Exoplanet REsearch (SPHERE; \citealt{bezuit2008}) instrument at the Very Large Telescope (VLT), Subraru Coronographic Extreme Adaptive Optics (SCExAO; \citealt{martinache2014}), MagAO-X (\citealt{males2018}), Gemini Planet Imager (GPI; \citealt{macintosh2014}), and the Near-Infrared Coronagraphic Imager (NICI; \citealt{chun2008}) at Gemini-South have made imaging brown dwarfs and exoplanet companions more accessible. With the development of these specialized instruments, several brown dwarfs and planets have been discovered with direct imaging (e.g. \citealt{marois2008}; \citealt{biller2010}; \citealt{macintosh2015}), but a detailed study of their atmospheres is difficult because they reside so close to their host stars.
The upcoming development of larger telescopes such as the \textit{James Webb Space Telescope}, European Extremely Large Telescope, and the Giant Magellan Telescope will be able to probe even closer separations and discover lower--mass planets to characterize  their fundamental properties. However, detailed, high signal-to-noise ratio (SNR) spectroscopy of these objects will be difficult, underscoring  the value of wide substellar companions which are free of  contamination from their host stars.

Nearby young moving groups represent some of the best regions to identify intermediate-age benchmark brown dwarfs and low-mass stars. In these associations, the metallicities and ages of substellar companions are known because their members can generally be studied in detail (e.g. \citealt{naud2013}; \citealt{biller2013}). Members of these young, loose associations are ideal targets for directly imaging exoplanets in contrast to young star-forming clusters and unassociated nearby young stars, as the distances of young moving groups (YMGs) are closer ($\lesssim$ 100 pc) and the ages are well established (e.g. \citealt{zuckerman&song2004}: \citealt{torres2008}; \citealt{kraus2014new}).
\par

The $\beta$ Pictoris Moving Group ($\beta$PMG) is an intermediate age (23 $\pm$ 3 Myr) cluster of stars that lies in the solar neighborhood (\citealt{zuckerman2001}; \citealt{torres2008}; \citealt{mamajek&bell2014}). This association has a large spatial distribution in the sky, but its members share similar kinematics, which allows for stars to be identified as a common cluster in velocity space. \cite{schlieder2010} proposed several new members of the $\beta$PMG  based on proper motions, signatures of youth (e.g. H$\alpha$ and X-ray emission), and radial velocity measurements. In this study we examine one system in detail, 2M0443+3723 AB, which comprises an early M-type primary (2MASS J04435686+3723033; hereinafter 2M0443+3723 A) and a mid-to-late-M secondary (2MASS J04435750+3723031; hereinafter 2M0443+3723 B).

\begin{figure*}
    \includegraphics[width=7in]{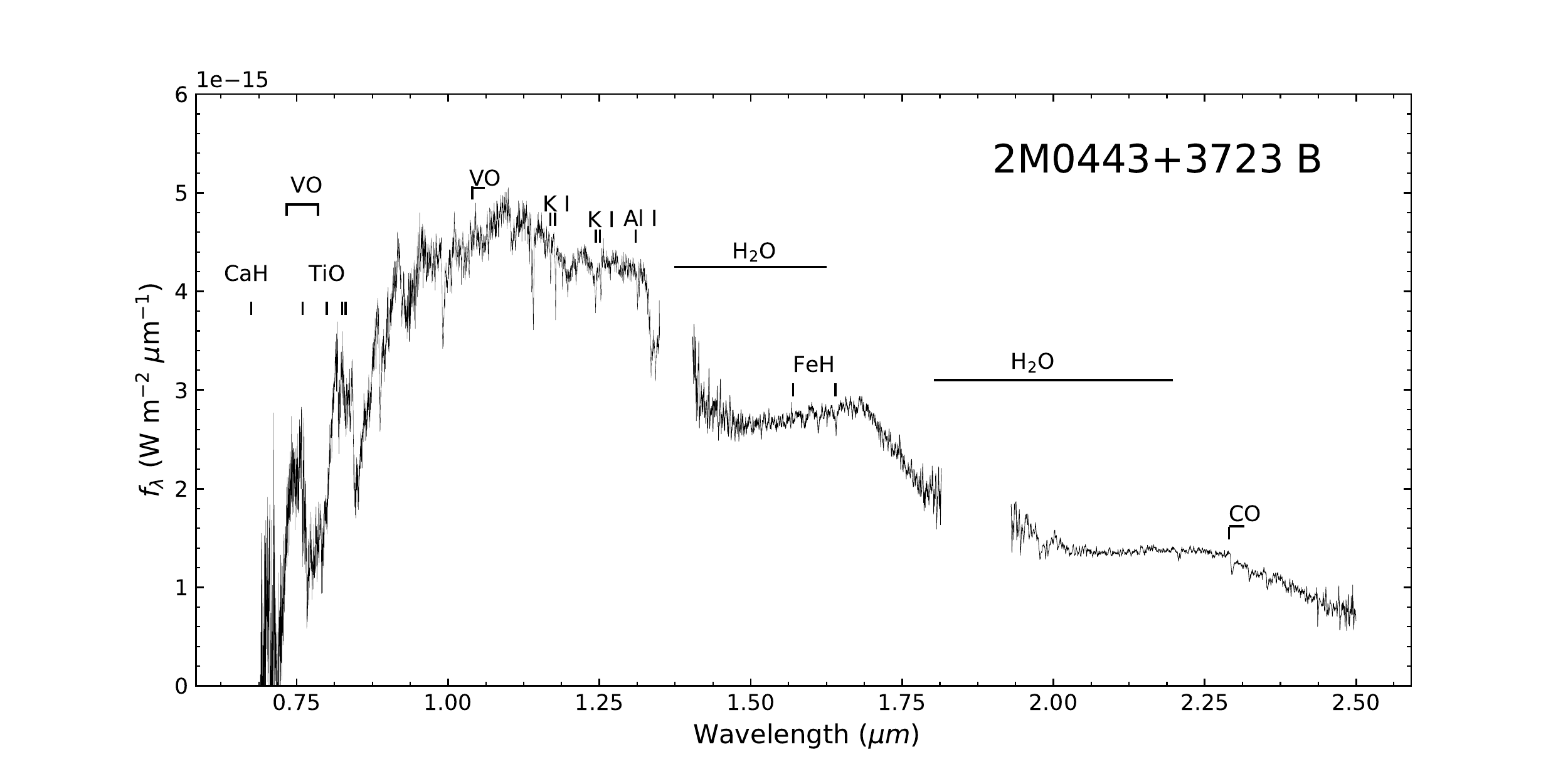}
    \caption{Our medium-resolution (\textit{R} $\sim$ 750) flux-calibrated 0.7--2.4 $\mu$m spectrum of 2M0443+3723 B from IRTF/SpeX. We derive a M6 $\pm$ 1 spectral type following \cite{aller&liu2013}. The low S/N regions have been removed.}
    \label{fig:2M0443_B}
\end{figure*}

\par
The companion, 2M0443+3723 B, is separated by 7$\farcs$6 (550 AU) from the M2 host and a spectral type of M5 is estimated by \cite{schlieder2010} based on the inferred absolute magnitudes. In a study of low-mass objects in the $\beta$PMG, \cite{messina2017} find that the 2M0443+3723 AB system deviated ($>$3$\sigma$) from other confirmed $\beta$PMG members in two components of the space velocity. However, \cite{shkolnik2017}  include the system as members of the $\beta$PMG in their ACRONYM survey. Because it is a potential new benchmark brown dwarf companion, our goal is to understand its properties and reassess its membership in the $\beta$PMG and other moving groups. 
\par
 In Section \ref{sec:observations} we describe the observations and data reduction. We summarize
our results in Section \ref{sec:results}, assess the membership in Section \ref{sec:discussion}, and conclude in Section \ref{sec:summaryandconclusions}.


\section{Observations}
\label{sec:observations}
\subsection{IRTF/SpeX}
We obtained a  near-infrared spectrum of 2M0443+3723 B using the short-wavelength cross-dispersed (SXD) mode on the SpeX spectrograph \citep{rayner2003}, which is located on the NASA Infrared Telescope Facility (IRTF) on Maunakea, Hawai'i. The SXD mode covers the 0.7--2.4 $\mu$m wavelength range and has a resolving power of \textit{R} $\sim$ 2000 for a 0$\farcs$3 slit width (or R $\sim$ 750 for a  0$\farcs$8 slit). The data were obtained on 2015 November 28 UT using the 0$\farcs$8 $\times$ 15\arcsec  slit with a total exposure time of 717 s. The observations were taken using consecutive ABBA nods and reduced using the \texttt{SpeXTool} package with the A0V standard HD 22859 (\citealt{vacca2003}; \citealt{cushing2004}). This reduction package extracts the spectra, performs wavelength calibration, corrects for telluric features using an A0V star, and merges the SXD orders. Our final medium-resolution (\textit{R} $\sim$ 750)
merged spectrum of 2M0443+3723 B is shown in Figure \ref{fig:2M0443_B}.

\subsection{Discovery Channel Telescope/IGRINS}
High resolution near-infrared  spectra of 2M0443+3723 A and 2M0443+3723 B were obtained with IGRINS (Immersion Grating Infrared Spectrometer; \citealt{park2014}) at the 4.3m Discovery Channel Telescope on 2017 September 02 UT. Observing conditions were partly cloudy. IGRINS simultaneously covers \textit{H} and \textit{K} bands with \textit{R} $\sim$ 45,000. 2M0443+3723 A was observed in an ABBA quad sequence with individual exposure times of 300 s and a total integration time of 1200 s. 2M0443+3723 B was observed in an ABBAAB pattern with each exposure lasting 900 s for a total integration time of 5400 s.  The data were reduced using the IGRINS pipeline package \citep{igrins} which rectifies the 2D image and optimally extracts the spectra, performs wavelength calibration with OH lines, and corrects for telluric features with standard A0V stars (in this case HR 1692 and HR 1237).

\subsection{Keck II/NIRC2 Adaptive Optics Imaging}
\label{sec:keck_imaging_data}
Natural guide star (NGS) adaptive optics (AO) images of the 2M0443+3723 AB system  were obtained on 2017 October 10 UT  with the NIRC2 near--infrared imaging camera mounted on the Keck II 10m telescope. The companion was positioned in the top left quadrant of the detector, as the bottom left quadrant has increased noise levels. The host star was positioned in the bottom right quadrant of the detector. We obtained a total of 6 images of the system in the Mauna Kea Observatories (MKO) \textit{H}--band filter 
(\citealt{tokunaga&simons2002}; \citealt{tokunga2002}) in narrow camera mode, which produces a plate scale of 9.971 $\pm$ 0.004 mas pixel$^{-1}$ \citep{service2016} and a field of view of 10$\farcs$2 $\times$ 10$\farcs$2. Each image is comprised of 10 unsaturated coadds with 2 seconds per coadd, resulting in a total on-source integration time of 20 seconds.
\par
\begin{figure}
\begin{center}

 \includegraphics[width=1.2\columnwidth]{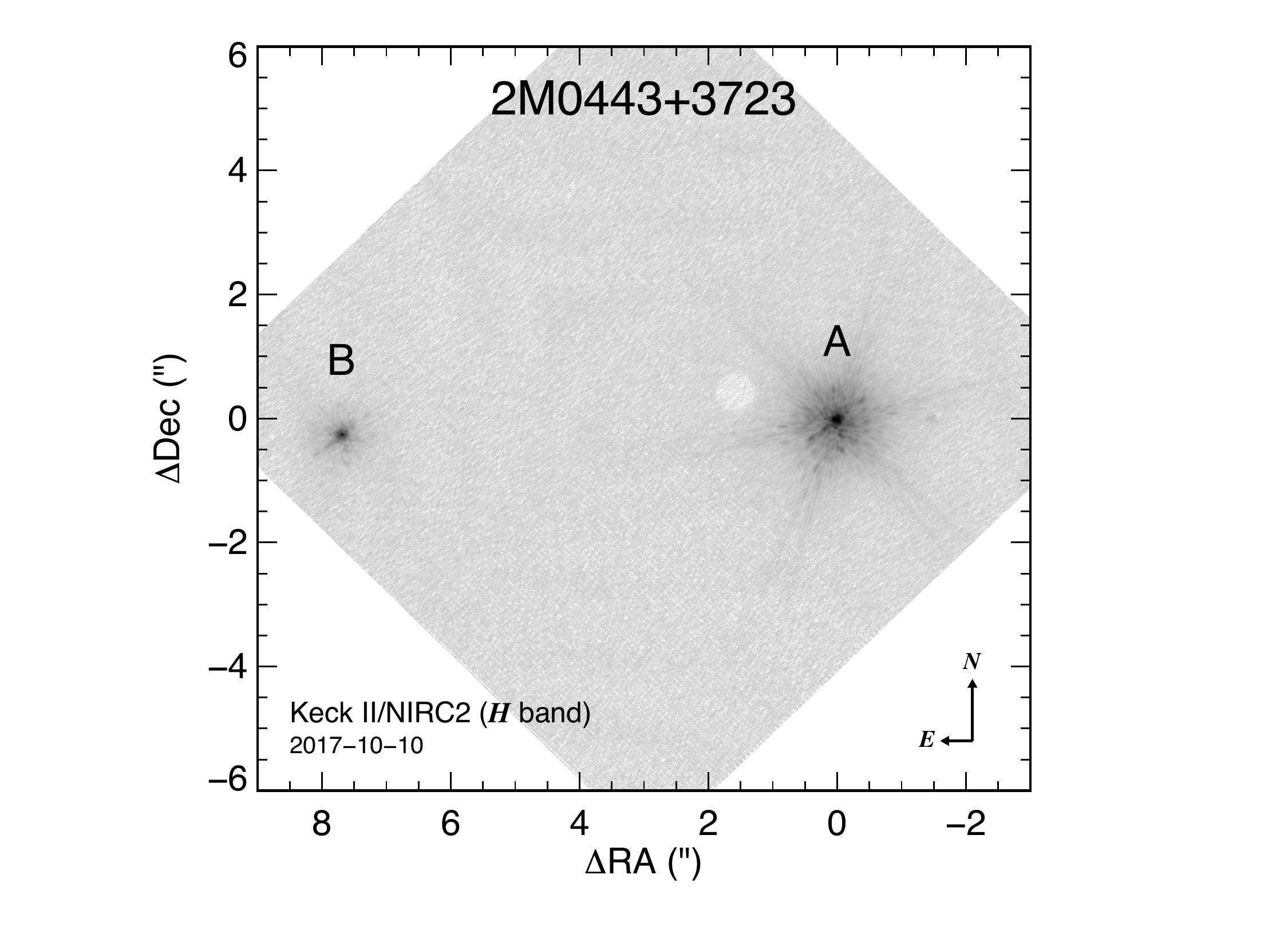}
    \caption{Keck II/NIRC2  AO image of 2M0443+3723 AB in \textit{H}-band.}
 \label{fig:keck_image}
\end{center}
\end{figure}

We perform basic image reduction including bias subtraction, flat field division, bad pixel correction, and removal of cosmic rays using the \texttt{cosmics.py} package \citep{dokkum2001} in \textsc{Python}, which utilizes Laplacian edge detection to identify and remove cosmic rays in the raw science frames. An example of a reduced image of the pair is shown in Figure \ref{fig:keck_image}.

\section{Properties of the 2M0443+3723 AB system}
\label{sec:results}


\subsection{Overview of 2M0443+3723 A}
\begin{figure}
    \includegraphics[width=1.1\columnwidth]{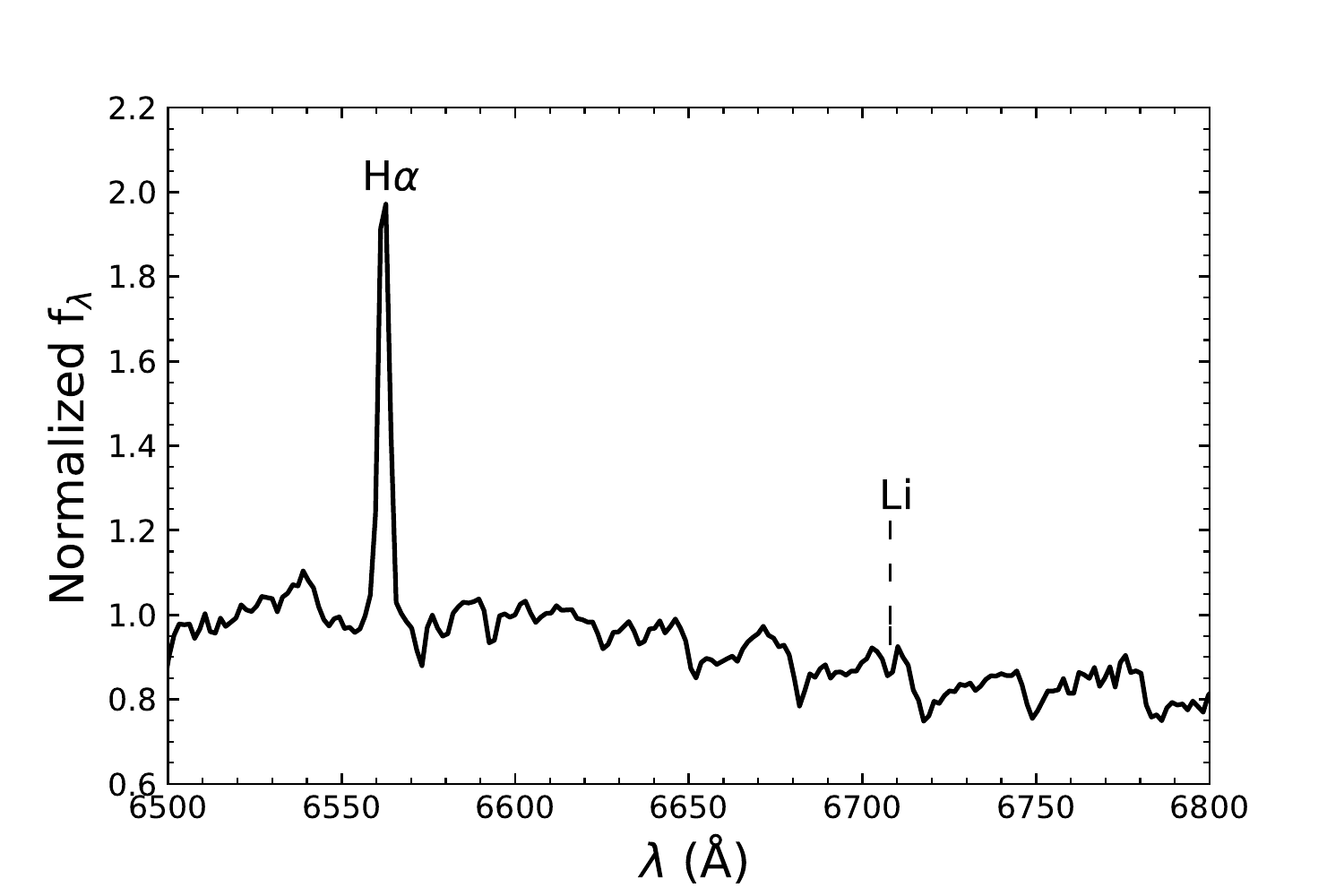}
    \caption{Optical spectrum of 2M0443+3723 A obtained from the RC-Spec instrument at the Mayall telescope, from \cite{bowler2019}. The spectrum has been normalized around 6600 $\mathrm{\AA}$. The H$\alpha$ emission  indicates signs of chromospheric activity and serves as a necessary but not sufficient condition for youth. The lithium absorption feature at 6708 $\mathrm{\AA}$ is also present, which unambiguously confirms the young age of the system.}
    \label{fig:host_star}
\end{figure}

The host star, 2M0443+3723 A, has a spectral type of M2 $\pm$ 1 as measured by \cite{schlieder2010} and a distance of 71.6 $\pm$ 0.3 pc (\citealt{gaia}) A similar photometric spectral type of M3  was determined by \cite{shkolnik2017}  in their ACRONYM survey. \cite{messina2017} measured an H$\alpha$ emission equivalent width (EW) of --4.60 $\pm$  0.21 $\mathrm{\AA}$. \cite{malo} report a 194 $\pm$ 4 m$\mathrm{\AA}$ EW of the \ion{Li}{1} 6708 $\mathrm{\AA}$ absorption feature. \cite{bowler2019} find similar values of EW(H$\alpha$) = --4.3 $\pm$ 0.9 $\mathrm{\AA}$ and EW(\ion{Li}{1}) = 0.12 $\pm$ 0.02 $\mathrm{\AA}$ (Figure \ref{fig:host_star}).

\subsection{Empirical Comparison to Cool and Ultracool Dwarfs}
To determine the spectral type of  2M0443+3723 B, we compare our SXD spectrum to the SpeX prism library of M, L, and T dwarfs \citep{burgasser2014}.  We convolve our medium--resolution spectrum (\textit{R} $\sim$ 750) with a 1D Gaussian kernel to match the resolving power  of the SpeX prism library (\textit{R} $\sim 250$) and resample them to a common wavelength grid. Each prism spectrum was optimally scaled to our SXD spectrum via reduced  $\chi^{2}$ minimization,
\begin{equation} \chi^{2}=\sum_{i=0}^{n-1} \frac{(O_{i}-cE_{i})^{2}}{\sigma_{i}^{2}},
\end{equation}
where  $O_{i}$ is the flux density of the  2M0443+3723 B spectrum at pixel \textit{i}, $E_{i}$ is the flux density from the SpeX prism templates, and  $\sigma_{i}$ is the uncertainty associated with our 2M0443+3723 B spectrum. For each spectrum the optimal scale factor, $c$, is found following \cite{cushing2008} to scale the prism templates to our SXD spectrum:
$$
    c=\frac{\sum O_{i}E_{i}/\sigma_{i}^{2}}{\sum E_{i}^{2}/ \sigma _{i}^{2}}.
$$

\par
\begin{figure}
    \centering
    \includegraphics[width=1.1\columnwidth]{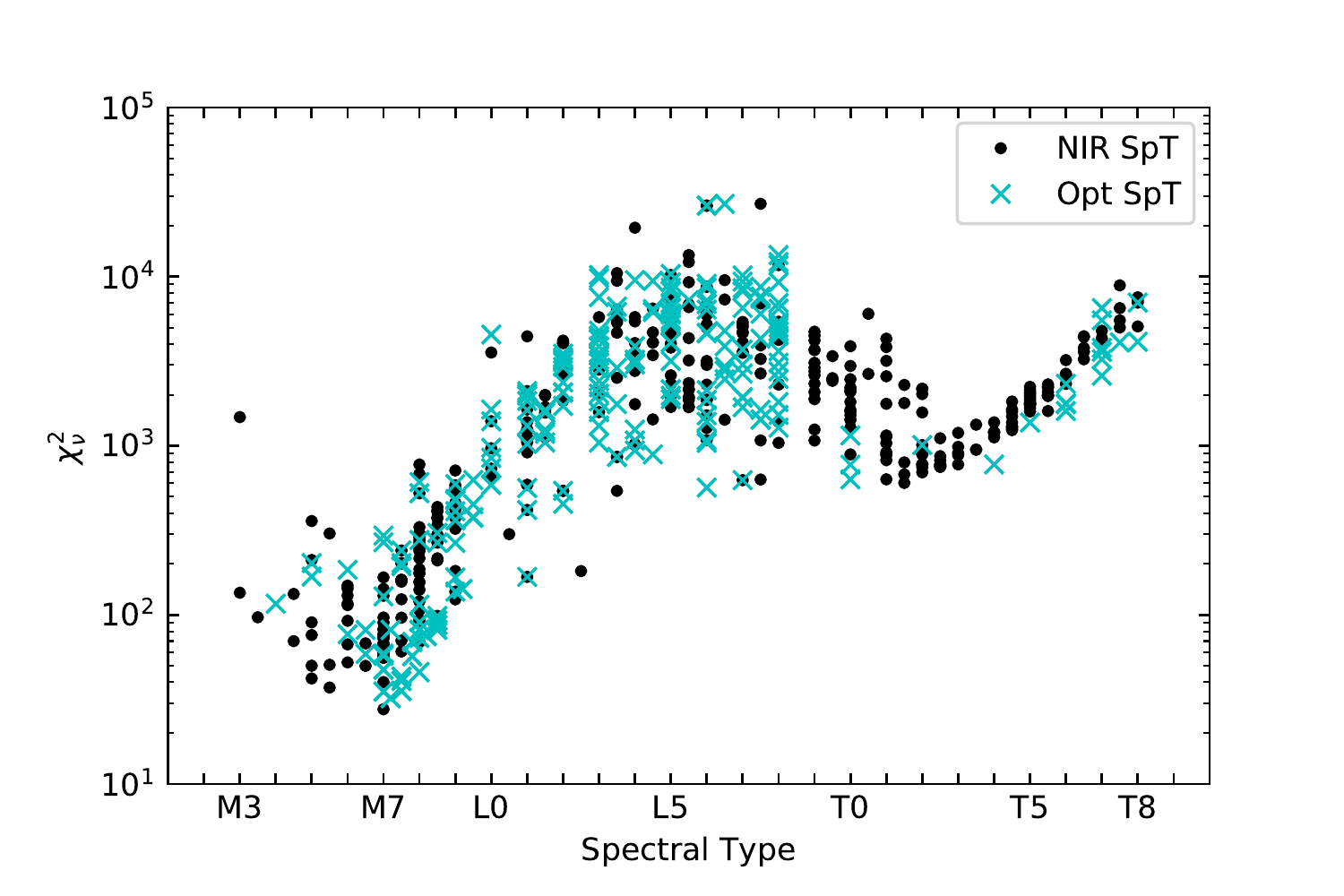}
    \caption{$\chi^{2}_{\nu}$ comparison between cool and ultracool  ($\geq$ M6)  dwarfs from the SpeX prism library and  our spectrum of 2M0443+3723 B. Optical spectral types are shown with cyan symbols, and the NIR spectral types are shown with black circles.  The closest match is that of 2MASS  J00013044+1010146, which has an  M7 NIR spectral type \citep{burgasser2004}.}
    \label{fig:chi_squared_minimization_spt}
\end{figure}
\begin{figure}
    \centering
    \includegraphics[width=1.1\columnwidth]{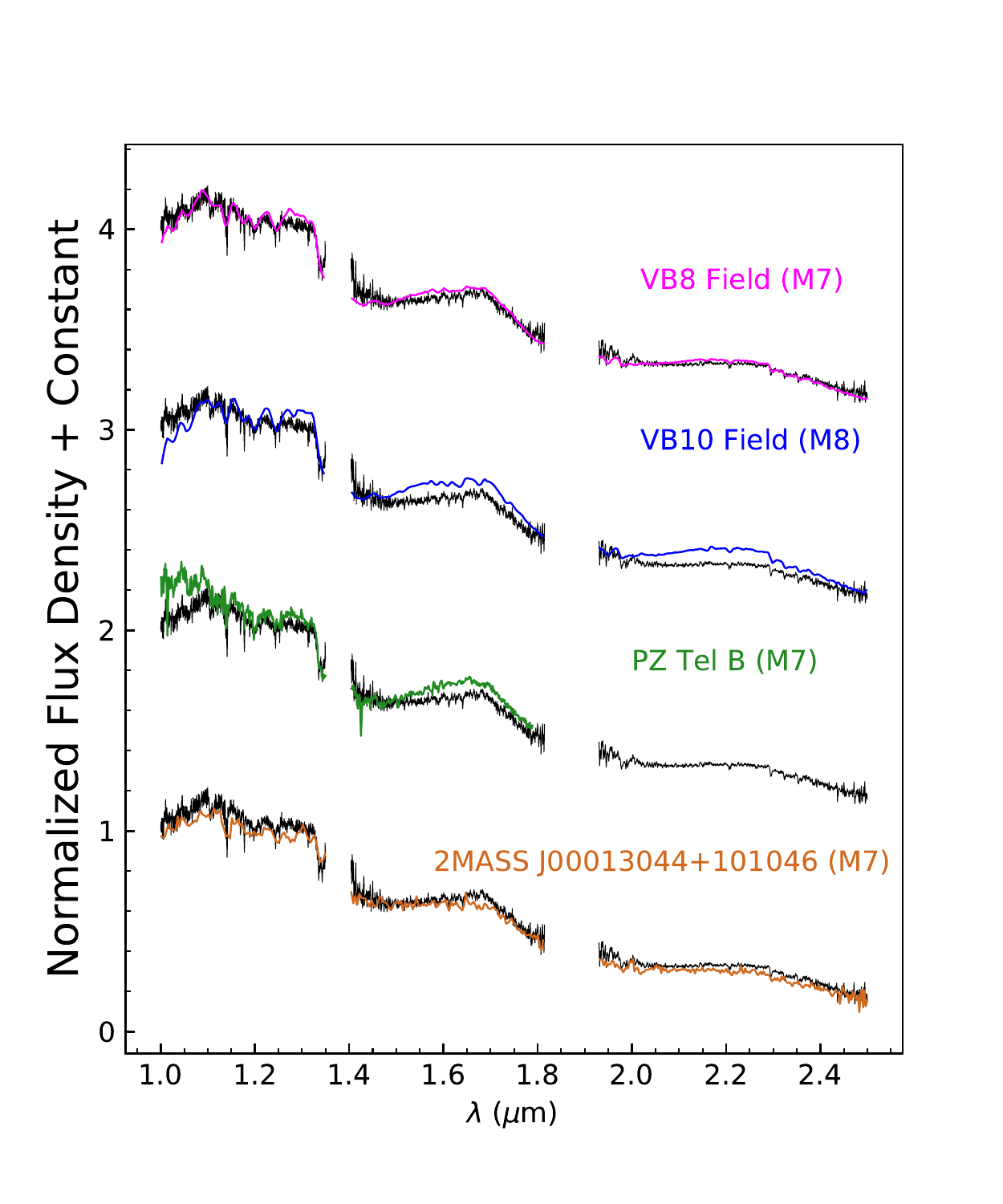}
 
    \caption{Comparison between our SpeX spectrum of 2M0443+3723 B (black)  and both young and field ultracool dwarfs. The best fitting spectrum from the $\chi_{\nu}^{2}$ comparison with M, L, and T dwarfs from the SpeX Prism Library is shown in orange (2MASS J0013044+1010146; \citealt{burgasser2004}), while comparison to the young brown dwarf PZ Tel B \citep{maire2015} is  displayed green. A field M7 dwarf (VB 8; \citealt{burgasser2008}) and field M8 dwarf (VB 10; \citealt{burgasser2014}) are also plotted for comparison. 2M0443+3723 B is a close match to the VB 8 field M7 dwarf \citep{burgasser2008} as well as the young M7 dwarf 2MASS J00013044+101046.} 
    \label{fig:comparision_best_fit_prism}
\end{figure}

\noindent 
In order to determine the goodness of fit between the prism library and our SXD spectrum we utilize the reduced $\chi^{2}$ statistic, $\chi_{\nu}^{2} \equiv \chi^{2}/\nu$, where $\nu$ represents the number of degrees of freedom (Figure \ref{fig:chi_squared_minimization_spt}). The best fitting spectral type found from this analysis is M7 $\pm$ 1. The best fitting object from this method is 2MASS J00013044+1010146 (Figure \ref{fig:comparision_best_fit_prism}; $\chi_{\nu}^{2}$ = 27.67) \citep{burgasser2004}. \cite{witte2011} used the DRIFT-PHOENIX atmospheric  models to find the best fitting parameters for 2MASS J00013044+1010146  which suggested it has a low surface gravity of log $g$ = 4.5 dex, $T\mathrm{_{eff}}$ = 2900 K, and M/H = +0.3 dex. \cite{gagne2014} found a NIR spectral type of M6 and noted this object as being potentially  young due to low gravity spectroscopic signs.

\subsubsection{Index-Based Spectral Type  and Surface Gravity}
\begin{figure}
     \centering
     \includegraphics[width=1.1\columnwidth]{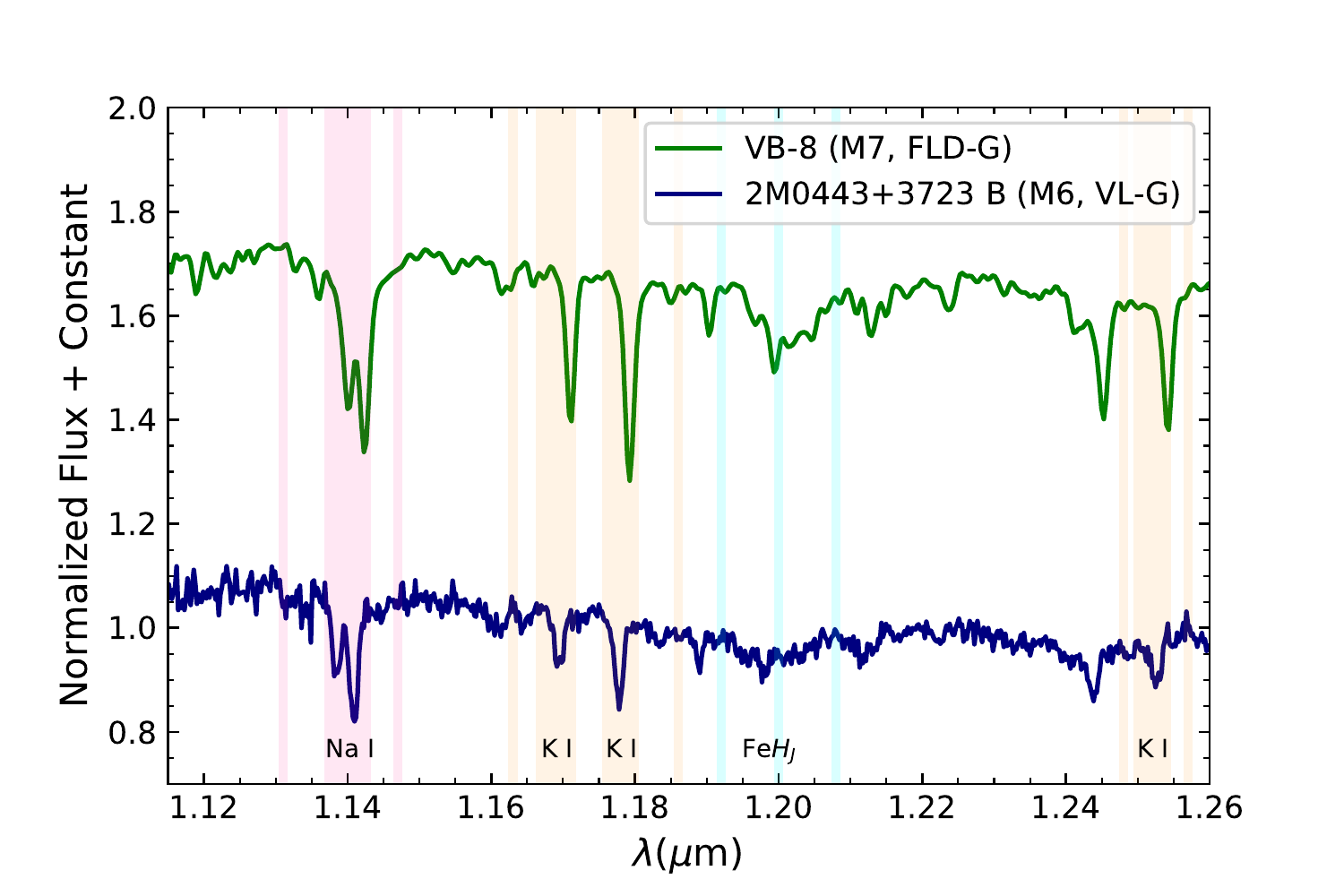}
     \caption{Our \textit{J}--band IRTF/SpeX spectrum of 2M0443+3723 B (blue) compared to a FLD-G standard from the IRTF Spectral Library, VB-8 (green; \citealt{cushing2005}). The \ion{Na}{1}, \ion{K}{1}, and  FeH  features of our companion are weaker than the FLD-G object, which indicates 2M0443+3723 B is a young, low gravity object. Note that the spectrum of VB-8 ($R$ $\sim$ 2000) has been smoothed to a comparable resolution as our SXD spectrum.}
     \label{fig:indices_comparision}
 \end{figure}

 Many features in the near-infrared spectra  of brown dwarfs are influenced by gravity and age, for example FeH (0.99, 1.20, 1.55 $\mu$m), VO (1.06 $\mu$m), \ion{K}{1} (1.17 $\mu$m), \ion{Na}{1} (1.14, 2.21 $\mu$m), and the \textit{H}-band continuum shape (\citealt{aller&liu2013}, hereinafter AL13). AL13 developed an index-based spectral  classification method based on visual classification  as well as flux ratios of near-infrared indices centered on spectroscopic features influenced by gravity and age. The visual classification method yields a spectral type of M6 $\pm$ 1 from both the \textit{J}-band and \textit{H}-band regions. The overall spectral type is M6 $\pm$ 1, with   a gravity class of \enquote{VL-G} (very low gravity), which supports a young age for 2M0443+3723 B. We adopt the index-based spectral type from AL13 because the results are anchored to quantitative definitions, whereas the $\chi^{2}$ spectral type depends on the author's spectral classification. Following the AL13 method, a score of 0--2 is assigned for objects with the following designations: 0 for field  gravity dwarfs ({\sc FLD-G}; $\gtrsim$ 200 Myr), 1 for intermediate--age gravity dwarfs ({\sc INT-G}; $\sim$50--200 Myr), and 2 for  young low gravity dwarfs ({\sc VL-G}; $\sim$ 10--30 Myr). The final gravity score measured is 2 for 2M0443+3723 B, although individual gravity scores from the FeH$_{z}$ and \ion{K}{1}$_\mathrm{{J}}$ yield values of 1, corresponding to {\sc INT-G} (Figure \ref{fig:indices_comparision}). Gravity scores have been shown to correlate with age (e.g. \citealt{liu2016}); most {\sc VL-G} objects are  young ($<$ 30 Myr).

\subsection{Atmospheric Properties of 2M0443+3723 B}
\label{sec:atmospheric_modeling_2M0443B}
 We utilize the BT-Settl \enquote{CIFIST 2011-2015} grids \footnote{\url{http://perso.ens-lyon.fr/france.allard/}}\citep{allard2012} to determine the effective temperature and surface gravity of 2M0443+3723 B. These models span an effective temperature range of 1200--7000 K, with increments of 100 K, and surface gravities from 2.5--5.5 dex in increments of 0.5 dex. The models assume solar metallicity and do not include alpha-enhancement.

\begin{figure}
    \centering
    \includegraphics[width=1.1\columnwidth]{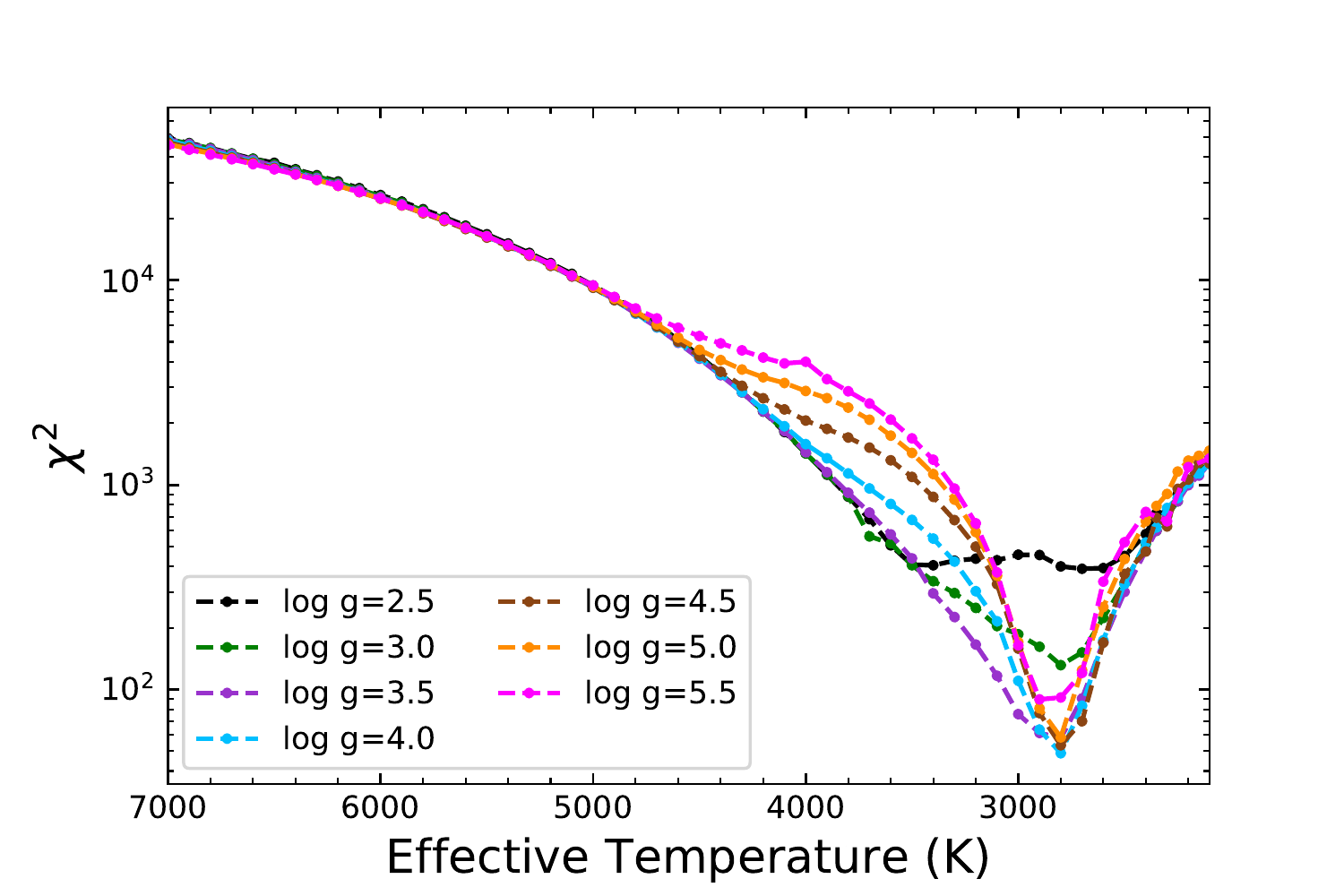}
    \caption{$\chi^{2}$  values for fits to our SXD spectrum values across the effective temperature range of the BT-Settl \enquote{CIFIST 2011-2015} model grid. There is a convergence to a minimum at a temperature of 2800 K and log $g$ of 4.0 dex.}
    \label{fig:chi_atmosphericmodels}
\end{figure}

\begin{figure}
\includegraphics[width=1.1\columnwidth]{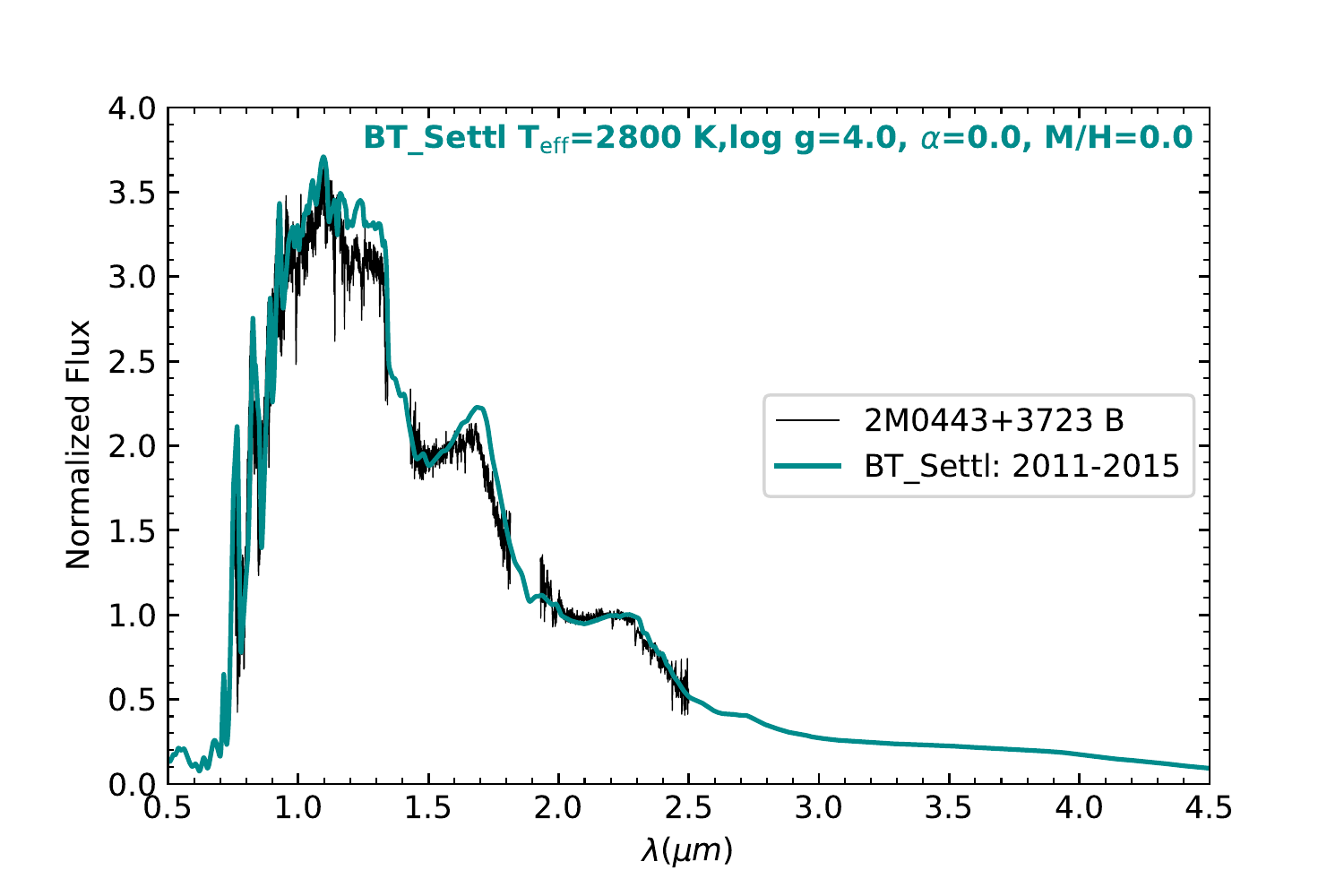}

\caption{2M0443+3723 B (black) SED compared to the best fitting  spectrum from the BT-Settl \enquote{CIFIST 2011--2015} atmospheric models.}
\label{fig:best_fitting_model_sxxd}
\end{figure}

 Our SXD data cover the entire NIR (0.7--2.4 $\mu$m) range. We remove the low S/N regions spanning the 1.4 $\mu$m and 1.9 $\mu$m water bands of our medium-resolution spectrum, which are not used in the analysis. Each synthetic spectrum was smoothed to the resolving power of our medium-resolution SXD spectrum (\textit{R} $\sim$ 750) through convolution with a 1D Gaussian kernel. We normalize the synthetic spectra and our SXD spectrum at a common wavelength, re-sample the model grids to a common wavelength grid, and then optimally scale them to the flux-calibrated SXD spectrum by minimizing the $\chi^{2}$ value between the model and data.  We select the best-fitting physical parameters by locating the minimum $\chi^{2}$ value from our SXD (Figure \ref{fig:chi_atmosphericmodels}) spectral fit. For our SXD spectrum we find the best fit to be $T_\mathrm{{eff}}$ = 2800 $\pm$ 100 K and log $g$ = 4.0 $\pm$ 0.5 dex (Figure \ref{fig:best_fitting_model_sxxd} and Table \ref{tab:atmospheric_parameters}).
 
\begin{figure}
    \centering
    \includegraphics[width=1.1\columnwidth]{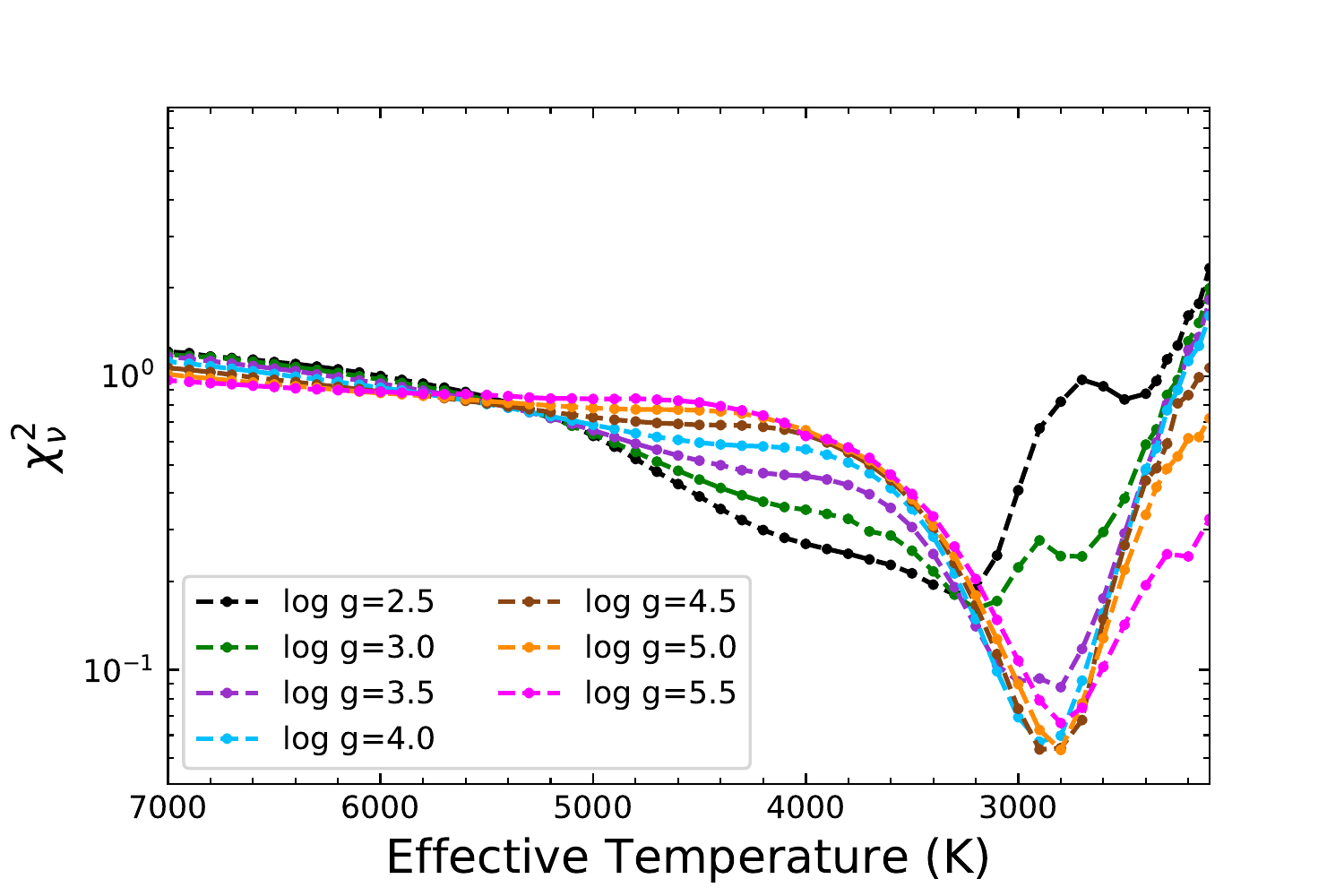}
    \caption{$\chi^{2}_{\nu}$  values for fits to our IGRINS $H$-band spectrum values across the effective temperature range of the BT-Settl \enquote{CIFIST 2011-2015} model grid. There is a convergence to a minimum at a temperature of 2900 K and log $g$ of 4.5 dex.}
    \label{fig:chi_atmospherIGRINS}
\end{figure}

 We use the \textit{H}-band region (1.550--1.72 $\mu$m) from our higher-resolution IGRINS (\textit{R} $\sim$ 45,000) spectrum of 2M0443+3723 B for our model fits. This portion of the  \textit{H}--band  spectrum comprises 14 separate orders and contains the temperature-sensitive Al and Fe lines \footnote{ We employ the $H$-band IGRINS spectrum because it has more temperature sensitive lines (Fe and Al), which are not degenerate with gravity like those in the K-band \citep{ricardo2019}.}.  We fit the model grids to each separate order to determine the temperature and gravity, then adopt the average values across 14 orders. We select the best-fitting physical parameters by locating the minimum $\chi^{2}$ value  from our IGRINS spectral fit (Figure \ref{fig:chi_atmospherIGRINS}).
  The best fitting parameters from the IGRINS $H$-band region fit are $T_\mathrm{{eff}}$ = 2900 $\pm$ 100 K and log $g$ = 4.5 $\pm$ 0.5 dex for the 1.55--1.72 $\mu$m region.
  \par
  The SpeX and IGRINS spectra both produce similar parameters from the model fits within 1$\sigma$ error from each other. The IRTF/SpeX  spectrum samples a wider spectral grasp (the entire NIR region) compared to the \textit{H}-band IGRINS spectrum, so we adopt those best fitting parameters for this study. 


\begin{table*}
    \centering
     \caption{Summary of the best-fitting parameters found from our fits of the near-infrared spectra of 2M0443+3723 B using the BT-Settl \enquote{CIFIST 2011-2015} grids. }
    \begin{tabular}{c c c c c c c}
         \hline \hline
         Object&Telescope/&Region&T$_\mathrm{{eff}}$&log(g)&$\chi_{\nu}^{2}$&$\chi^{2}$\\
           &Instrument&&(K)&(dex)\\
           \hline
           
      2M0443+3723 B&IRTF/SpeX (SXD)&0.68--2.4 $\mu$m&2800&4.0&0.016&37.35\\
       2M0443+3723 B&DCT/IGRINS&1.550--1.72 $\mu$m&2900&4.5&0.052&$\cdots$\\
       
        \hline

    \end{tabular}
    \tablecomments{Results from fitting the BT-Settl \enquote{CIFIST 2011--2015} atmospheric model grids to 2M0443+3723 B.  The best solutions are shown for each of the regions used. Note that the $\chi^{2}$ for the SpeX fit is 37.35 which produce best fit parameters of $T_\mathrm{{eff}}$ = 2800 K and log $g$ = 4.0 dex.}
    \label{tab:atmospheric_parameters}
    \end{table*}
 \subsection{Bolometric Luminosity}
 \label{sec:bolometric_luminosity}
 The bolometric luminosity of 2M0443+3723 B is derived using the BT-Settl atmospheric models \citep{allard2012}  for bolometric corrections at short wavelengths ($\lambda$ $<$ 0.68 $\mu$m) and long wavelengths (2.4--500 $\mu$m).  We flux calibrate the $T_{\mathrm{eff}}$ = 2800 K and log $\textit{g}$ = 4.0 dex BT-Settl model using the 2MASS \textit{H}--band apparent magnitude of 2M0443+3723 B and  scale the synthetic spectrum  to the data using the optimal scaling factor (\textit{C$_{H}$}) following \cite{cushing2008}:
 \begin{equation}
 C_{{H}}=10^{-0.4m_{H}}\frac{\int \lambda f_{\lambda}^{Vega}{T}_{H}(\lambda) \textit{d} \lambda}{\int \lambda f_{\lambda}^{obs}(\lambda){T}_{H}(\lambda) \textit{d} \lambda}.
 \end{equation}

 \noindent Here ${T}_{H}(\lambda)$ is the transmission profile of the 2MASS \textit{H}-band, $f_{\lambda}^{Vega}$ is the  flux density for Vega, $\lambda$ is the wavelength array for 2M0443+3723 B,  $f_{\lambda}^{obs}$ is the flux density of the science target, and $m_{H}$ is the 2M0443+3723 B \textit{H}--band magnitude from 2MASS. 
 \par
The bolometric luminosity is then $$L_{bol}=4\pi d^{2} \int_{0.125\mu m}^{500 \mu m} F_{\lambda} d\lambda,$$

\par
\noindent where $F_{\lambda}$ is the flux calibrated spectrum  and $d$ represents the  distance to 2M0443+3723 B. The bolometric flux we find for 2M0443+3723 B is 4.17 $\times$ 10$^{-14}$ W m$^{-2}$. We utilize a  Monte Carlo method to calculate the bolometric luminosity  by taking into account the uncertainties from the 2MASS \textit{H}-band photometry, measurement errors from our SXD spectrum of  2M0443+3723 B, and the distance. A distance of 72.4 $\pm$ 0.8 pc is used from the \textit{Gaia} DR2 parallax measurement for 2M0443+3723 B. This yields a bolometric luminosity of $L_\mathrm{bol}$=  --2.16 $\pm$ 0.02 dex.
\subsection{Mass}
Mass is a fundamental property  used to distinguish brown dwarfs from gas giants and low--mass stars. Evolutionary models  are generally necessary to infer masses using bolometric luminosities and ages of substellar objects (e.g. \citealt{burrows2001}). Assuming membership in the $\beta$MPG (see Section \ref{sec:membership_analysis}), we adopt an age of 23 $\pm$ 3 Myr \citep{mamajek&bell2014} for 2M0443+3723 B. The mass is then determined following a Monte Carlo approach with 10$^{6}$ trials using the bolometric luminosity and age for 2M0443+3723 B. 
The median and standard deviation of this mass distribution  is 99 $\pm$ 5 M$_\mathrm{Jup}$ if 2M0443+3723 B is single. We also derived the mass for scenarios in which this companion is not single (Section \ref{sec:is_2M0043 a binary}) or is not a member of the $\beta$PMG (Section \ref{sec:membership_analysis}). We find a mass of  52 $\pm$ 3 M$_\mathrm{Jup}$ if it is an equal-flux binary, and 30--110 M$_\mathrm{Jup}$ if it is a single young field object ($\lesssim$ 30 Myr).
 \begin{figure}
     \includegraphics[width=1.1\columnwidth]{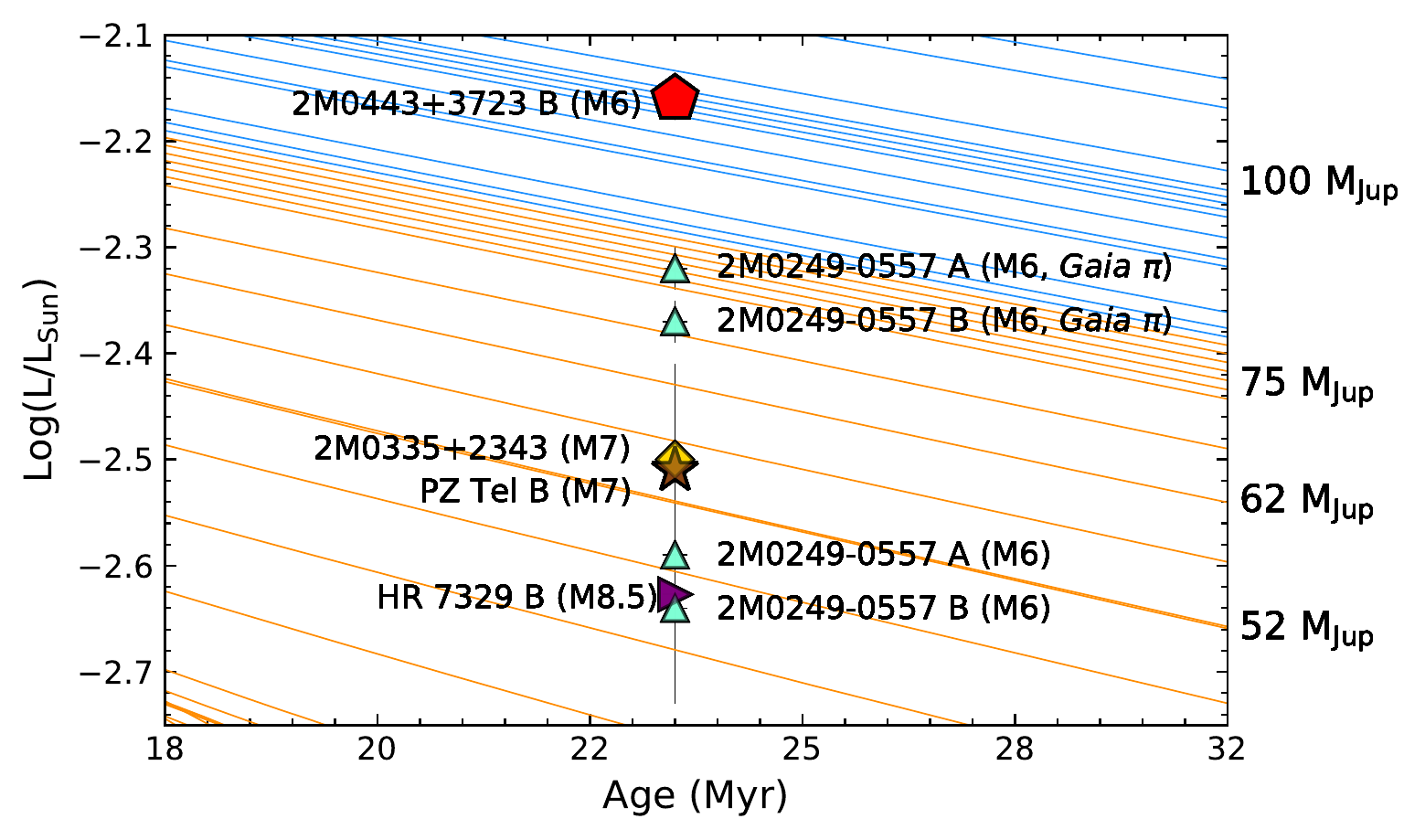}
     \caption{Late M-type members of the $\beta$PMG compared to evolutionary \enquote{iso-mass} model tracks from \cite{burrows2001}. The orange lines represent brown dwarfs and the blue lines represent low--mass stars.  
     Known brown dwarf companions in the $\beta$PMG are labeled and shown alongside 2M0443+3723 B (red pentagon). Note that we also include the 2M0249-0557 AB bolometric luminosity values  found by \cite{dupuy2018} using the $Gaia$ DR2 parallax (aquamarine triangles) in addition to the values from the Hawaii Infrared Parallax Program. PZ Tel B (brown star) and HR 7329 B (purple triangle) are clearly within  the brown dwarf regime, while 2M0443+3723 B lies near the hydrogen-burning limit, which is surprising if it is a single member of this group because it has a similar spectral type as 2M0249-0557 AB and PZ Tel B. }
     \label{fig:mass_evolutionary}
 \end{figure}
\label{sec:mass}
 
\begin{table*}
  
    \caption{Properties of the 2M0443+3723 AB System}

    \centering
    \begin{tabular}{c c c c c}
    \hline \hline
         Property&2M0443+3723 A&&2M0443+3723 B&References\\
         \hline
         &Positions and Kinematics\\
         \hline
         R.A. (Degrees)&70.98707&&70.98975&6 \\
         Dec. (Degrees)&37.38400&&37.38394& 6 \\
         Parallax (mas)&13.9572$\pm0.0518$&&13.8198$\pm$0.1585& 6 \\
         Distance (pc)\tablenotemark{a}&71.6$\pm$0.3  &&72.4$\pm$0.8&1 \\
         $\mu_{\alpha}$cos$\delta$ (mas yr$^{-1}$)&22.869$\pm$0.097&&24.204$\pm$0.261& 6 \\
         $\mu_{\delta}$ (mas yr$^{-1}$)&--61.837$\pm$0.060&&--61.670$\pm$0.156& 6 \\
         RV (km s$^{-1}$)&7.0$\pm$0.2&&5.8$\pm$0.2& 1 \\
         vsin$i$ (kms$^{-1}$)&13.3$\pm$0.8&&15.2$\pm$0.8& 1 \\
         $X$ (pc)&--69.07$\pm$0.25&&--69.75$\pm$0.84& 1 \\
         $Y$ (pc)&17.72$\pm$0.08&&17.89$\pm$0.24& 1 \\
         $Z$ (pc)&--6.92$\pm$0.05&&--6.98$\pm$0.15& 1 \\
         $U$ (km s$^{-1}$)&--11.25$\pm$0.19&&--10.23$\pm$0.20& 1 \\
         $V$ (km s$^{-1}$)&--18.79$\pm$0.09&&--19.53$\pm$0.24& 1 \\
         $W$ (km s$^{-1}$)&--8.41$\pm$0.04&&--7.98$\pm$0.121& 1  \\

         \hline
         &Photometry\\
         \hline 
         
         $J_\mathrm{{2MASS}}$ (mag)&9.71$\pm$0.02&&12.22$\pm$0.03& 2 \\
         $H_\mathrm{{2MASS}}$ (mag)&9.03$\pm$0.02&&11.80$\pm$0.04& 2 \\
         $K_\mathrm{{S,2MASS}}$ (mag)&8.80$\pm$0.02&&11.46$\pm$0.03& 2 \\
         $B$ (mag)&15.35&&$\cdots$ &10 \\
         $V$ (mag)&13.3$\pm$0.09&&$\cdots$&9 \\
         $G_{Gaia}$ (mag)&12.327$\pm$0.002&&16.160$\pm$0.001& 6 \\
          $g$ (mag)&14.00$\pm$0.09&&$\cdots$&9  \\
         $r$ (mag)&12.68$\pm$0.06&&$\cdots$&9  \\
         $i$ (mag)&11.56$\pm$0.12&&$\cdots$&9  \\
         $W1$ (mag)&8.66$\pm$0.02&&11.92$\pm$0.07&3 \\
         $W2$ (mag)&8.56$\pm$0.02&&10.93$\pm$0.05&3 \\
         $W3$ (mag)&8.35$\pm$0.02&&10.56$\pm$0.13&3 \\
         $W4$ (mag)&7.551$\pm$0.16&&8.191&3 \\
         
         $M_{J,\mathrm{2MASS}}$ (mag)&5.41$\pm$0.03&&7.92$\pm$0.03&1 \\
         $M_{H,\mathrm{2MASS}}$ (mag)&4.73$\pm$0.003&&7.50$\pm$0.04&1 \\
         $M_{K,\mathrm{2MASS}}$ (mag)&4.502$\pm$0.04&&7.16$\pm$0.03&1 \\

         $J$--$H$ (mag)&0.68&&0.42&1  \\
         $H$--$K$ (mag)&0.228&&0.35&1  \\
         $J$--$K$ (mag)&0.908&&0.77&1 \\
         \hline
         &Fundamental Properties\\
         
         \hline
         log($\frac{L_{bol}}{L_{\odot}}$)(dex)&0.73$\pm$ 0.02&&--2.16$\pm$0.02 &4,1\\
         Spectral Type&M2$\pm$1&&M6$\pm$1&5,1\\
         Mass (M$_{\mathrm{Jup}})$\tablenotemark{b}&$\cdots$&&99$\pm$5 & 1                             \\
         Age (Myr)\tablenotemark{b}&23$\pm$3&&23$\pm$3&7\\
         $T\mathrm{_{eff}}$ (K)&3700&&2800$\pm$100&8,1\\
         log $g$ (dex)&5.0&&4.0$\pm$0.5&8,1\\
         Separation ($\arcsec$)&$\cdots$&&7$\farcs$6&5\\
         Separation (AU)&$\cdots$&&550&5\\
         Radius (R$_{\odot}$)&0.68$\pm$0.22&&0.35$\pm$0.03&4,1
    \\
    \hline
 \end{tabular}
 \tablenotetext{a}{\cite{bailer-jones} find consistent values of 71.5 $\pm$ 0.26 pc and 72.4 $\pm$ 0.8 pc for 2M0443A and 2M0443B respectively using probabilistic inference from parallax measurements. }
 \tablenotetext{b}{Masses and ages assume membership in the $\beta$PMG.
 }
    
    \tablerefs{(1) This work (2) 2MASS \citep{cutri2003}, (3) $\textit{WISE}$ \citep{cutri2012}, (4) \cite{messina2017}, (5) \cite{schlieder2010}, (6) $Gaia$ \citep{gaia}, (7) \cite{mamajek&bell2014}, (8) \cite{malo}, (9) APASS \citep{zacharias2012}, (10) \cite{norton2007}}
 
    \label{tab:summary of 2M0443+3723AB}
   
\end{table*}

\subsection{Radial Velocities}
\label{sec:radial_and_rotational_velocities}

We measure radial velocities (RV) of  2M0443+3723 A and 2M0443+3723 B  following  \cite{mann2016}. We cross-correlate $>$250 order segments of the IGRINS ($R$ $\sim$ 45,000)  $H$ and $K$-band spectra. For each segment we cross-correlate the telluric spectrum to find offsets in the wavelength solution between epochs of observation, and the target spectrum pixel offset was converted into a radial velocity using the instrument dispersion solution. The measured radial velocity is the median of the $>$250 segment measurements compared with $>$150 M2--M6 templates with known RVs. Our reported RV was barycenter corrected and shifted to the absolute scale using the radial velocities of the M2--M6 templates. Uncertainties in the radial velocities are the standard deviation of the mean added in quadrature with the absolute scale zero-point uncertainty of $\sim$150 m s$^{-1}$. For 2M0443+3723 A we find a RV$_{A}$ =  7.0 $\pm$ 0.2 km s$^{-1}$. For 2M0443+3723 B we find a  RV$_{B}$ =  5.8 $\pm$ 0.2 km s$^{-1}$.
These values are similar to previous measurements of the host star, 2M0443+3723 A: the radial velocity  of 2M0443+3723 A has been measured to be  6.0 $\pm$ 2.0 km s$^{-1}$ from \cite{schlieder2010}, and 6.4 $\pm$ 0.3 km s$^{-1}$ from \cite{shkolnik2017}.

\newpage
 \subsection{Radius}
 The radii of brown dwarfs can provide  additional evidence of youth. During the course of their lifetimes, brown dwarfs continuously contract as they dissipate their leftover energy from formation. As brown dwarfs age their radii eventually settle at  $\sim$ 1 R$_\mathrm{Jup}$ as degeneracy pressure sets in. As a result, an inflated radius can serve as a youth indicator  \citep{filippazzo2015}.
 \par
 We previously determined an effective temperature of 2800 $\pm$ 100 K from atmospheric models and a bolometric flux ($F_\mathrm{{Bol}}$) from  integrating our flux-calibrated spectrum.  We also have a precise measured distance from ${Gaia}$ DR2. The Stefan-Boltzmann law can be used to determine the radius of 2M0443+3723 B:  $$ R=\Big(\frac{{F}_{\mathrm{Bol}} \textit{d}^{2}}{\sigma T^{4}}\Big)^{\frac{1}{2}}.$$
 
\noindent Radius uncertainties are computed using a Monte Carlo approach with the associated errors from the effective temperature, bolometric flux, and distance. We find an inflated radius  0.35 $\pm$ 0.03 R$_{\odot}$ (3.5 R$_\mathrm{Jup}$), an additional indication that 2M0443+3723 B is young. For comparison, the old field object 2MASS J02530084+1652532  has a radius of  1.20 $\pm$ 0.09 R$_\mathrm{Jup}$ with a spectral type of M7 and T$_\mathrm{{eff}}$ = 2688 $\pm$ 212 K \citep{filippazzo2015}.


\section{Discussion}
\label{sec:discussion}
\subsection{Moving Group Membership Assessment}
\label{sec:membership_analysis}
\begin{figure*}[htp]
    \centering
     \vspace{-0.1in}
    \includegraphics[width=.3\textwidth]{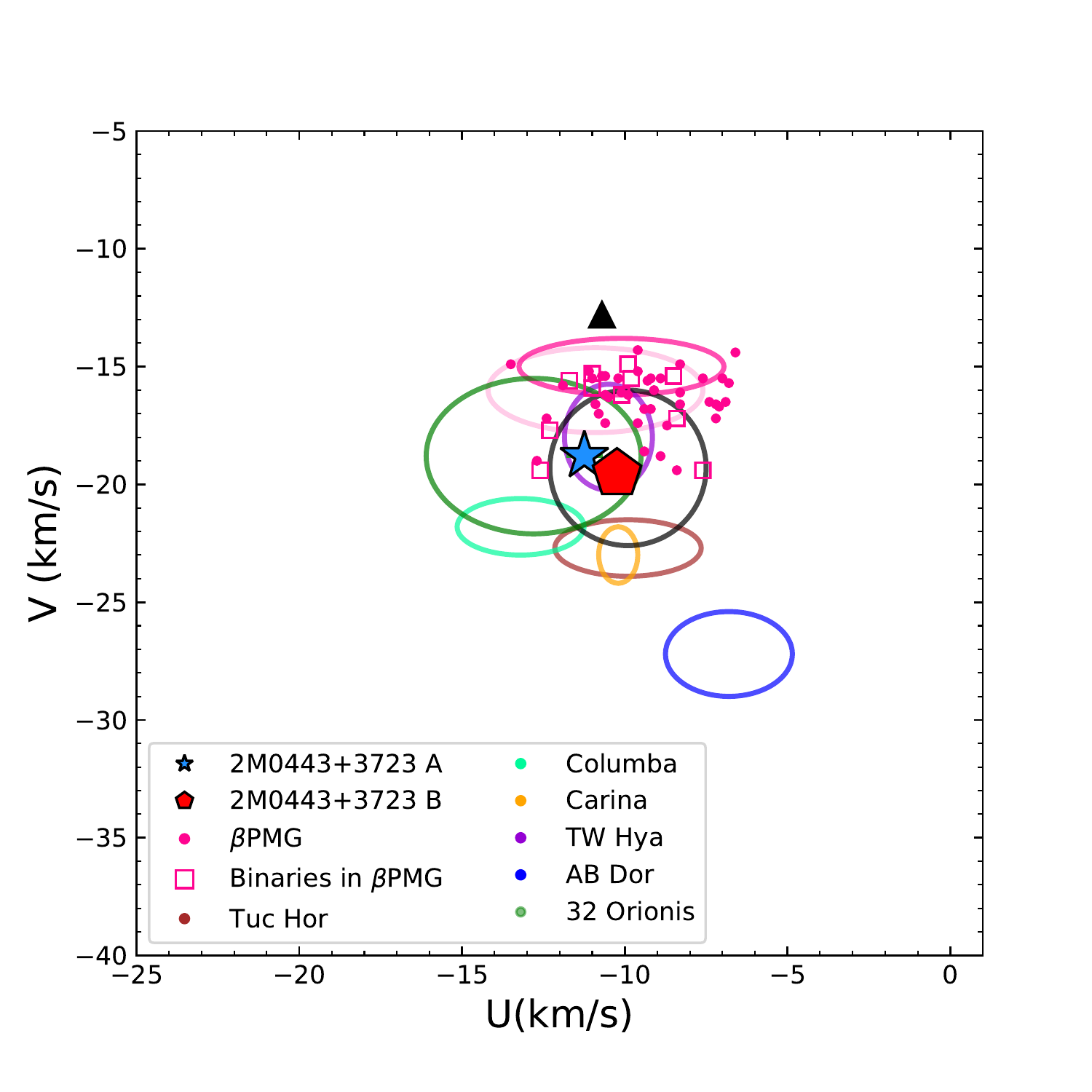}
    \includegraphics[width=.3\textwidth]{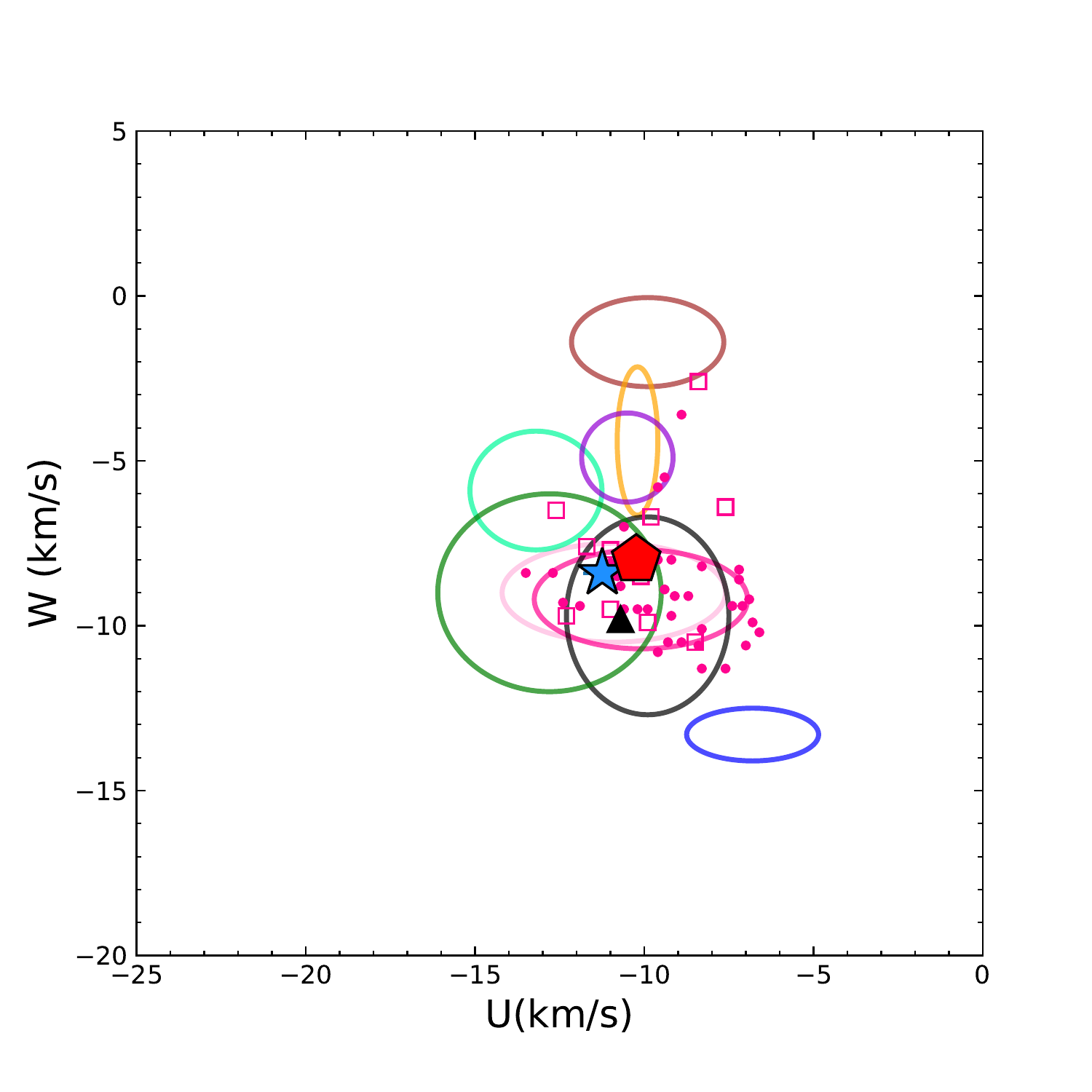}
    \includegraphics[width=.3\textwidth]{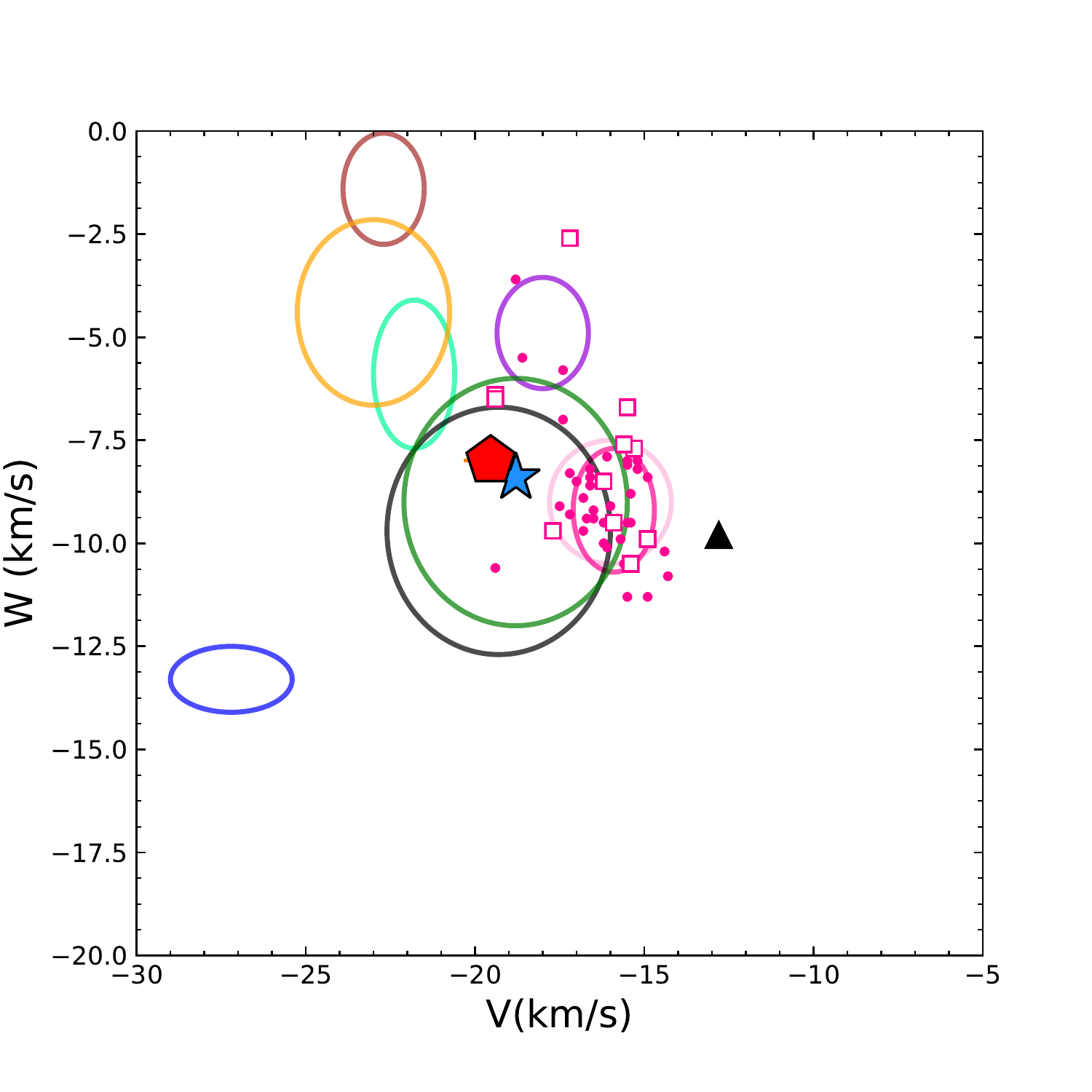}
    \includegraphics[width=.3\textwidth]{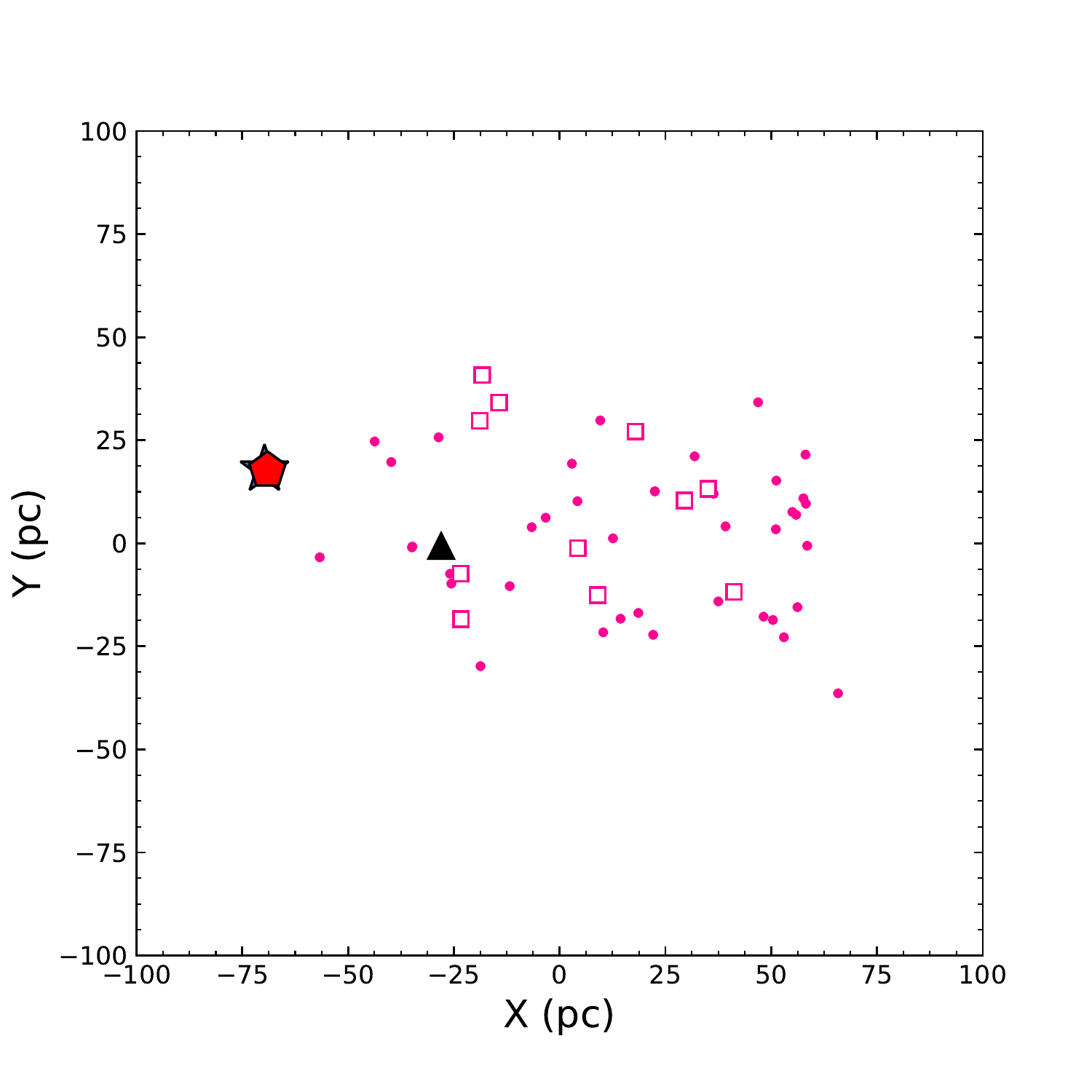}
    \includegraphics[width=.3\textwidth]{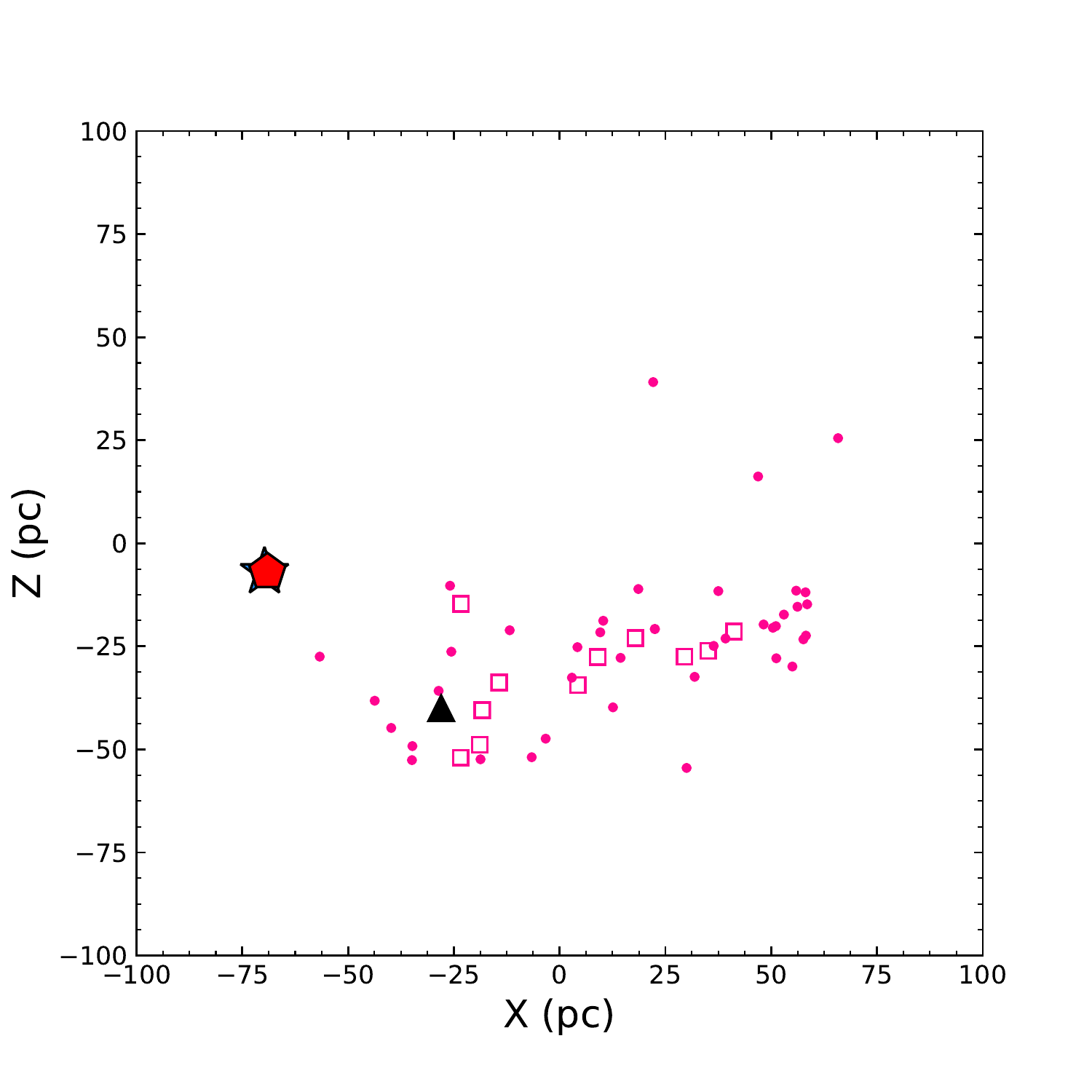}
    \includegraphics[width=.3\textwidth]{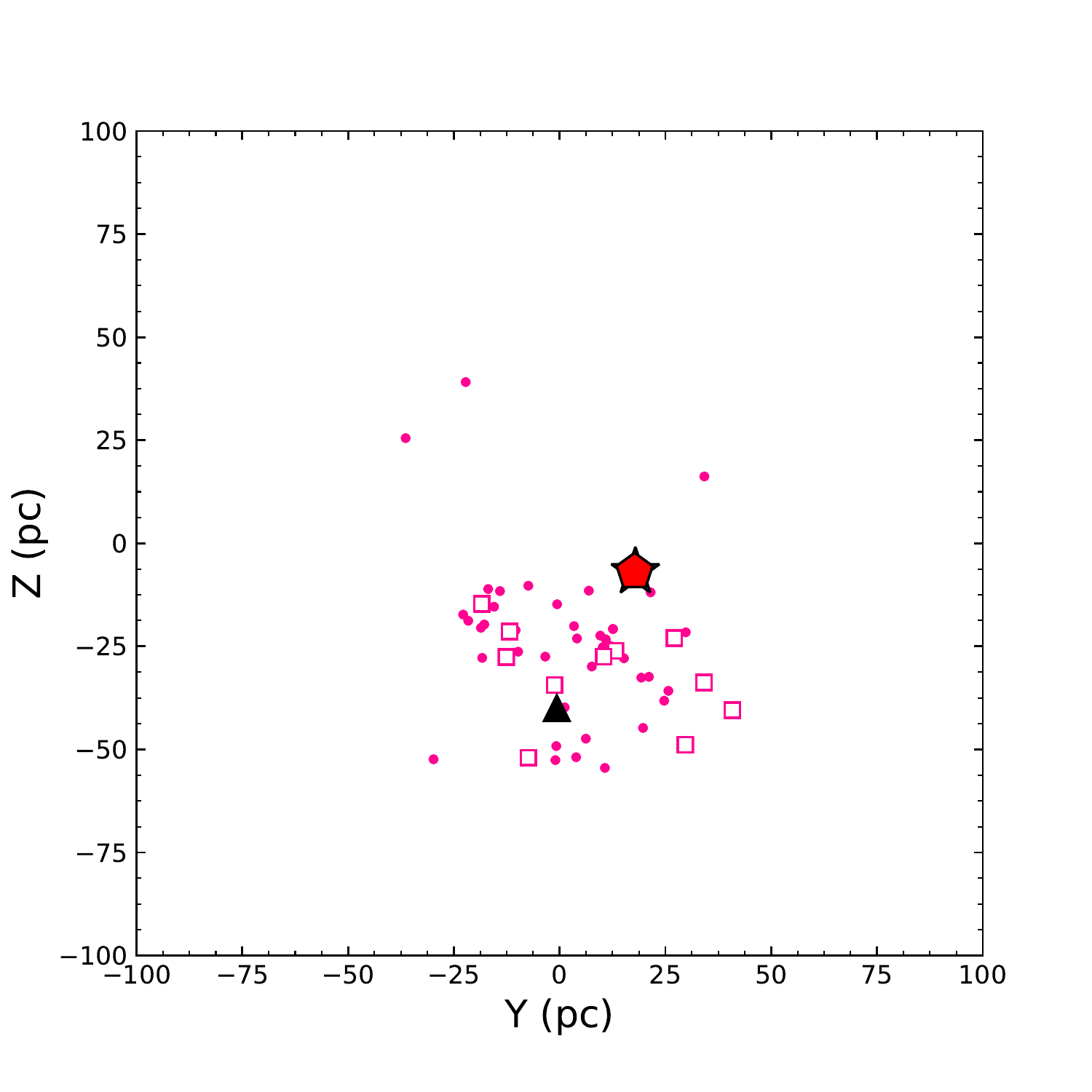}
    \caption{\textit{UVW} galactic space velocities and $XYZ$ heliocentric positions of  2M0443+3723 A and 2M0443+3723 B (blue star and red pentagon respectively) and those of nearby young moving groups. The 3$\sigma$ dispersion in $UVW$ velocities for the $\beta$PMG from \cite{torres2008} is shown with dark pink circles and from \citealt{gagne2018} is shown in light pink for comparison.
$\beta$PMG members are from \cite{shkolnik2017}. These are the available M dwarf members  with confirmed membership in this group (pink circles). Those designated with a \enquote{N}  or \enquote{Y?} from that study have been excluded. Binaries from the $\beta$PMG  are shown as pink squares and the  substellar companion, 2MASS J0249-0557 c, is shown as a black triangle. Ellipses represent the 3$\sigma$ confidence limits of the moving groups. $\beta$PMG members are  generally clustered together in \textit{UVW} space, but are dispersed over many tens of  parsecs as shown in the $XYZ$ galactic position plots. If 2M0443+3723 AB is a member of this group then it is a kinematic and spatial outlier relative to the current population.
 }
\label{fig:young Moving Groups_UVW_XYZ}
\end{figure*}

BANYAN (Bayesian Analysis for Nearby Young AssociatioNs) $\Sigma$ is a tool developed by \cite{gagne2018} that analyzes galactic positions and  space velocities to determine membership probabilities in nearby young associations. BANYAN $\Sigma$ achieves a 90$\%$ recovery rate of known members if full kinematic parameters are provided.  \cite{gagne2018} use proper motions, radial velocities, parallaxes, and  sky coordinates to determine the likelihood that an object belongs to  any of the 27 associations spanning ages 1--800 Myr  or the field population. 
\par
The 2M0443+3723 AB system has been proposed as a member of the $\beta$PMG in several studies (\citealt{schlieder2010}; \citealt{malobanyan}; \citealt{shkolnik2017}).
Recently, \cite{leesong2019} developed BAMG (Bayesian Analysis of Moving Group), a four stage moving group membership tool. This new tool lists 2M0443+3723 A as a highly likely member with a 86.2$\%$ membership probability for the $\beta$PMG. Yet,  BANYAN $\Sigma$  yields a membership probability  for 2M0443+3723 A of 0.4$\%$ for $\beta$PMG and 99.6$\%$ for the field. This discrepancy may be a result of restrictive kinematic and spatial priors from BANYAN $\Sigma$ and/or the iterative membership approach used by \cite{leesong2019}.
For 2M0443+3723 B, BANYAN $\Sigma$ gives 0.0$\%$ for the $\beta$PMG  and 99.9$\%$ for the field. 
\par
Using updated astrometric and kinematic data from \textit{Gaia} DR2, we  determine the $XYZ$ positions and \textit{UVW} space motions of the pair along with uncertainties (Figure \ref{fig:young Moving Groups_UVW_XYZ} and Table \ref{tab:summary of 2M0443+3723AB}). 
The $\beta$PMG has average $XYZ$ positions of $X$,$Y,$$Z$$_{\beta PMG}$ = \{4.1 $\pm$ 29.3, --6.7 $\pm$ 14.0, --15.7 $\pm$ 9.0 pc\}, and space velocities $U$,$V$,$W$$_{\beta PMG}$ = \{--10.9 $\pm$ 2.2, --16.0 $\pm$ 0.3, --9.2 $\pm$ 0.3 km s$^{-1}$\}\citep{gagne2018}.
\par

\begin{figure}
    
     \includegraphics[width=1.1\columnwidth]{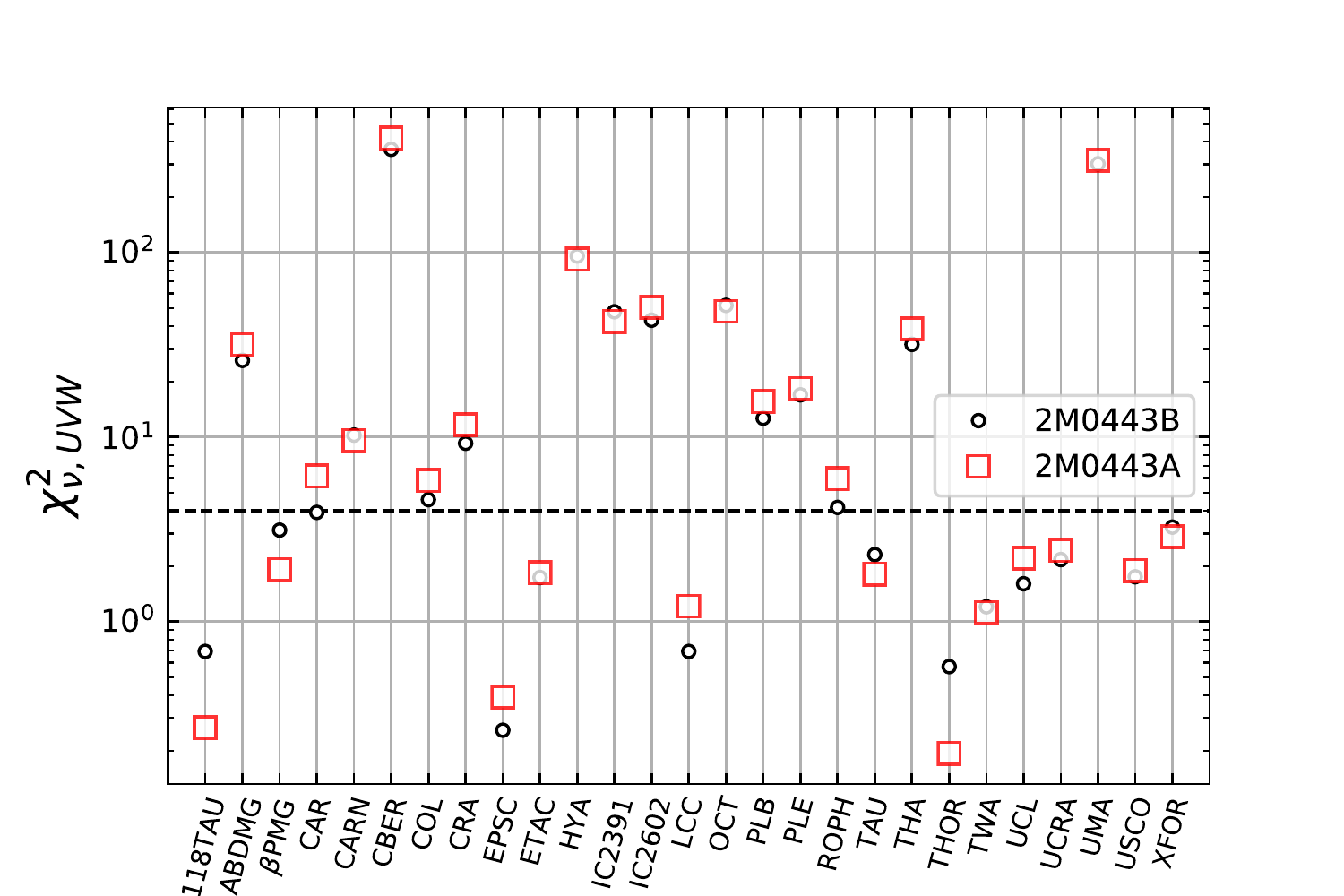}
     \caption{$\chi_{\nu,UVW}^{2}$ values for 27 known groups used in BANYAN $\Sigma$ \citep{gagne2018}. The full names of young associations are: 118 Tau (118TAU), AB Doradus (ABDMG), $\beta$ Pictoris ($\beta$PMG), Carina (CAR), Carina-Near (CARN), Coma Berenices (CBER), Columba (COL), Corona Australis (CRA), $\epsilon$ Chamaeleontis (EPSC), $\eta$ Chamaeleontis (ETAC), the Hyades cluster (HYA), Lower Centaurus Crux (LCC), Octans (OCT), Platais 8 (PL8), the Pleiades cluster (PLE), $\rho$ Ophiuchi (ROPH), the Tucana-Horologium association (THA), 32 Orionis (THOR), TW Hya (TWA), Upper Centaurus Lupus (UCL), Upper CrA (UCRA), the core of the Ursa Major cluster (UMA), Upper Scorpius (USCO), Taurus (TAU), and $\chi^{1}$ For (XFOR). 2M0443+3723 B is shown as black circles and 2M0443+3723 A is shown with red squares. Several groups are good $UVW$ matches to 2M0443+3723 AB but disagree when distance and declination are considered. Altogether,  2M0443+3723 AB is most consistent with the $\beta$PMG, although if it is a member it would be a kinematic outlier ($\Delta \nu$ = 3.7 km s$^{-1}$).}
     \label{fig:chi_squared_UVW}
 \end{figure}

We explored another quantitative approach to determine membership properties by comparing the 3D velocity difference between 2M0443+3723 AB and known young moving groups using a reduced $\chi^{2}$ metric following \cite{shkolnik2012}; see their Equation 14. This method allows for a membership  examination that solely relies on galactic velocities and neglects $XYZ$ parameters which can affect Bayesian methods if the prior is too restrictive. Here we adopt a cutoff value of $\chi_{\nu,UVW}^{2}$ $\leq$ 4 for consideration. A visualization of this method is shown in Figure  \ref{fig:chi_squared_UVW}. For each group we calculate the following two goodness-of-fit metrics: 

\begin{equation}
\begin{split}
    \tilde{\chi}{_{UVW}^{2}}=\frac{1}{3}\Bigg[\frac{(U_{\star}-U_{MG})^{2}}{(\sigma_{U\star}^{2}+\sigma_{U,MG}^{2})}+\frac{(V_{\star}-V_{MG})^{2}}{(\sigma_{V\star}^{2}+\sigma_{V,MG}^{2})}\\
     +\frac{(W_{\star}-W_{MG})^{2}}{(\sigma_{W\star}^{2}+\sigma_{W,MG}^{2})}\Bigg]
    \label{eq:chi_uvw}
\end{split}
\end{equation}
\begin{equation}
\begin{split}
   \tilde{\chi}{_{XYZ}^{2}}=\frac{1}{3}\Bigg[\frac{(X_{\star}-X_{MG})^{2}}{(\sigma_{X\star}^{2}+\sigma_{X,MG}^{2})}+\frac{(Y_{\star}-Y_{MG})^{2}}{(\sigma_{Y\star}^{2}+\sigma_{Y,MG}^{2})}\\+\frac{(Z_{\star}-Z_{MG})^{2}}{(\sigma_{Z\star}^{2}+\sigma_{Z,MG}^{2})}\Bigg]
     \label{eq:chi_xyz}
\end{split}
\end{equation}

\noindent Here the subscript $\star$ represents the values for the object in question and the subscript $MG$ represents a selected moving group.  \cite{shkolnik2012} assumed a constant velocity dispersion ($\sigma$) of 2 km s$^{-1}$ for the $UVW$ motions to avoid biasing their results in favor of moving groups with  larger velocity dispersions. Here we use the locus and dispersion for each of the 27 associations included in \cite{gagne2018} in this analysis.
\par
\begin{figure}
     \includegraphics[width=1.1\columnwidth]{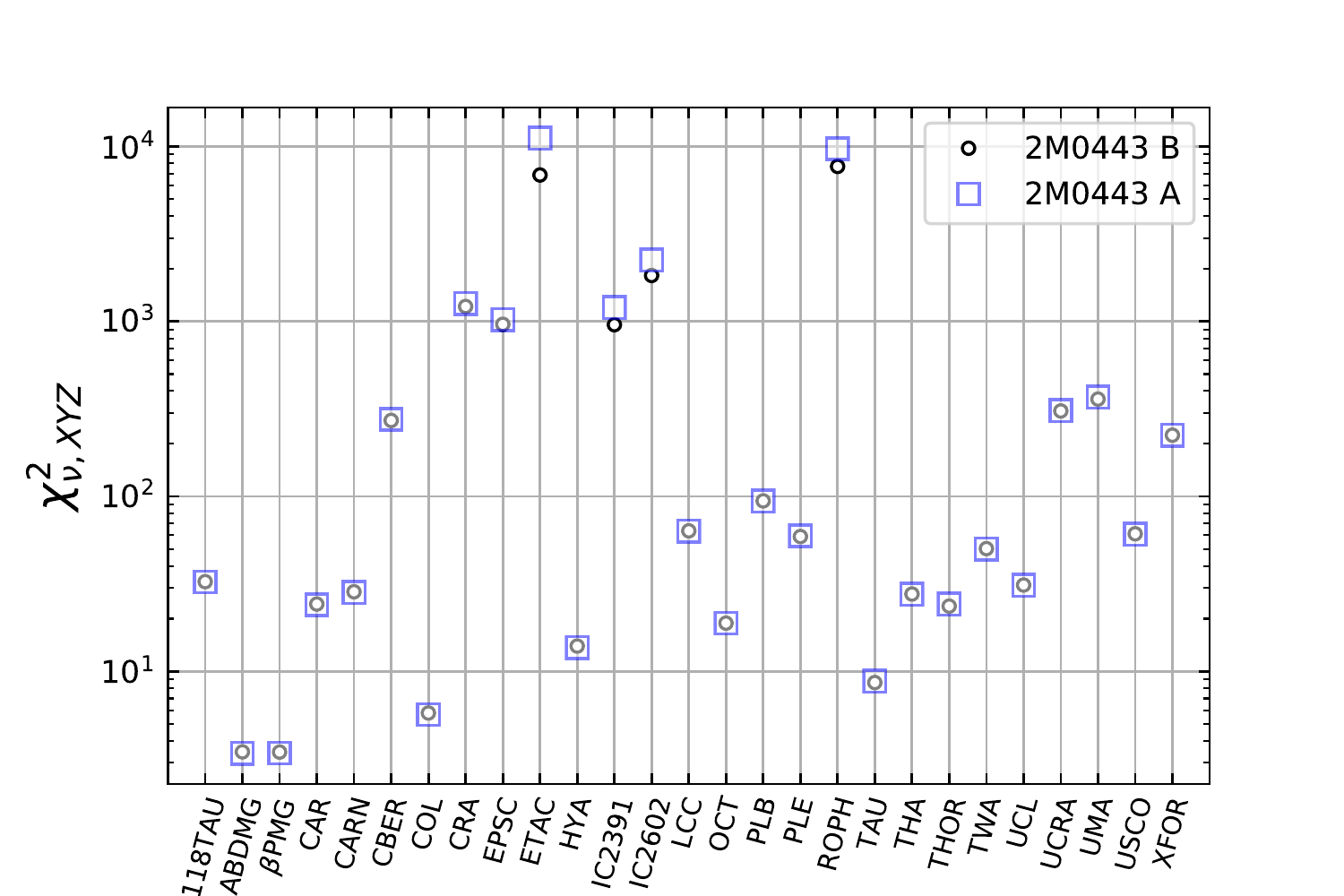}
     \caption{$\chi_{\nu,XYZ}^{2}$ values for 27 known groups used in BANYAN $\Sigma$.  
     Groups names are the same as Figure \ref{fig:chi_squared_UVW}.} 
     \label{fig:chi_square_xyz}
 \end{figure}

12 associations have  $\chi_{\nu,UVW}^{2} \leq$ 4: 118TAU, $\beta$PMG, EPSC, ETAC, LCC, TAU, THOR, TWA, UCL, UCRA, USCO, and XFOR. Nearly all of these have inconsistent distances or sky positions relative to 2M0443+3723 AB (Figure \ref{fig:chi_square_xyz}). 
 2M0443+3723 A has the best match to the THOR (32 Orionis) association ($\chi_{\nu, UVW}^{2}$ = 0.6). THOR  has an age of $\sim$ 22 Myr \citep{bell2015}, and a typical distance  of 96 $\pm$ 2 pc. This disagrees with the $Gaia$ parallactic distance of  72.4 $\pm$ 0.8 pc.
 The  EPSC ($\epsilon$ Chamaeleontis) group has an average age of $\sim$ 3.7 Myr \citep{murphy2013} and lies at  an average distance  of 102 $\pm$ 4 pc primarily in the southern hemisphere.  This distance excludes EPSC  even though this group has the lowest $\chi_{\nu,UVW}^{2}$ for 2M0443+3723 B ($\chi_{\nu, UVW}^{2}$ = 0.26). 
 \par
TAU (Taurus) is a $\sim$ 1-2 Myr \citep{kenyonhartman1995} star forming region;  
its members lie at an average distance of 120 $\pm$ 10 pc which makes it too distant to host the 2M0443+3723 AB system. 118 TAU is a young $\sim$ 10 Myr association located at an average distance of 145 $\pm$ 15 pc \citep{galli2018}, which is  likewise inconsistent with the distance of 2M0443+3723 AB (72.4 $\pm$ 0.8 pc). TWA (TW Hya) has a younger age of $\sim$ 10 Myr \citep{bell2015}. This system would be a probable match for 2M0443+3723 AB given the distance distribution (60 $\pm$ 10 pc),  but its members are tightly clustered in the southern hemisphere.
\par

At $\sim$ 500 Myr \citep{pohnlpaunzen2010}, XFOR is one of the oldest  associations 
located at an average distance of 100 $\pm$ 6 pc. Similarly, UCRA and ETAC are located too far south. The Sco-Cen star-forming region hosts UCL, LCC, and USCO; it can collectively be eliminated due to the distances of $\sim$ 110--154 pc.

 The velocity modulus (Equation \ref{eq:velocity_modulus}) can also be calculated to further assess membership probabilities for nearby young moving groups:
 \begin{equation}
    \Delta \nu=\sqrt{(U_{\star}-U_{MG})^{2}+(V_{\star}-V_{MG})^{2}+(W_{\star}-W_{MG})^{2}}.
    \label{eq:velocity_modulus}
 \end{equation}
 
\begin{figure}
    \centering
    \includegraphics[width=1.1\columnwidth]{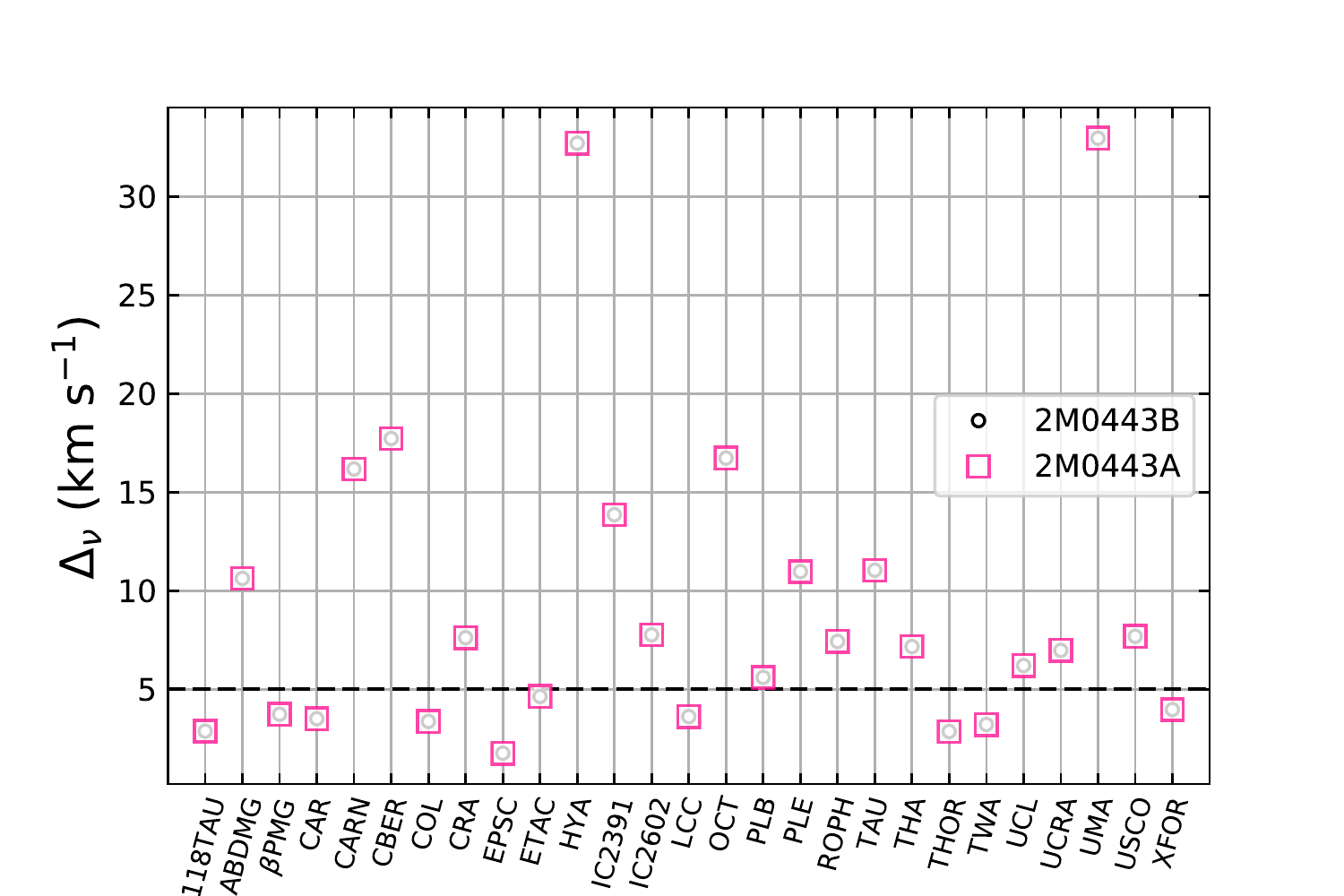}
    \caption{Velocity  modulus of 27 known moving groups and associations. A cutoff value of 5 km s$^{-1}$ is shown with a dotted line.}
    \label{fig:velocity_modulus}
\end{figure}

\noindent The advantage of this metric is that it accounts for moving groups with large \textit{UVW} velocity dispersions by just focusing on the velocity locus. We adopt a cutoff $\Delta \nu$ of 5 km s$^{-1}$ (following \citealt{shkolnik2012}; Figure \ref{fig:velocity_modulus}). 10 associations have  $\Delta \nu \leq$ 5 km s$^{-1}$: 118TAU, $\beta$PMG, CAR, COL, EPSC, ETAC, LCC, THOR, TWA, and XFOR. As previously discussed, most of these associations are not a good match due to discrepancies with their distances declinations (Figure \ref{fig:chi_square_xyz}).
\par
For the $\beta$PMG, $\chi_{\nu, UVW}^{2}$ = 3.13 for 2M0443+3723 B and $\chi_{\nu, UVW}^{2}$ = 1.92 for 2M0443+3723 A. The $\chi_{\nu, XYZ}^{2}$  values  for 2M0443+3723 B and  2M0443+3723 A are 3.72 and 2.87, respectively. The $\beta$PMG does not have the lowest $\chi_{\nu, UVW}^{2}$ but we conclude that it is the most probable moving group to host 2M0443+3723 AB. Similarly, $\chi_{\nu, XYZ}^{2}$ and $\Delta \nu$ values for 2M0443+3723 AB
indicate that the $\beta$PMG is the most likely of the known moving groups to host 2M0443+3723 AB.

\subsection{Comparison to ultracool companions in the $\beta$ Pictoris Moving Group}

\begin{figure}
    \centering
    \includegraphics[width=1.1\columnwidth]{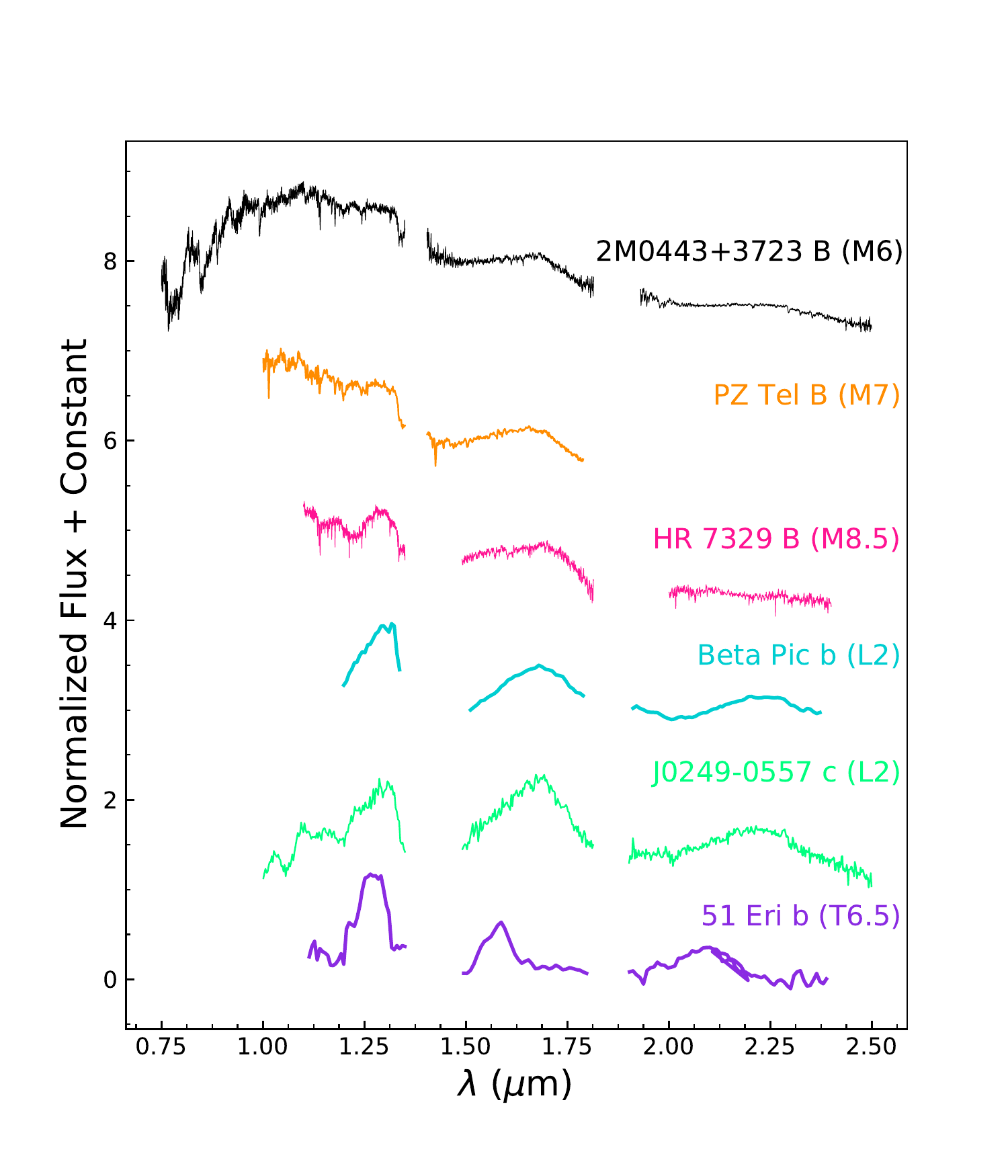}
    \caption{Spectral sequence  of known substellar companions in the $\beta$PMG  along with our medium-resolution SXD spectrum of the candidate benchmark brown dwarf, 2M0443+3723 B. PZ Tel B and HR 7329 B have masses above the deuterium-burning limit ($\sim$ 13 M$_\mathrm{Jup}$), while J0249--0557 c, $\beta$ Pic b, and 51 Eri b have masses below the deuterium-burning limit (\citealt{lowrance2000}; \citealt{lagrange2010}; \citealt{biller2010}; \citealt{mugrauer2010}; \citealt{macintosh2015}; \citealt{dupuy2018}). Spectra of known substellar companions are from \cite{bonnefoy2014}; \cite{maire2015}; \cite{chilcote2017}; \cite{dupuy2018}; and \cite{rajan2017}.}
  
    \label{fig:substellar_companions}
\end{figure}

Over one thousand isolated brown dwarfs are known, but brown dwarf companions are much rarer. The frequency of BDs that are companions to stars is 2--4$\%$; these occurrence  rates are independent of stellar mass or spectral type \citep{bowlerandnielsen2018}. To date there have only been a handful of ultracool companions discovered in the $\beta$PMG (see \citealt{dupuy2018} Table 5): HR 7329 B \citep{lowrance2000}, PZ Tel B (\citealt{biller2010}; \citealt{mugrauer2010}), $\beta$ Pic b \citep{lagrange2010}, 51 Eri b \citep{macintosh2015}, and 2MASS J0249-0557 c \citep{dupuy2018}. The spectral types of these companions range from M7--T6.5 and their masses span 11--72 M$_\mathrm{Jup}$.  The spectral sequence of these objects is shown in Figure \ref{fig:substellar_companions}. In addition, the membership compilation by \citealt{shkolnik2017} (their Table 4) lists one additional ultracool companion, the M6 object GJ 3076 B.\footnote{Note that both 2MASS~J01365516--0647379 and 2MASS~J08224744--5726530 are listed in Table 4 of \citealt{shkolnik2017} as having companions with spectral types $>$L0, but photometry from \cite{janson2012}  indicates that the companion to 2MASS~J08224744--5726530 is $\approx$M5.5, and deep imaging of  2MASS~J01365516--0647379 from \cite{bowler2015} did not reveal any comoving companions.}

\subsubsection{Planetary--Mass Companions:\\ 2MASS J0249-0557 c, $\beta$ Pic b and 51 Eri b}

There are currently three known planetary-mass companions in the $\beta$PMG: 2MASS J0249-0557 c, $\beta$ Pic b, and 51 Eri b. Here we provide a brief overview of each system.

\par
2MASS J0249-0557 c  \citep{dupuy2018} has a spectral type of L2 $\pm$ 1, luminosity of log(\textit{L/L$_\odot$})= --4.00 $\pm$ 0.09 dex, and an estimated mass of 11.6$_{-1.0}^{+1.3}$ M$_\mathrm{Jup}$. It lies at a separation of 40\arcsec (1950 AU) from its host, which is a tight equal-flux ratio brown dwarf binary with an integrated-light spectral type of M6 $\pm$ 1.
\par
51 Eri b \citep{macintosh2015} has a spectral type of T6.5 $\pm$ 1.5, an estimated mass of 2--12 M$_\mathrm{Jup}$, and an effective temperature of 605--737 K. \cite{rajan2017} report a bolometric luminosity of --5.83$_{-0.12}^{+0.15}$ to  --5.93$_{-0.14}^{+0.19}$ dex. 51 Eri b orbits its  massive F0\mathrmnum{4} host star at a separation of 0$\farcs$45 (13.2 AU).

\par
$\beta$ Pic b \citep{lagrange2010} has a spectral type of L2 $\pm$ 1 and orbits the namesake member of the $\beta$PMG, the A6V type star, $\beta$ Pic, at a separation of 0$\farcs$13 (9 AU; \citealt{lagrange2019}). \cite{dupuy2019} find a model independent dynamical mass of 13 $\pm$ 3 M$_\mathrm{Jup}$, comparable with results from \cite{snellenbrown2018}. \cite{chilcote2017} find an effective temperature of 1700 K and  a bolometric luminosity of log(\textit{L/L$_{\odot}$})= --3.76 $\pm$ 0.02 dex.  

\subsubsection{Brown Dwarf Companions: HR 7329 B and \\ PZ Tel B}

HR 7329 B \citep{lowrance2000} was the first brown dwarf companion  identified in the $\beta$PMG. It orbits at a  separation of 4\arcsec (220 AU) from its massive A0V host star, HR 7329 A. \cite{bonnefoy2014} derive a spectral type of M8.5 $\pm$ 0.5, with an effective temperature of 2200--2500 K, a luminosity of log(\textit{L/L$_\odot$})= --2.627 $\pm$ 0.087 dex, and a model-dependent mass of $\sim$ 20--50 M$_\mathrm{Jup}$ \citep{neuhauser2011}. 
\par
The substellar companion PZ Tel B (\citealt{biller2010}; \citealt{mugrauer2010}) has a separation of 0$\farcs$5 ($\sim$ 25 AU) from its G9\mathrmnum{4} host star. PZ Tel B has a spectral type of M7 $\pm$ 1, an effective temperature of 2700 $\pm$ 100 K, a luminosity of log(\textit{L/L$_\odot$})= --2.51 $\pm$ 0.10 dex, and a  model--dependent mass of 59$_{-8}^{+13}$ M$_\mathrm{Jup}$ --- well within the brown dwarf mass regime  \citep{maire2015}.

\subsection{Color Magnitude Diagrams}
\label{sec: color_magnitude_diagrams}
\begin{figure*}[h]
    \centering
     \vspace{-4.0in}
     \plottwo{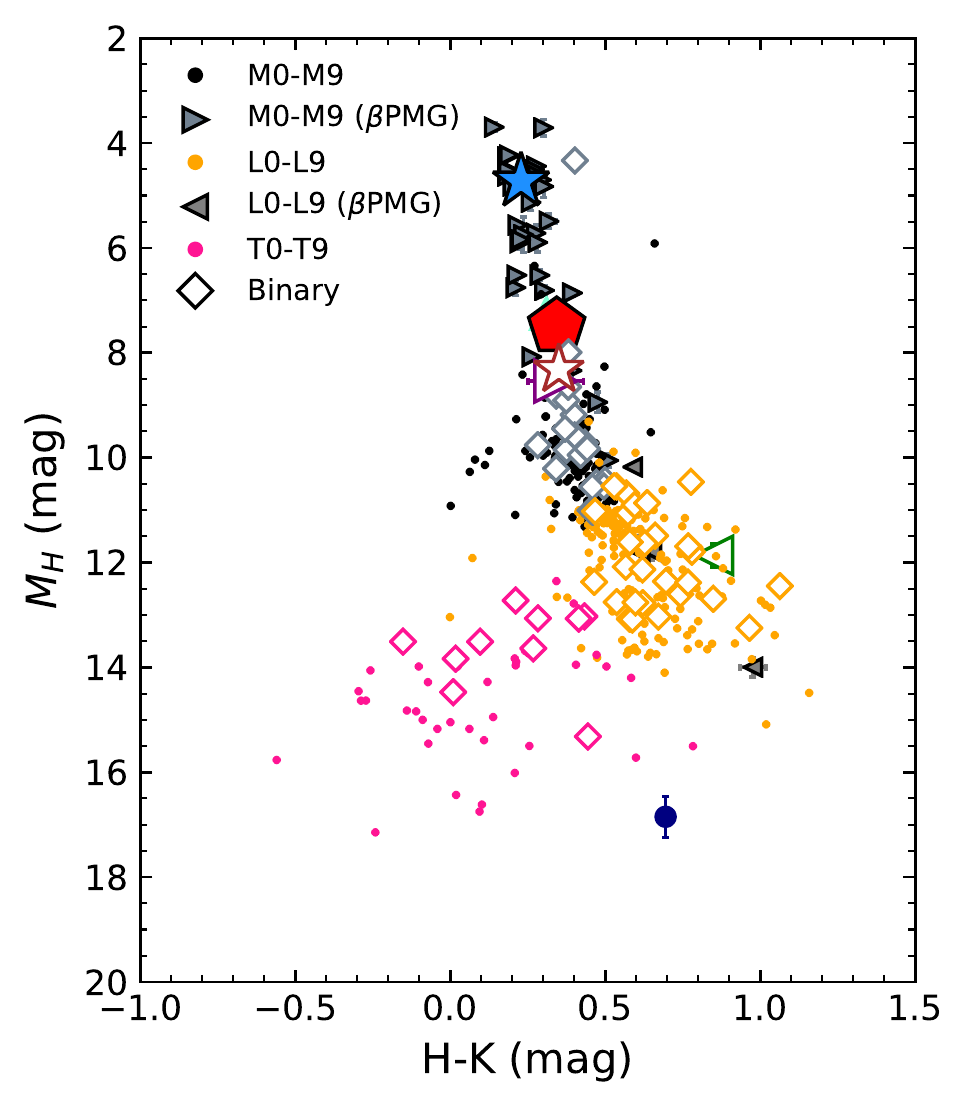}{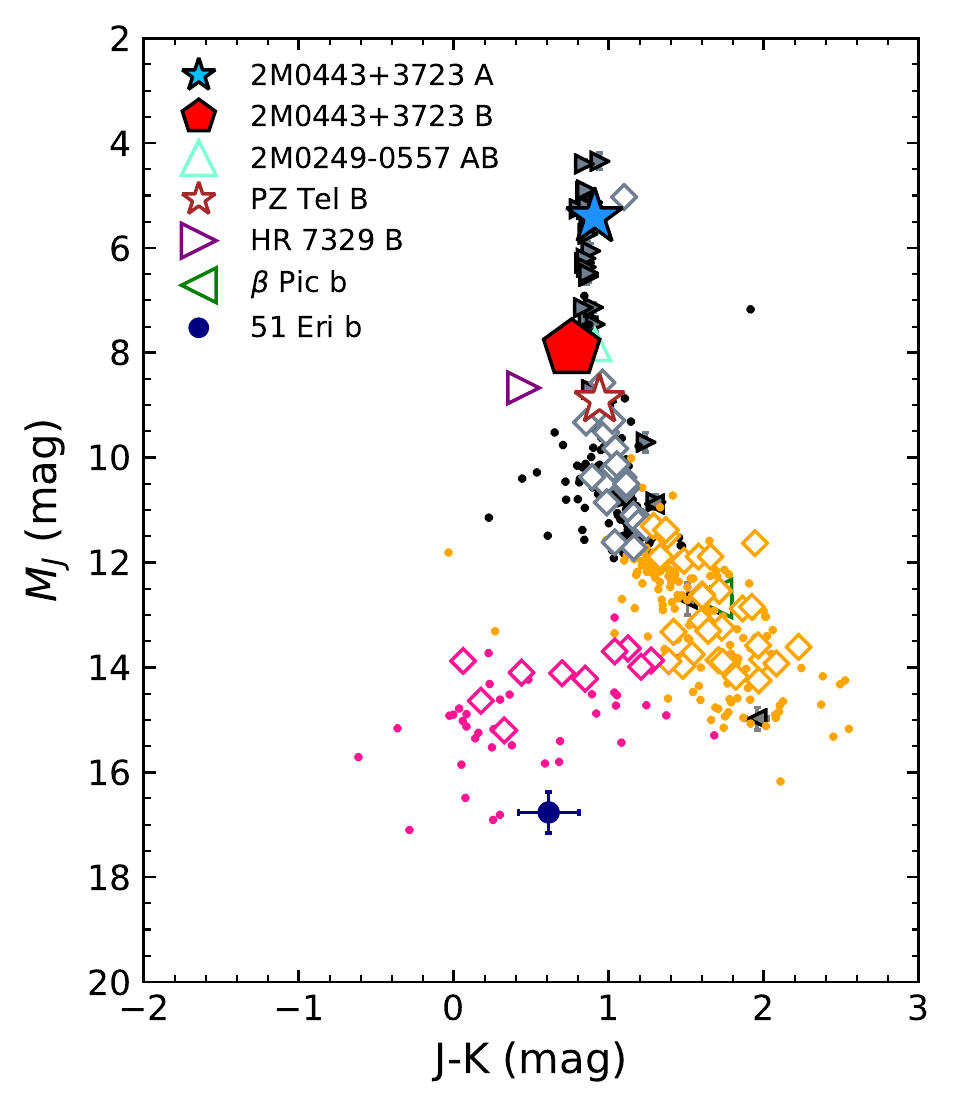}
    
    \caption{$M_{H}$ vs (\textit{H-K})(left) and $M_{J}$ vs (\textit{J-K})(right) color magnitude diagrams of brown dwarfs and gas giants (M--T spectral types) from the $\beta$PMG and other field ultracool dwarfs compared to 2M0443+3723 B (red
    pentagon). 2M0443+3723 B is brighter than other single brown dwarf companion, PZ Tel B (M7 $\pm$ 1). All photometry is on the 2MASS system. Photometry is shown for young M dwarfs from the $\beta$PMG (right-facing blue triangles) and L dwarfs from the $\beta$PMG (left-facing grey triangles). 2M0443+3723 A is shown with a blue star. Binaries are shown (diamonds) and substellar companions in the $\beta$PMG are labeled: PZ Tel B, HR 7329 B, $\beta$ Pic b, and 51 Eri b. Photometry are from \citealt{dupuyliu2012}; \citealt{dupuyandkraus2013}; \citealt{dupuyallers2016}; and \citealt{dupuy2018}. }
    \label{fig:color_magnitude_diagrams}
\end{figure*}

In Figure \ref{fig:color_magnitude_diagrams}, color magnitude diagrams are used to compare the photometry of 2M0443+3723 B to ultracool objects in $\beta$PMG  and the field. We combine the parallactic distance to 2M0443+3723 B from $Gaia$ DR2 with 2MASS photometry to derive absolute magnitudes (Table \ref{tab:summary of 2M0443+3723AB}).
 
2M0443+3723 B has a slightly bluer $J$--$K$ color and is about 0.9 mag brighter in  $M_{H}$  than PZ Tel B. However, it is nearly identical to the M6 pair 2M0249-0557 A and B when using their parallatic distance as measured by $Gaia$.

\subsection{Is 2M0443+3723 B a Close Binary?}
 \label{sec:is_2M0043 a binary}
 
 The empirical analysis of 2M0443+3723 B implies a higher bolometric luminosity for a single object in $\beta$PMG compared to other M6--M7 members  (Figure \ref{fig:mass_evolutionary}). This may mean that this companion is itself a close binary or that it is not a member of the $\beta$PMG.
 \par
\begin{figure}
    \includegraphics[width=\columnwidth]{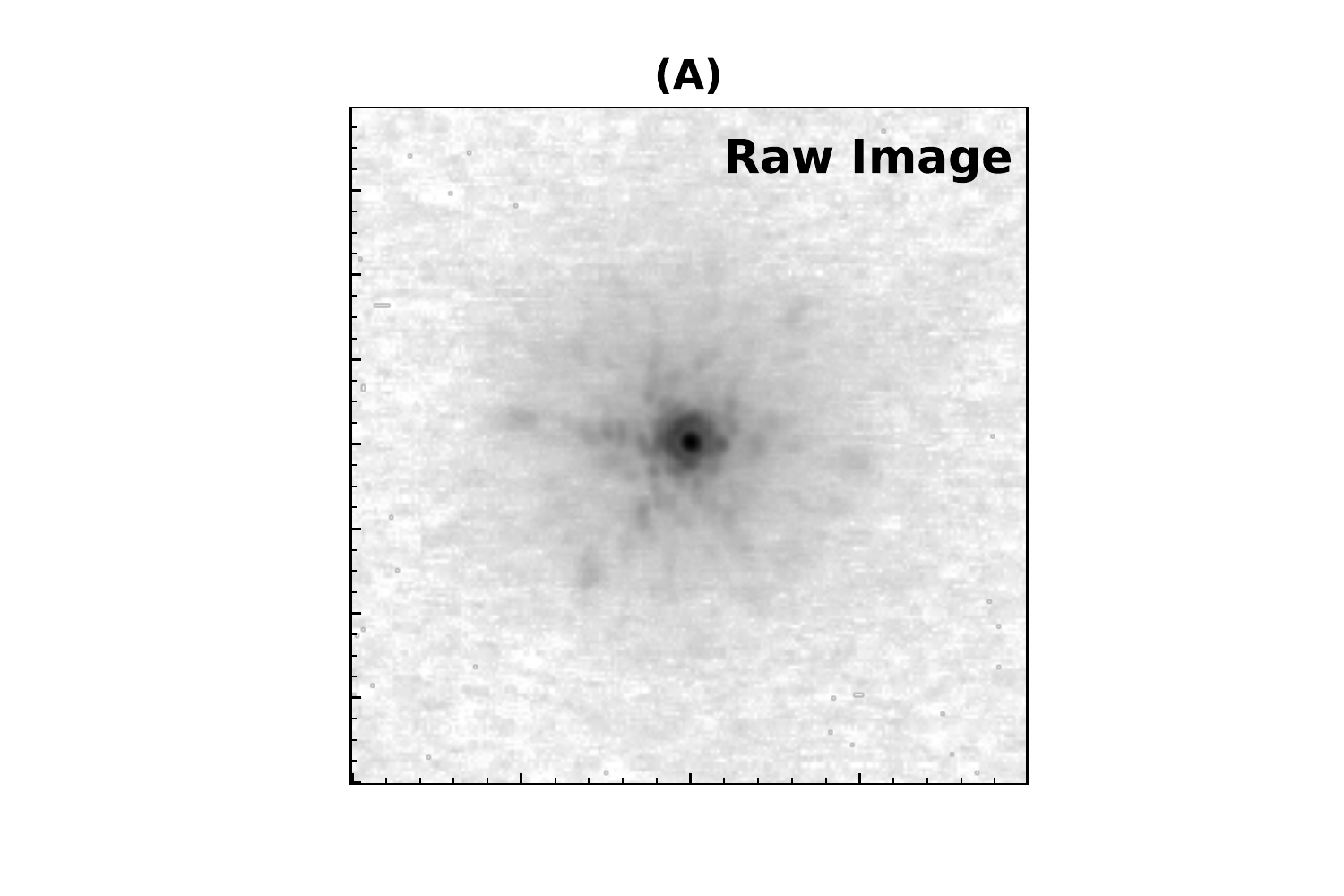}
    \includegraphics[width=\columnwidth]{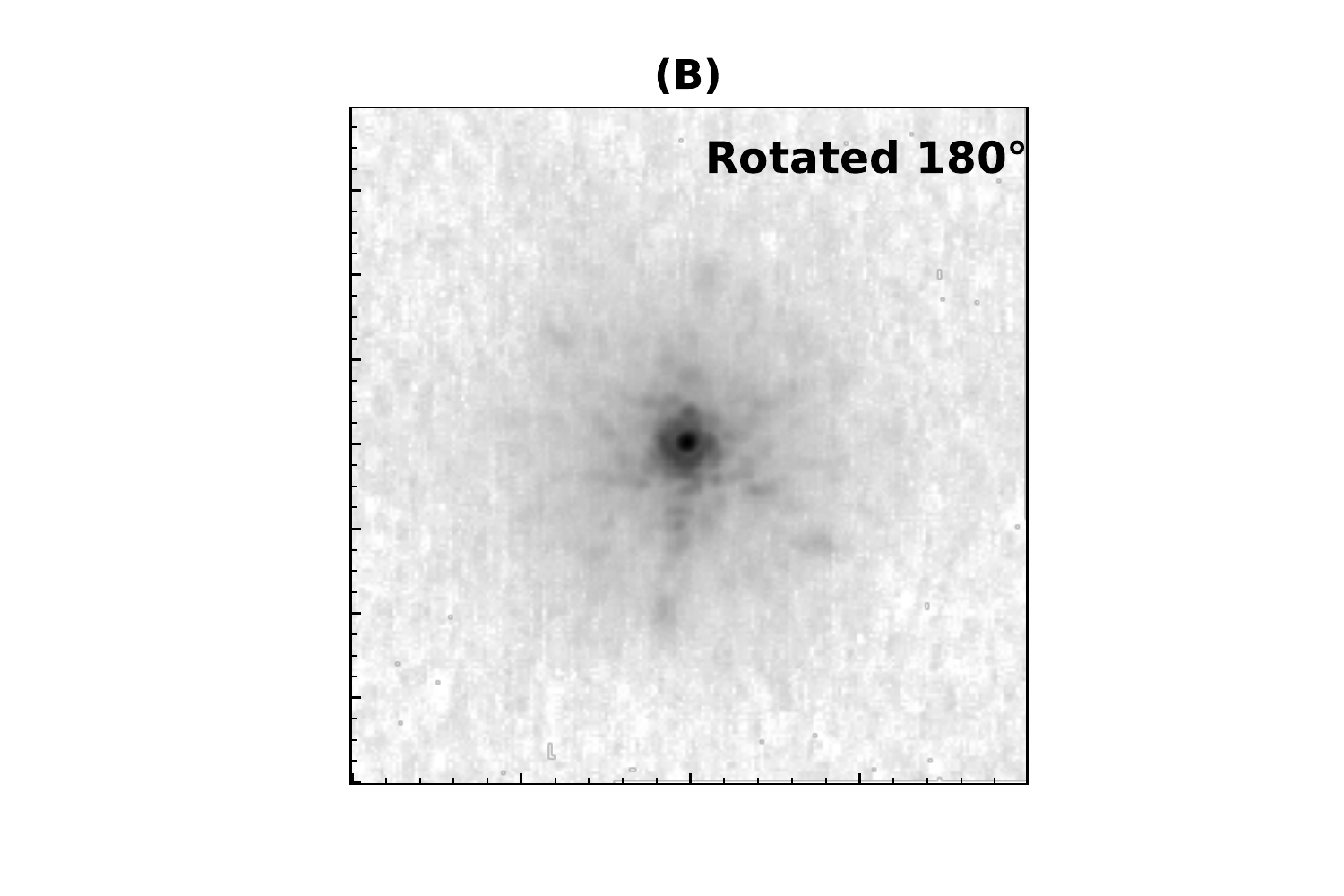}
    \centering
    \includegraphics[width=.57\columnwidth]{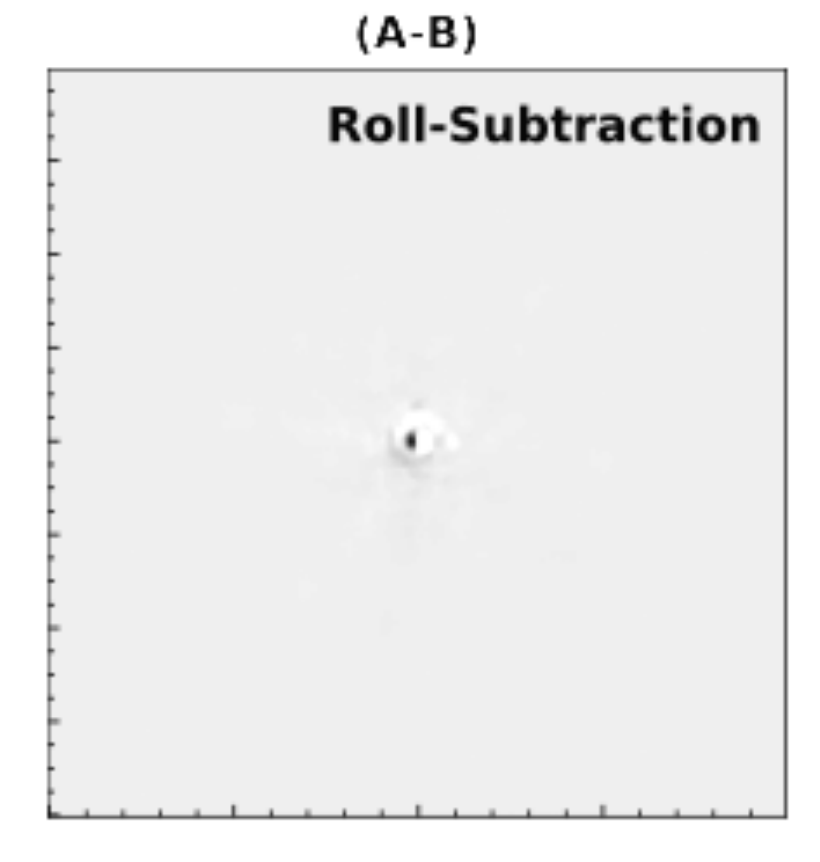}
    \caption{Steps involving  classical \enquote{roll angle subtraction}  PSF subtraction.  Top: The final median--combined 2\arcsec$\times$2\arcsec \hspace{0.01cm}Keck II/NIRC2 image of 2M0443+3723 B. Middle: the roll-subtraction  method  using a 180$^{\circ}$ roll angle. Bottom: Difference between image A and B.  A candidate companion is visible in the A-B image, but this is likely a speckle because a similar feature is present in the host as well.}
    \label{fig:roll_subtraction}
    
\end{figure}

 As described in Section \ref{sec:keck_imaging_data}, we acquired AO images with Keck II/NIRC2 to investigate the possibility that 2M0443+3723 B  is itself a close binary. After the basic image reduction, the dithered science frames were  registered, shifted to the common centroid position of the companion, and median-combined to produce the co-added image of the companion shown in Fig  \ref{fig:roll_subtraction}. There are no obvious indications that 2M0443+3723 B is a close binary from these diffraction-limited images.

\begin{figure}

    \plottwo{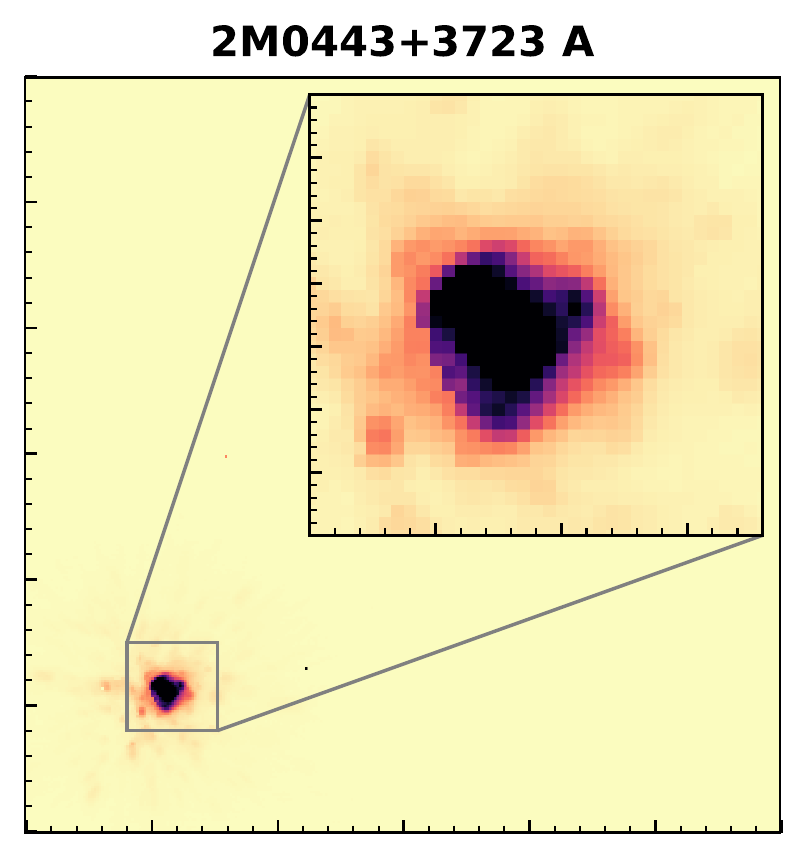}{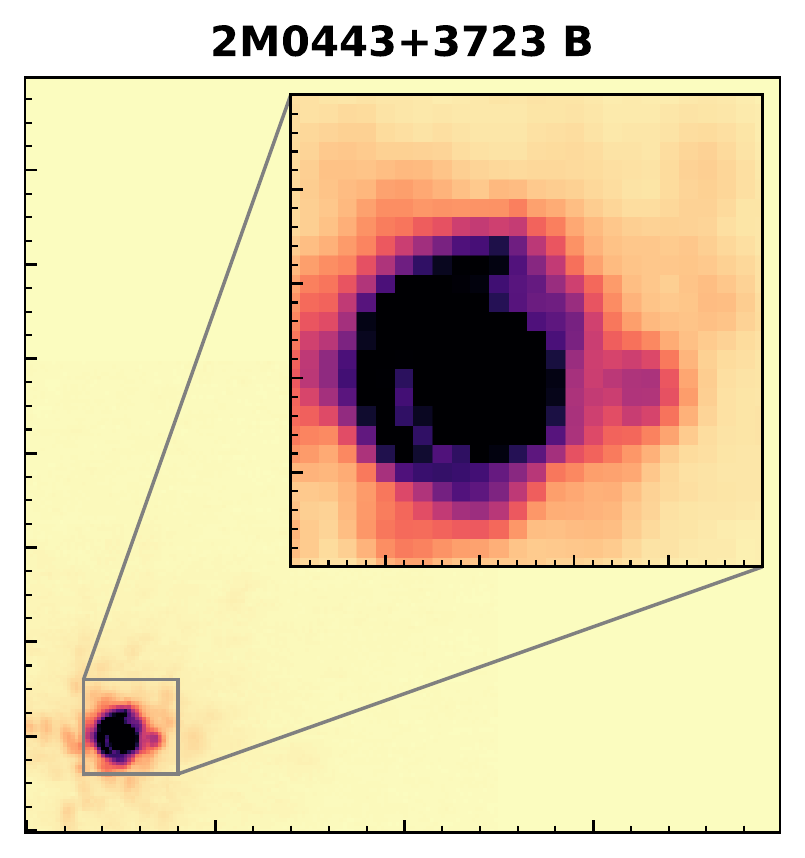}
    \caption{Visual inspection of the point source resembling a close companion to 2M0443+3723 B. Right: magnified image of the host star, 2M0443+3723 A in a 3\arcsec $\times$ 3\arcsec cutout. Left: magnified median--combined of Keck II/NIRC2 image 2M0443+3723 B in a 2\arcsec $\times$ 2\arcsec cutout.  The point source is present near 2M0443+3723 B and 2M0443+3723 A. This indicates that the source is part of the speckle pattern and not a companion.}
    \label{fig:inspection_point_source}
\end{figure}

We carried out basic PSF \enquote{roll angle subtraction}  to further assess whether 2M0443+3723 B is a close, marginally resolved binary.  To mimic \enquote{roll-subtraction} (\citealt{liu2004}; \citealt{song2006}) the image of the companion was rotated and subtracted from the non-rotated reference image (Figure \ref{fig:roll_subtraction}) to search for potential companions that could account for the abnormally high luminosity of 2M0443+3723 B, assuming membership in the $\beta$PMG. The same procedure was carried out for 36 equally-spaced roll angles spanning 10$^{\circ}$--350$^{\circ}$. After performing PSF subtraction, we noted a point source adjacent to 2M0443+3723 B; however, this same source is visible in the image of 2M0443+3723 A, which indicates this residual artifact is part of the speckle pattern and not a true companion (Figure \ref{fig:inspection_point_source}).
\par
We analyzed the IGRINS spectra of 
2M0443+3723 A and 2M0443+3723 B to search for signs of spectroscopic binarity by utilizing a cross correlation analysis of  2M0443+3723 A and 2M0443+3723 B against one another. This cross correlation confirms the radial velocity difference and shows no trace of a third component in the system. However, we can not rule out the presence of a significantly lower mass companion in this system.

\section{Conclusions}
\label{sec:summaryandconclusions}
We have presented a detailed characterization of 2M0443+3723 B, a wide companion candidate brown dwarf to a lithium-rich M dwarf that may belong to the $\beta$PMG. We obtained a near-infrared spectrum from IRFT/SpeX and find a   M6 $\pm$ 1 spectral type and VL-G gravity classification using the index-based system of  \cite{aller&liu2013}. 2M0443+3723 B has an inflated radius of 3.5 R$\mathrm{_{Jup}}$ and BT-Settl atmospheric model fits yield  $T_{\mathrm{eff}}$ = 2800 $\pm$ 100 K and log $g$ = 4.0 $\pm$ 0.5 dex. We also calculate a bolometric luminosity of --2.16 $\pm$ 0.02 dex. This luminosity combined with the age of the $\beta$PMG, 23 $\pm$ 3 Myr \citep{mamajek&bell2014}, implies a model-dependent mass of 99 $\pm$ 5 M$\mathrm{_{Jup}}$. This mass and luminosity are higher than other M6 members of the $\beta$PMG.
\par
We acquired AO images with Keck II/NIRC2 to explore the possibility that the companion could be a close binary, which could explain this anomalously high luminosity and corresponding higher mass. PSF subtraction does not reveal any close, modest flux ratio point sources.  We reassess whether the 2M0443+3723 AB system is a member of the $\beta$PMG as has been proposed in past studies (\citealt{schlieder2010}; \citealt{malobanyan}, \citealt{shkolnik2017}; \citealt{leesong2019}). We utilize the membership analysis tool BANYAN $\Sigma$ \citep{gagne2018} and find low probability of membership in the $\beta$PMG of 0.0$\%$ and 0.4$\%$ for 2M0443+3723 B and 2M0443+3723 A, respectively. This is contrary to the 86.2$\%$ membership probability in the $\beta$PMG for 2M0443+3723 A found by \cite{leesong2019} using their BAMG membership analysis tool. We explored other analytical methods to evaluate membership and found the $\beta$PMG to be the most likely young moving group, should this system belong to a known association, although it may also be a young field object. 

Regardless of its status in the $\beta$PMG, 2M0443+3723 B is certainly young, and likely $\lesssim$ 30 Myr. Our analysis shows that this system has various indicators of youth: an inflated radius for 2M0443+3723 B, lithium in the host, and VL-G classification. We expect the status of this system to be clarified as the spatial and kinematic distributions of the $\beta$PMG continue to be refined with parallaxes from $Gaia$ and additional radial velocities of candidate members in the future.

\facilities{Keck:II (NIRC2), IRTF (SpeX) DCT (IGRINS)}
\software{IGRINS Pipeline \citep{plppipeline}, SpeXTool (\citealt{vacca2003};  \citealt{cushing2004}), cosmics.py package \citep{dokkum2001}, BANYAN $\Sigma$ \citep{gagne2018}}

\clearpage

\acknowledgments
We thank the anonymous referee for their time providing helpful comments which improved the quality of this paper.
The authors wish to thank Michael Bonnefoy, Trent Dupuy, Abhijith Rajan, and Jeffrey Chilcote for providing spectra that were used in this work.
The Database of Ultracool Parallaxes is maintained by Trent Dupuy.
This research has benefited from the SpeX Prism Library (and/or SpeX Prism Library Analysis Toolkit), maintained by Adam Burgasser at \url{http://www.browndwarfs.org/spexprism}.

This work was supported by a NASA Keck PI Data Award, administered by the NASA Exoplanet Science Institute. Data presented herein were obtained at the W. M. Keck Observatory from telescope time allocated to the National Aeronautics and Space Administration through the agency’s scientific partnership with the California Institute of Technology and the University of California. The Observatory was made possible by the generous financial support of the W. M. Keck Foundation. The authors wish to recognize and acknowledge the very significant cultural role and reverence that the summit of Mauna Kea has always had within the indigenous Hawaiian community. We are most fortunate to have the opportunity to conduct observations from this mountain. We utilized data acquired with the  SpeX instrument at the IRTF, which is operated by the University of Hawaii under contract NNH14CK55B with the National Aeronautics and Space Administration.

This work used the Immersion Grating Infrared Spectrometer (IGRINS) that was developed under a collaboration between the University of Texas at Austin and the Korea Astronomy and Space Science Institute (KASI) with the financial support of the US National Science Foundation under grant AST-1229522, of the University of Texas at Austin, and of the Korean GMT Project of KASI.
These results made use of the Discovery Channel Telescope at Lowell Observatory. Lowell is a private, non-profit institution dedicated to astrophysical research and public appreciation of astronomy and operates the DCT in partnership with Boston University, the University of Maryland, the University of Toledo, Northern Arizona University and Yale University.

This publication makes use of data products from the Two Micron All Sky Survey, which is a joint project of the University of Massachusetts and the Infrared Processing and Analysis Center/California Institute of Technology, funded by the National Aeronautics and Space Administration and the National Science Foundation.

NASA's Astrophysics Data System Bibliographic Services together with the VizieR catalogue access tool and SIMBAD database operated at CDS, Strasbourg, France, were invaluable resources for this work. 

This work has made use of data from the European Space Agency (ESA) mission Gaia \url{(https://www.cosmos.esa.int/gaia)}, processed by the Gaia Data Processing and Analysis Consortium (DPAC, \url{https://www.cosmos.esa.int/web/ gaia/dpac/consortium)}. Funding for the DPAC has been provided by national institutions, in particular the institutions participating in the Gaia Multilateral Agreement.

BPB acknowledges support from the National Science Foundation grant AST-1909209.

Caprice Phillips thanks the LSSTC Data Science Fellowship Program, which is funded by LSSTC, NSF Cybertraining Grant $\#$1829740, the Brinson Foundation, and the Moore Foundation; her participation in the program has benefited this work.

\bibliographystyle{aasjournal}
\bibliography{references}

\begin{thebibliography}{}
\expandafter\ifx\csname natexlab\endcsname\relax\def\natexlab#1{#1}\fi
\providecommand{\url}[1]{\href{#1}{#1}}

\bibitem[{{Allard} {et~al.}(2012){Allard}, {Homeier}, \&
  {Freytag}}]{allard2012}
{Allard}, F., {Homeier}, D., \& {Freytag}, B. 2012, Philosophical Transactions
  of the Royal Society of London Series A, 370, 2765

\bibitem[{{Allers} \& {Liu}(2013)}]{aller&liu2013}
{Allers}, K.~N., \& {Liu}, M.~C. 2013, \apj, 772, 79

\bibitem[{Bailer-Jones {et~al.}(2018)Bailer-Jones, Rybizki, Fouesneau,
  Mantelet, \& Andrae}]{bailer-jones}
Bailer-Jones, C. A.~L., Rybizki, J., Fouesneau, M., Mantelet, G., \& Andrae, R.
  2018, 1804.10121

\bibitem[{{Basri} {et~al.}(1996){Basri}, {Marcy}, \& {Graham}}]{basri1996}
{Basri}, G., {Marcy}, G.~W., \& {Graham}, J.~R. 1996, The Astrophysical
  Journal, 458, 600

\bibitem[{{Bell} {et~al.}(2015){Bell}, {Mamajek}, \& {Naylor}}]{bell2015}
{Bell}, C.~P.~M., {Mamajek}, E.~E., \& {Naylor}, T. 2015, \mnras, 454, 593

\bibitem[{{Bell} {et~al.}(2017){Bell}, {Murphy}, \& {Mamajek}}]{bell2017}
{Bell}, C.~P.~M., {Murphy}, S.~J., \& {Mamajek}, E.~E. 2017, \mnras, 468, 1198

\bibitem[{{Beuzit} {et~al.}(2008){Beuzit}, {Feldt}, {Dohlen}, {Mouillet},
  {Puget}, {Wildi}, {Abe}, {Antichi}, {Baruffolo}, {Baudoz}, {Boccaletti},
  {Carbillet}, {Charton}, {Claudi}, {Downing}, {Fabron}, {Feautrier},
  {Fedrigo}, {Fusco}, {Gach}, {Gratton}, {Henning}, {Hubin}, {Joos}, {Kasper},
  {Langlois}, {Lenzen}, {Moutou}, {Pavlov}, {Petit}, {Pragt}, {Rabou}, {Rigal},
  {Roelfsema}, {Rousset}, {Saisse}, {Schmid}, {Stadler}, {Thalmann}, {Turatto},
  {Udry}, {Vakili}, \& {Waters}}]{bezuit2008}
{Beuzit}, J.-L., {Feldt}, M., {Dohlen}, K., {et~al.} 2008, in \procspie, Vol.
  7014, Ground-based and Airborne Instrumentation for Astronomy II, 701418

\bibitem[{{Biller} {et~al.}(2010){Biller}, {Liu}, {Wahhaj}, {Nielsen}, {Close},
  {Dupuy}, {Hayward}, {Burrows}, {Chun}, {Ftaclas}, {Clarke}, {Hartung},
  {Males}, {Reid}, {Shkolnik}, {Skemer}, {Tecza}, {Thatte}, {Alencar},
  {Artymowicz}, {Boss}, {de Gouveia Dal Pino}, {Gregorio-Hetem}, {Ida},
  {Kuchner}, {Lin}, \& {Toomey}}]{biller2010}
{Biller}, B.~A., {Liu}, M.~C., {Wahhaj}, Z., {et~al.} 2010, \apjl, 720, L82

\bibitem[{{Biller} {et~al.}(2013){Biller}, {Liu}, {Wahhaj}, {Nielsen},
  {Hayward}, {Males}, {Skemer}, {Close}, {Chun}, {Ftaclas}, {Clarke}, {Thatte},
  {Shkolnik}, {Reid}, {Hartung}, {Boss}, {Lin}, {Alencar}, {de Gouveia Dal
  Pino}, {Gregorio-Hetem}, \& {Toomey}}]{biller2013}
---. 2013, \apj, 777, 160

\bibitem[{{Bonnefoy} {et~al.}(2014){Bonnefoy}, {Chauvin}, {Lagrange}, {Rojo},
  {Allard}, {Pinte}, {Dumas}, \& {Homeier}}]{bonnefoy2014}
{Bonnefoy}, M., {Chauvin}, G., {Lagrange}, A.-M., {et~al.} 2014, \aap, 562,
  A127

\bibitem[{{Bowler}(2016)}]{bowler2016}
{Bowler}, B.~P. 2016, \pasp, 128, 102001

\bibitem[{{Bowler} {et~al.}(2009){Bowler}, {Liu}, \& {Cushing}}]{bowler2009}
{Bowler}, B.~P., {Liu}, M.~C., \& {Cushing}, M.~C. 2009, \apj, 706, 1114

\bibitem[{{Bowler} {et~al.}(2015){Bowler}, {Liu}, {Shkolnik}, \&
  {Tamura}}]{bowler2015}
{Bowler}, B.~P., {Liu}, M.~C., {Shkolnik}, E.~L., \& {Tamura}, M. 2015, \apjs,
  216, 7

\bibitem[{{Bowler} \& {Nielsen}(2018)}]{bowlerandnielsen2018}
{Bowler}, B.~P., \& {Nielsen}, E.~L. 2018, {Occurrence Rates from Direct
  Imaging Surveys}, 155

\bibitem[{{Bowler} {et~al.}(2019){Bowler}, {Hinkley}, {Ziegler}, {Baranec},
  {Gizis}, {Law}, {Liu}, {Shah}, {Shkolnik}, {Riaz}, \& {Riddle}}]{bowler2019}
{Bowler}, B.~P., {Hinkley}, S., {Ziegler}, C., {et~al.} 2019, arXiv e-prints,
  arXiv:1903.06303

\bibitem[{{Brandt} {et~al.}(2018){Brandt}, {Dupuy}, \& {Bowler}}]{brandt2018}
{Brandt}, T.~D., {Dupuy}, T.~J., \& {Bowler}, B.~P. 2018, arXiv e-prints,
  arXiv:1811.07285

\bibitem[{{Burgasser}(2014)}]{burgasser2014}
{Burgasser}, A.~J. 2014, in Astronomical Society of India Conference Series,
  Vol.~11, Astronomical Society of India Conference Series

\bibitem[{{Burgasser} {et~al.}(2008){Burgasser}, {Liu}, {Ireland}, {Cruz}, \&
  {Dupuy}}]{burgasser2008}
{Burgasser}, A.~J., {Liu}, M.~C., {Ireland}, M.~J., {Cruz}, K.~L., \& {Dupuy},
  T.~J. 2008, \apj, 681, 579

\bibitem[{{Burgasser} {et~al.}(2004){Burgasser}, {McElwain}, {Kirkpatrick},
  {Cruz}, {Tinney}, \& {Reid}}]{burgasser2004}
{Burgasser}, A.~J., {McElwain}, M.~W., {Kirkpatrick}, J.~D., {et~al.} 2004,
  \aj, 127, 2856

\bibitem[{{Burrows} {et~al.}(2001){Burrows}, {Hubbard}, {Lunine}, \&
  {Liebert}}]{burrows2001}
{Burrows}, A., {Hubbard}, W.~B., {Lunine}, J.~I., \& {Liebert}, J. 2001,
  Reviews of Modern Physics, 73, 719

\bibitem[{{Chilcote} {et~al.}(2017){Chilcote}, {Pueyo}, {De Rosa}, {Vargas},
  {Macintosh}, {Bailey}, {Barman}, {Bauman}, {Bruzzone}, {Bulger}, {Burrows},
  {Cardwell}, {Chen}, {Cotten}, {Dillon}, {Doyon}, {Draper}, {Duch{\^e}ne},
  {Dunn}, {Erikson}, {Fitzgerald}, {Follette}, {Gavel}, {Goodsell}, {Graham},
  {Greenbaum}, {Hartung}, {Hibon}, {Hung}, {Ingraham}, {Kalas}, {Konopacky},
  {Larkin}, {Maire}, {Marchis}, {Marley}, {Marois}, {Metchev},
  {Millar-Blanchaer}, {Morzinski}, {Nielsen}, {Norton}, {Oppenheimer},
  {Palmer}, {Patience}, {Perrin}, {Poyneer}, {Rajan}, {Rameau},
  {Rantakyr{\"o}}, {Sadakuni}, {Saddlemyer}, {Savransky}, {Schneider}, {Serio},
  {Sivaramakrishnan}, {Song}, {Soummer}, {Thomas}, {Wallace}, {Wang},
  {Ward-Duong}, {Wiktorowicz}, \& {Wolff}}]{chilcote2017}
{Chilcote}, J., {Pueyo}, L., {De Rosa}, R.~J., {et~al.} 2017, \aj, 153, 182

\bibitem[{{Chun} {et~al.}(2008){Chun}, {Toomey}, {Wahhaj}, {Biller}, {Artigau},
  {Hayward}, {Liu}, {Close}, {Hartung}, {Rigaut}, \& {Ftaclas}}]{chun2008}
{Chun}, M., {Toomey}, D., {Wahhaj}, Z., {et~al.} 2008, in \procspie, Vol. 7015,
  Adaptive Optics Systems, 70151V

\bibitem[{{Crepp} {et~al.}(2012){Crepp}, {Johnson}, {Fischer}, {Howard},
  {Marcy}, {Wright}, {Isaacson}, {Boyajian}, {von Braun}, {Hillenbrand},
  {Hinkley}, {Carpenter}, \& {Brewer}}]{crepp2012}
{Crepp}, J.~R., {Johnson}, J.~A., {Fischer}, D.~A., {et~al.} 2012, \apj, 751,
  97

\bibitem[{{Cushing} {et~al.}(2005){Cushing}, {Rayner}, \&
  {Vacca}}]{cushing2005}
{Cushing}, M.~C., {Rayner}, J.~T., \& {Vacca}, W.~D. 2005, \apj, 623, 1115

\bibitem[{{Cushing} {et~al.}(2004){Cushing}, {Vacca}, \&
  {Rayner}}]{cushing2004}
{Cushing}, M.~C., {Vacca}, W.~D., \& {Rayner}, J.~T. 2004, \pasp, 116, 362

\bibitem[{{Cushing} {et~al.}(2008){Cushing}, {Marley}, {Saumon}, {Kelly},
  {Vacca}, {Rayner}, {Freedman}, {Lodders}, \& {Roellig}}]{cushing2008}
{Cushing}, M.~C., {Marley}, M.~S., {Saumon}, D., {et~al.} 2008, \apj, 678, 1372

\bibitem[{{Cutri} {et~al.}(2003){Cutri}, {Skrutskie}, {van Dyk}, {Beichman},
  {Carpenter}, {Chester}, {Cambresy}, {Evans}, {Fowler}, {Gizis}, {Howard},
  {Huchra}, {Jarrett}, {Kopan}, {Kirkpatrick}, {Light}, {Marsh}, {McCallon},
  {Schneider}, {Stiening}, {Sykes}, {Weinberg}, {Wheaton}, {Wheelock}, \&
  {Zacarias}}]{cutri2003}
{Cutri}, R.~M., {Skrutskie}, M.~F., {van Dyk}, S., {et~al.} 2003, VizieR Online
  Data Catalog, 2246

\bibitem[{{Cutri} {et~al.}(2012){Cutri}, {Wright}, {Conrow}, {Bauer},
  {Benford}, {Brandenburg}, {Dailey}, {Eisenhardt}, {Evans}, {Fajardo-Acosta},
  {Fowler}, {Gelino}, {Grillmair}, {Harbut}, {Hoffman}, {Jarrett},
  {Kirkpatrick}, {Leisawitz}, {Liu}, {Mainzer}, {Marsh}, {Masci}, {McCallon},
  {Padgett}, {Ressler}, {Royer}, {Skrutskie}, {Stanford}, {Wyatt}, {Tholen},
  {Tsai}, {Wachter}, {Wheelock}, {Yan}, {Alles}, {Beck}, {Grav}, {Masiero},
  {McCollum}, {McGehee}, {Papin}, \& {Wittman}}]{cutri2012}
{Cutri}, R.~M., {Wright}, E.~L., {Conrow}, T., {et~al.} 2012, {Explanatory
  Supplement to the WISE All-Sky Data Release Products}, Tech. rep.

\bibitem[{{Dahn} {et~al.}(2017){Dahn}, {Harris}, {Subasavage}, {Ables},
  {Canzian}, {Guetter}, {Harris}, {Henden}, {Leggett}, {Levine}, {Luginbuhl},
  {Monet}, {Monet}, {Munn}, {Pier}, {Stone}, {Vrba}, {Walker}, \&
  {Tilleman}}]{dahn2007}
{Dahn}, C.~C., {Harris}, H.~C., {Subasavage}, J.~P., {et~al.} 2017, \aj, 154,
  147

\bibitem[{{David} {et~al.}(2015){David}, {Stauffer}, {Hillenbrand}, {Cody},
  {Conroy}, {Stassun}, {Pope}, {Aigrain}, {Gillen}, {Collier Cameron},
  {Barrado}, {Rebull}, {Isaacson}, {Marcy}, {Zhang}, {Riddle}, {Ziegler},
  {Law}, \& {Baranec}}]{david2015}
{David}, T.~J., {Stauffer}, J., {Hillenbrand}, L.~A., {et~al.} 2015, \apj, 814,
  62

\bibitem[{{Dupuy} {et~al.}(2019){Dupuy}, {Brandt}, {Kratter}, \&
  {Bowler}}]{dupuy2019}
{Dupuy}, T.~J., {Brandt}, T.~D., {Kratter}, K.~M., \& {Bowler}, B.~P. 2019,
  \apjl, 871, L4

\bibitem[{{Dupuy} \& {Kraus}(2013)}]{dupuyandkraus2013}
{Dupuy}, T.~J., \& {Kraus}, A.~L. 2013, Science, 341, 1492

\bibitem[{{Dupuy} \& {Liu}(2012)}]{dupuyliu2012}
{Dupuy}, T.~J., \& {Liu}, M.~C. 2012, \apjs, 201, 19

\bibitem[{{Dupuy} {et~al.}(2009){Dupuy}, {Liu}, \& {Ireland}}]{dupuy2009}
{Dupuy}, T.~J., {Liu}, M.~C., \& {Ireland}, M.~J. 2009, \apj, 692, 729

\bibitem[{{Dupuy} {et~al.}(2018){Dupuy}, {Liu}, {Allers}, {Biller}, {Kratter},
  {Mann}, {Shkolnik}, {Kraus}, \& {Best}}]{dupuy2018}
{Dupuy}, T.~J., {Liu}, M.~C., {Allers}, K.~N., {et~al.} 2018, \aj, 156, 57

\bibitem[{{Faherty} {et~al.}(2010){Faherty}, {Burgasser}, {West}, {Bochanski},
  {Cruz}, {Shara}, \& {Walter}}]{faherty2010}
{Faherty}, J.~K., {Burgasser}, A.~J., {West}, A.~A., {et~al.} 2010, \aj, 139,
  176

\bibitem[{{Feiden}(2016)}]{feiden2016}
{Feiden}, G.~A. 2016, \aap, 593, A99

\bibitem[{{Filippazzo} {et~al.}(2015){Filippazzo}, {Rice}, {Faherty}, {Cruz},
  {Van Gordon}, \& {Looper}}]{filippazzo2015}
{Filippazzo}, J.~C., {Rice}, E.~L., {Faherty}, J., {et~al.} 2015, \apj, 810,
  158

\bibitem[{{Gagn{\'e}} {et~al.}(2014){Gagn{\'e}}, {Lafreni{\`e}re}, {Doyon},
  {Malo}, \& {Artigau}}]{gagne2014}
{Gagn{\'e}}, J., {Lafreni{\`e}re}, D., {Doyon}, R., {Malo}, L., \& {Artigau},
  {\'E}. 2014, \apj, 783, 121

\bibitem[{{Gagn{\'e}} {et~al.}(2018){Gagn{\'e}}, {Mamajek}, {Malo}, {Riedel},
  {Rodriguez}, {Lafreni{\`e}re}, {Faherty}, {Roy-Loubier}, {Pueyo}, {Robin}, \&
  {Doyon}}]{gagne2018}
{Gagn{\'e}}, J., {Mamajek}, E.~E., {Malo}, L., {et~al.} 2018, \apj, 856, 23

\bibitem[{{Gaia Collaboration} {et~al.}(2018){Gaia Collaboration}, {Brown},
  {Vallenari}, {Prusti}, {de Bruijne}, {Babusiaux}, {Bailer-Jones}, {Biermann},
  {Evans}, {Eyer}, {Jansen}, {Jordi}, {Klioner}, {Lammers}, {Lindegren},
  {Luri}, {Mignard}, {Panem}, {Pourbaix}, {Randich}, {Sartoretti}, {Siddiqui},
  {Soubiran}, {van Leeuwen}, {Walton}, {Arenou}, {Bastian}, {Cropper},
  {Drimmel}, {Katz}, {Lattanzi}, {Bakker}, {Cacciari}, {Casta{\~n}eda},
  {Chaoul}, {Cheek}, {De Angeli}, {Fabricius}, {Guerra}, {Holl}, {Masana},
  {Messineo}, {Mowlavi}, {Nienartowicz}, {Panuzzo}, {Portell}, {Riello},
  {Seabroke}, {Tanga}, {Th{\'e}venin}, {Gracia-Abril}, {Comoretto},
  {Garcia-Reinaldos}, {Teyssier}, {Altmann}, {Andrae}, {Audard},
  {Bellas-Velidis}, {Benson}, {Berthier}, {Blomme}, {Burgess}, {Busso},
  {Carry}, {Cellino}, {Clementini}, {Clotet}, {Creevey}, {Davidson}, {De
  Ridder}, {Delchambre}, {Dell'Oro}, {Ducourant},
  {Fern{\'a}ndez-Hern{\'a}ndez}, {Fouesneau}, {Fr{\'e}mat}, {Galluccio},
  {Garc{\'\i}a-Torres}, {Gonz{\'a}lez-N{\'u}{\~n}ez}, {Gonz{\'a}lez-Vidal},
  {Gosset}, {Guy}, {Halbwachs}, {Hambly}, {Harrison}, {Hern{\'a}ndez},
  {Hestroffer}, {Hodgkin}, {Hutton}, {Jasniewicz}, {Jean-Antoine-Piccolo},
  {Jordan}, {Korn}, {Krone-Martins}, {Lanzafame}, {Lebzelter}, {L{\"o}ffler},
  {Manteiga}, {Marrese}, {Mart{\'\i}n-Fleitas}, {Moitinho}, {Mora}, {Muinonen},
  {Osinde}, {Pancino}, {Pauwels}, {Petit}, {Recio-Blanco}, {Richards},
  {Rimoldini}, {Robin}, {Sarro}, {Siopis}, {Smith}, {Sozzetti}, {S{\"u}veges},
  {Torra}, {van Reeven}, {Abbas}, {Abreu Aramburu}, {Accart}, {Aerts},
  {Altavilla}, {{\'A}lvarez}, {Alvarez}, {Alves}, {Anderson}, {Andrei},
  {Anglada Varela}, {Antiche}, {Antoja}, {Arcay}, {Astraatmadja}, {Bach},
  {Baker}, {Balaguer-N{\'u}{\~n}ez}, {Balm}, {Barache}, {Barata}, {Barbato},
  {Barblan}, {Barklem}, {Barrado}, {Barros}, {Barstow}, {Bartholom{\'e}
  Mu{\~n}oz}, {Bassilana}, {Becciani}, {Bellazzini}, {Berihuete}, {Bertone},
  {Bianchi}, {Bienaym{\'e}}, {Blanco-Cuaresma}, {Boch}, {Boeche}, {Bombrun},
  {Borrachero}, {Bossini}, {Bouquillon}, {Bourda}, {Bragaglia}, {Bramante},
  {Breddels}, {Bressan}, {Brouillet}, {Br{\"u}semeister}, {Brugaletta},
  {Bucciarelli}, {Burlacu}, {Busonero}, {Butkevich}, {Buzzi}, {Caffau},
  {Cancelliere}, {Cannizzaro}, {Cantat-Gaudin}, {Carballo}, {Carlucci},
  {Carrasco}, {Casamiquela}, {Castellani}, {Castro-Ginard}, {Charlot},
  {Chemin}, {Chiavassa}, {Cocozza}, {Costigan}, {Cowell}, {Crifo}, {Crosta},
  {Crowley}, {Cuypers}, {Dafonte}, {Damerdji}, {Dapergolas}, {David}, {David},
  {de Laverny}, {De Luise}, {De March}, {de Martino}, {de Souza}, {de Torres},
  {Debosscher}, {del Pozo}, {Delbo}, {Delgado}, {Delgado}, {Di Matteo},
  {Diakite}, {Diener}, {Distefano}, {Dolding}, {Drazinos}, {Dur{\'a}n},
  {Edvardsson}, {Enke}, {Eriksson}, {Esquej}, {Eynard Bontemps}, {Fabre},
  {Fabrizio}, {Faigler}, {Falc{\~a}o}, {Farr{\`a}s Casas}, {Federici},
  {Fedorets}, {Fernique}, {Figueras}, {Filippi}, {Findeisen}, {Fonti},
  {Fraile}, {Fraser}, {Fr{\'e}zouls}, {Gai}, {Galleti}, {Garabato},
  {Garc{\'\i}a-Sedano}, {Garofalo}, {Garralda}, {Gavel}, {Gavras}, {Gerssen},
  {Geyer}, {Giacobbe}, {Gilmore}, {Girona}, {Giuffrida}, {Glass}, {Gomes},
  {Granvik}, {Gueguen}, {Guerrier}, {Guiraud}, {Guti{\'e}rrez-S{\'a}nchez},
  {Haigron}, {Hatzidimitriou}, {Hauser}, {Haywood}, {Heiter}, {Helmi}, {Heu},
  {Hilger}, {Hobbs}, {Hofmann}, {Holland}, {Huckle}, {Hypki}, {Icardi},
  {Jan{\ss}en}, {Jevardat de Fombelle}, {Jonker}, {Juh{\'a}sz}, {Julbe},
  {Karampelas}, {Kewley}, {Klar}, {Kochoska}, {Kohley}, {Kolenberg},
  {Kontizas}, {Kontizas}, {Koposov}, {Kordopatis}, {Kostrzewa-Rutkowska},
  {Koubsky}, {Lambert}, {Lanza}, {Lasne}, {Lavigne}, {Le Fustec}, {Le
  Poncin-Lafitte}, {Lebreton}, {Leccia}, {Leclerc}, {Lecoeur-Taibi},
  {Lenhardt}, {Leroux}, {Liao}, {Licata}, {Lindstr{\o}m}, {Lister}, {Livanou},
  {Lobel}, {L{\'o}pez}, {Managau}, {Mann}, {Mantelet}, {Marchal}, {Marchant},
  {Marconi}, {Marinoni}, {Marschalk{\'o}}, {Marshall}, {Martino}, {Marton},
  {Mary}, {Massari}, {Matijevi{\v{c}}}, {Mazeh}, {McMillan}, {Messina},
  {Michalik}, {Millar}, {Molina}, {Molinaro}, {Moln{\'a}r}, {Montegriffo},
  {Mor}, {Morbidelli}, {Morel}, {Morris}, {Mulone}, {Muraveva}, {Musella},
  {Nelemans}, {Nicastro}, {Noval}, {O'Mullane}, {Ord{\'e}novic},
  {Ord{\'o}{\~n}ez-Blanco}, {Osborne}, {Pagani}, {Pagano}, {Pailler},
  {Palacin}, {Palaversa}, {Panahi}, {Pawlak}, {Piersimoni}, {Pineau}, {Plachy},
  {Plum}, {Poggio}, {Poujoulet}, {Pr{\v{s}}a}, {Pulone}, {Racero}, {Ragaini},
  {Rambaux}, {Ramos-Lerate}, {Regibo}, {Reyl{\'e}}, {Riclet}, {Ripepi}, {Riva},
  {Rivard}, {Rixon}, {Roegiers}, {Roelens}, {Romero-G{\'o}mez}, {Rowell},
  {Royer}, {Ruiz-Dern}, {Sadowski}, {Sagrist{\`a} Sell{\'e}s}, {Sahlmann},
  {Salgado}, {Salguero}, {Sanna}, {Santana-Ros}, {Sarasso}, {Savietto},
  {Schultheis}, {Sciacca}, {Segol}, {Segovia}, {S{\'e}gransan}, {Shih},
  {Siltala}, {Silva}, {Smart}, {Smith}, {Solano}, {Solitro}, {Sordo}, {Soria
  Nieto}, {Souchay}, {Spagna}, {Spoto}, {Stampa}, {Steele},
  {Steidelm{\"u}ller}, {Stephenson}, {Stoev}, {Suess}, {Surdej}, {Szabados},
  {Szegedi-Elek}, {Tapiador}, {Taris}, {Tauran}, {Taylor}, {Teixeira},
  {Terrett}, {Teyssand ier}, {Thuillot}, {Titarenko}, {Torra Clotet}, {Turon},
  {Ulla}, {Utrilla}, {Uzzi}, {Vaillant}, {Valentini}, {Valette}, {van Elteren},
  {Van Hemelryck}, {van Leeuwen}, {Vaschetto}, {Vecchiato}, {Veljanoski},
  {Viala}, {Vicente}, {Vogt}, {von Essen}, {Voss}, {Votruba}, {Voutsinas},
  {Walmsley}, {Weiler}, {Wertz}, {Wevers}, {Wyrzykowski}, {Yoldas},
  {{\v{Z}}erjal}, {Ziaeepour}, {Zorec}, {Zschocke}, {Zucker}, {Zurbach}, \&
  {Zwitter}}]{gaia}
{Gaia Collaboration}, {Brown}, A.~G.~A., {Vallenari}, A., {et~al.} 2018, \aap,
  616, A1

\bibitem[{{Galli} {et~al.}(2018){Galli}, {Loinard}, {Ortiz-L{\'e}on},
  {Kounkel}, {Dzib}, {Mioduszewski}, {Rodr{\'\i}guez}, {Hartmann}, {Teixeira},
  \& {Torres}}]{galli2018}
{Galli}, P. A.~B., {Loinard}, L., {Ortiz-L{\'e}on}, G.~N., {et~al.} 2018, \apj,
  859, 33

\bibitem[{{Gilmozzi} \& {Spyromilio}(2008)}]{gilmozzi&spyromilio2008}
{Gilmozzi}, R., \& {Spyromilio}, J. 2008, in \procspie, Vol. 7012, Ground-based
  and Airborne Telescopes II, 701219

\bibitem[{{Janson} {et~al.}(2012){Janson}, {Hormuth}, {Bergfors}, {Brand ner},
  {Hippler}, {Daemgen}, {Kudryavtseva}, {Schmalzl}, {Schnupp}, \&
  {Henning}}]{janson2012}
{Janson}, M., {Hormuth}, F., {Bergfors}, C., {et~al.} 2012, \apj, 754, 44

\bibitem[{{Kenyon} \& {Hartmann}(1995)}]{kenyonhartman1995}
{Kenyon}, S.~J., \& {Hartmann}, L. 1995, \apjs, 101, 117

\bibitem[{{Kraus} {et~al.}(2015){Kraus}, {Cody}, {Covey}, {Rizzuto}, {Mann}, \&
  {Ireland}}]{kraus2015}
{Kraus}, A.~L., {Cody}, A.~M., {Covey}, K.~R., {et~al.} 2015, \apj, 807, 3

\bibitem[{{Kraus} {et~al.}(2017){Kraus}, {Herczeg}, {Rizzuto}, {Mann},
  {Slesnick}, {Carpenter}, {Hillenbrand}, \& {Mamajek}}]{kraus2017}
{Kraus}, A.~L., {Herczeg}, G.~J., {Rizzuto}, A.~C., {et~al.} 2017, \apj, 838,
  150

\bibitem[{{Kraus} {et~al.}(2014{\natexlab{a}}){Kraus}, {Ireland}, {Cieza},
  {Hinkley}, {Dupuy}, {Bowler}, \& {Liu}}]{kraus2014}
{Kraus}, A.~L., {Ireland}, M.~J., {Cieza}, L.~A., {et~al.} 2014{\natexlab{a}},
  \apj, 781, 20

\bibitem[{{Kraus} {et~al.}(2014{\natexlab{b}}){Kraus}, {Shkolnik}, {Allers}, \&
  {Liu}}]{kraus2014new}
{Kraus}, A.~L., {Shkolnik}, E.~L., {Allers}, K.~N., \& {Liu}, M.~C.
  2014{\natexlab{b}}, \aj, 147, 146

\bibitem[{{Lagrange} {et~al.}(2010){Lagrange}, {Bonnefoy}, {Chauvin}, {Apai},
  {Ehrenreich}, {Boccaletti}, {Gratadour}, {Rouan}, {Mouillet}, {Lacour}, \&
  {Kasper}}]{lagrange2010}
{Lagrange}, A.-M., {Bonnefoy}, M., {Chauvin}, G., {et~al.} 2010, Science, 329,
  57

\bibitem[{{Lagrange} {et~al.}(2019){Lagrange}, {Boccaletti}, {Langlois},
  {Chauvin}, {Gratton}, {Beust}, {Desidera}, {Milli}, {Bonnefoy}, {Cheetham},
  {Feldt}, {Meyer}, {Vigan}, {Biller}, {Bonavita}, {Baudino}, {Cantalloube},
  {Cudel}, {Daemgen}, {Delorme}, {D'Orazi}, {Girard}, {Fontanive}, {Hagelberg},
  {Janson}, {Keppler}, {Koypitova}, {Galicher}, {Lannier}, {Le Coroller},
  {Ligi}, {Maire}, {Mesa}, {Messina}, {M{\"u}eller}, {Peretti}, {Perrot},
  {Rouan}, {Salter}, {Samland}, {Schmidt}, {Sissa}, {Zurlo}, {Beuzit},
  {Mouillet}, {Dominik}, {Henning}, {Lagadec}, {M{\'e}nard}, {Schmid},
  {Turatto}, {Udry}, {Bohn}, {Charnay}, {Gomez Gonzales}, {Gry}, {Kenworthy},
  {Kral}, {Mordasini}, {Moutou}, {van der Plas}, {Schlieder}, {Abe}, {Antichi},
  {Baruffolo}, {Baudoz}, {Baudrand}, {Blanchard}, {Bazzon}, {Buey},
  {Carbillet}, {Carle}, {Charton}, {Cascone}, {Claudi}, {Costille}, {Deboulbe},
  {De Caprio}, {Dohlen}, {Fantinel}, {Feautrier}, {Fusco}, {Gigan}, {Giro},
  {Gisler}, {Gluck}, {Hubin}, {Hugot}, {Jaquet}, {Kasper}, {Madec}, {Magnard},
  {Martinez}, {Maurel}, {Le Mignant}, {M{\"o}ller-Nilsson}, {Llored}, {Moulin},
  {Orign{\'e}}, {Pavlov}, {Perret}, {Petit}, {Pragt}, {Szulagyi}, \&
  {Wildi}}]{lagrange2019}
{Lagrange}, A.-M., {Boccaletti}, A., {Langlois}, M., {et~al.} 2019, \aap, 621,
  L8

\bibitem[{{Lee} \& {Song}(2019)}]{leesong2019}
{Lee}, J., \& {Song}, I. 2019, \mnras, 998

\bibitem[{Lee(2015)}]{plppipeline}
Lee, J.-J. 2015, plp: Version 2.0, , , doi:10.5281/zenodo.18579.
\newblock \url{https://doi.org/10.5281/zenodo.18579}

\bibitem[{Lee {et~al.}(2017)Lee, Gullikson, \& Kaplan}]{igrins}
Lee, J.-J., Gullikson, K., \& Kaplan, K. 2017, igrins/plp 2.2.0,  Zenodo,
  doi:10.5281/zenodo.845059.
\newblock \url{https://doi.org/10.5281/zenodo.845059}

\bibitem[{{Liu}(2004)}]{liu2004}
{Liu}, M.~C. 2004, Science, 305, 1442

\bibitem[{{Liu} {et~al.}(2016{\natexlab{a}}){Liu}, {Dupuy}, \&
  {Allers}}]{liu2016}
{Liu}, M.~C., {Dupuy}, T.~J., \& {Allers}, K.~N. 2016{\natexlab{a}}, \apj, 833,
  96

\bibitem[{{Liu} {et~al.}(2016{\natexlab{b}}){Liu}, {Dupuy}, \&
  {Allers}}]{dupuyallers2016}
---. 2016{\natexlab{b}}, \apj, 833, 96

\bibitem[{{L{\'o}pez-Valdivia} {et~al.}(2019){L{\'o}pez-Valdivia}, {Mace},
  {Sokal}, {Hussaini}, {Kidder}, {Mann}, {Gosnell}, {Oh}, {Kesseli},
  {Muirhead}, {Johns-Krull}, \& {Jaffe}}]{ricardo2019}
{L{\'o}pez-Valdivia}, R., {Mace}, G.~N., {Sokal}, K.~R., {et~al.} 2019, \apj,
  879, 105

\bibitem[{{Lowrance} {et~al.}(2000){Lowrance}, {Schneider}, {Kirkpatrick},
  {Becklin}, {Weinberger}, {Zuckerman}, {Plait}, {Malmuth}, {Heap}, {Schultz},
  {Smith}, {Terrile}, \& {Hines}}]{lowrance2000}
{Lowrance}, P.~J., {Schneider}, G., {Kirkpatrick}, J.~D., {et~al.} 2000, \apj,
  541, 390

\bibitem[{{Macintosh} {et~al.}(2014){Macintosh}, {Graham}, {Ingraham},
  {Konopacky}, {Marois}, {Perrin}, {Poyneer}, {Bauman}, {Barman}, {Burrows},
  {Cardwell}, {Chilcote}, {De Rosa}, {Dillon}, {Doyon}, {Dunn}, {Erikson},
  {Fitzgerald}, {Gavel}, {Goodsell}, {Hartung}, {Hibon}, {Kalas}, {Larkin},
  {Maire}, {Marchis}, {Marley}, {McBride}, {Millar-Blanchaer}, {Morzinski},
  {Norton}, {Oppenheimer}, {Palmer}, {Patience}, {Pueyo}, {Rantakyro},
  {Sadakuni}, {Saddlemyer}, {Savransky}, {Serio}, {Soummer},
  {Sivaramakrishnan}, {Song}, {Thomas}, {Wallace}, {Wiktorowicz}, \&
  {Wolff}}]{macintosh2014}
{Macintosh}, B., {Graham}, J.~R., {Ingraham}, P., {et~al.} 2014, Proceedings of
  the National Academy of Science, 111, 12661

\bibitem[{{Macintosh} {et~al.}(2015){Macintosh}, {Graham}, {Barman}, {De Rosa},
  {Konopacky}, {Marley}, {Marois}, {Nielsen}, {Pueyo}, {Rajan}, {Rameau},
  {Saumon}, {Wang}, {Patience}, {Ammons}, {Arriaga}, {Artigau}, {Beckwith},
  {Brewster}, {Bruzzone}, {Bulger}, {Burningham}, {Burrows}, {Chen}, {Chiang},
  {Chilcote}, {Dawson}, {Dong}, {Doyon}, {Draper}, {Duch{\^e}ne}, {Esposito},
  {Fabrycky}, {Fitzgerald}, {Follette}, {Fortney}, {Gerard}, {Goodsell},
  {Greenbaum}, {Hibon}, {Hinkley}, {Cotten}, {Hung}, {Ingraham},
  {Johnson-Groh}, {Kalas}, {Lafreniere}, {Larkin}, {Lee}, {Line}, {Long},
  {Maire}, {Marchis}, {Matthews}, {Max}, {Metchev}, {Millar-Blanchaer},
  {Mittal}, {Morley}, {Morzinski}, {Murray-Clay}, {Oppenheimer}, {Palmer},
  {Patel}, {Perrin}, {Poyneer}, {Rafikov}, {Rantakyr{\"o}}, {Rice}, {Rojo},
  {Rudy}, {Ruffio}, {Ruiz}, {Sadakuni}, {Saddlemyer}, {Salama}, {Savransky},
  {Schneider}, {Sivaramakrishnan}, {Song}, {Soummer}, {Thomas}, {Vasisht},
  {Wallace}, {Ward-Duong}, {Wiktorowicz}, {Wolff}, \&
  {Zuckerman}}]{macintosh2015}
{Macintosh}, B., {Graham}, J.~R., {Barman}, T., {et~al.} 2015, Science, 350, 64

\bibitem[{{Maire} {et~al.}(2016){Maire}, {Bonnefoy}, {Ginski}, {Vigan},
  {Messina}, {Mesa}, {Galicher}, {Gratton}, {Desidera}, {Kopytova}, {Millward},
  {Thalmann}, {Claudi}, {Ehrenreich}, {Zurlo}, {Chauvin}, {Antichi},
  {Baruffolo}, {Bazzon}, {Beuzit}, {Blanchard}, {Boccaletti}, {de Boer},
  {Carle}, {Cascone}, {Costille}, {De Caprio}, {Delboulb{\'e}}, {Dohlen},
  {Dominik}, {Feldt}, {Fusco}, {Girard}, {Giro}, {Gisler}, {Gluck}, {Gry},
  {Henning}, {Hubin}, {Hugot}, {Jaquet}, {Kasper}, {Lagrange}, {Langlois}, {Le
  Mignant}, {Llored}, {Madec}, {Martinez}, {Mawet}, {Milli},
  {M{\"o}ller-Nilsson}, {Mouillet}, {Moulin}, {Moutou}, {Orign{\'e}}, {Pavlov},
  {Petit}, {Pragt}, {Puget}, {Ramos}, {Rochat}, {Roelfsema}, {Salasnich},
  {Sauvage}, {Schmid}, {Turatto}, {Udry}, {Vakili}, {Wahhaj}, {Weber}, \&
  {Wildi}}]{maire2015}
{Maire}, A.~L., {Bonnefoy}, M., {Ginski}, C., {et~al.} 2016, \aap, 587, A56

\bibitem[{{Males} {et~al.}(2018){Males}, {Close}, {Miller}, {Schatz},
  {Doelman}, {Lumbres}, {Snik}, {Rodack}, {Knight}, {Van Gorkom}, {Long},
  {Hedglen}, {Kautz}, {Jovanovic}, {Morzinski}, {Guyon}, {Douglas}, {Follette},
  {Lozi}, {Bohlman}, {Durney}, {Gasho}, {Hinz}, {Ireland}, {Jean}, {Keller},
  {Kenworthy}, {Mazin}, {Noenickx}, {Alfred}, {Perez}, {Sanchez}, {Sauve},
  {Weinberger}, \& {Conrad}}]{males2018}
{Males}, J.~R., {Close}, L.~M., {Miller}, K., {et~al.} 2018, in Society of
  Photo-Optical Instrumentation Engineers (SPIE) Conference Series, Vol. 10703,
  Adaptive Optics Systems VI, 1070309

\bibitem[{{Malo} {et~al.}(2014{\natexlab{a}}){Malo}, {Artigau}, {Doyon},
  {Lafreni{\`e}re}, {Albert}, \& {Gagn{\'e}}}]{malo2014a}
{Malo}, L., {Artigau}, {\'E}., {Doyon}, R., {et~al.} 2014{\natexlab{a}}, \apj,
  788, 81

\bibitem[{{Malo} {et~al.}(2014{\natexlab{b}}){Malo}, {Doyon}, {Feiden},
  {Albert}, {Lafreni{\`e}re}, {Artigau}, {Gagn{\'e}}, \& {Riedel}}]{malo}
{Malo}, L., {Doyon}, R., {Feiden}, G.~A., {et~al.} 2014{\natexlab{b}}, \apj,
  792, 37

\bibitem[{{Malo} {et~al.}(2014{\natexlab{c}}){Malo}, {Doyon}, {Feiden},
  {Albert}, {Lafreni{\`e}re}, {Artigau}, {Gagn{\'e}}, \& {Riedel}}]{malobanyan}
---. 2014{\natexlab{c}}, \apj, 792, 37

\bibitem[{Mamajek(2016)}]{mamajek2016}
Mamajek, E. 2016, doi:10.6084/m9.figshare.3122689.v1.
\newblock
  \url{https://figshare.com/articles/A_New_Candidate_Young_Stellar_Group_at_d_121_pc_Associated_with_118_Tauri/3122689}

\bibitem[{{Mamajek} \& {Bell}(2014)}]{mamajek&bell2014}
{Mamajek}, E.~E., \& {Bell}, C.~P.~M. 2014, \mnras, 445, 2169

\bibitem[{{Mann} {et~al.}(2016){Mann}, {Gaidos}, {Mace}, {Johnson}, {Bowler},
  {LaCourse}, {Jacobs}, {Vanderburg}, {Kraus}, {Kaplan}, \& {Jaffe}}]{mann2016}
{Mann}, A.~W., {Gaidos}, E., {Mace}, G.~N., {et~al.} 2016, \apj, 818, 46

\bibitem[{{Marois} {et~al.}(2008){Marois}, {Macintosh}, {Barman}, {Zuckerman},
  {Song}, {Patience}, {Lafreni{\`e}re}, \& {Doyon}}]{marois2008}
{Marois}, C., {Macintosh}, B., {Barman}, T., {et~al.} 2008, Science, 322, 1348

\bibitem[{{Martinache} {et~al.}(2014){Martinache}, {Guyon}, {Jovanovic},
  {Clergeon}, {Singh}, {Kudo}, {Currie}, {Thalmann}, {McElwain}, \&
  {Tamura}}]{martinache2014}
{Martinache}, F., {Guyon}, O., {Jovanovic}, N., {et~al.} 2014, \pasp, 126, 565

\bibitem[{{Messina} {et~al.}(2017{\natexlab{a}}){Messina}, {Lanzafame}, {Malo},
  {Desidera}, {Buccino}, {Zhang}, {Artemenko}, {Millward}, \&
  {Hambsch}}]{messina2017}
{Messina}, S., {Lanzafame}, A.~C., {Malo}, L., {et~al.} 2017{\natexlab{a}},
  \aap, 607, A3

\bibitem[{{Messina} {et~al.}(2017{\natexlab{b}}){Messina}, {Millward},
  {Buccino}, {Zhang}, {Medhi}, {Jofr{\'e}}, {Petrucci}, {Pi}, {Hambsch},
  {Kehusmaa}, {Harlingten}, {Artemenko}, {Curtis}, {Hentunen}, {Malo}, {Mauas},
  {Monard}, {Muro Serrano}, {Naves}, {Santallo}, {Savuskin}, \&
  {Tan}}]{messina2017low_mass}
{Messina}, S., {Millward}, M., {Buccino}, A., {et~al.} 2017{\natexlab{b}},
  \aap, 600, A83

\bibitem[{{Messina} {et~al.}(2017{\natexlab{c}}){Messina}, {Millward},
  {Buccino}, {Zhang}, {Medhi}, {Jofr{\'e}}, {Petrucci}, {Pi}, {Hambsch},
  {Kehusmaa}, {Harlingten}, {Artemenko}, {Curtis}, {Hentunen}, {Malo}, {Mauas},
  {Monard}, {Muro Serrano}, {Naves}, {Santallo}, {Savuskin}, \&
  {Tan}}]{messina2017b}
---. 2017{\natexlab{c}}, \aap, 600, A83

\bibitem[{{Montet} {et~al.}(2015){Montet}, {Bowler}, {Shkolnik}, {Deck},
  {Wang}, {Horch}, {Liu}, {Hillenbrand}, {Kraus}, \&
  {Charbonneau}}]{montet2015}
{Montet}, B.~T., {Bowler}, B.~P., {Shkolnik}, E.~L., {et~al.} 2015, \apjl, 813,
  L11

\bibitem[{{Mugrauer} {et~al.}(2010){Mugrauer}, {Vogt}, {Neuh{\"a}user}, \&
  {Schmidt}}]{mugrauer2010}
{Mugrauer}, M., {Vogt}, N., {Neuh{\"a}user}, R., \& {Schmidt}, T.~O.~B. 2010,
  \aap, 523, L1

\bibitem[{{Murphy} {et~al.}(2013){Murphy}, {Lawson}, \& {Bessell}}]{murphy2013}
{Murphy}, S.~J., {Lawson}, W.~A., \& {Bessell}, M.~S. 2013, \mnras, 435, 1325

\bibitem[{{Nakajima} {et~al.}(1995){Nakajima}, {Oppenheimer}, {Kulkarni},
  {Golimowski}, {Matthews}, \& {Durrance}}]{nakajima1995}
{Nakajima}, T., {Oppenheimer}, B.~R., {Kulkarni}, S.~R., {et~al.} 1995, \nat,
  378, 463

\bibitem[{{Naud} {et~al.}(2013){Naud}, {Artigau}, {Doyon}, {Malo}, {Albert},
  {Lafreni{\`e}re}, \& {Gagn{\'e}}}]{naud2013}
{Naud}, M.-E., {Artigau}, {\'E}., {Doyon}, R., {et~al.} 2013, in European
  Physical Journal Web of Conferences, Vol.~47, European Physical Journal Web
  of Conferences, 13004

\bibitem[{{Neuh{\"a}user} {et~al.}(2011){Neuh{\"a}user}, {Ginski}, {Schmidt},
  \& {Mugrauer}}]{neuhauser2011}
{Neuh{\"a}user}, R., {Ginski}, C., {Schmidt}, T.~O.~B., \& {Mugrauer}, M. 2011,
  \mnras, 416, 1430

\bibitem[{{Nielsen} {et~al.}(2016){Nielsen}, {De Rosa}, {Wang}, {Rameau},
  {Song}, {Graham}, {Macintosh}, {Ammons}, {Bailey}, {Barman}, {Bulger},
  {Chilcote}, {Cotten}, {Doyon}, {Duch{\^e}ne}, {Fitzgerald}, {Follette},
  {Greenbaum}, {Hibon}, {Hung}, {Ingraham}, {Kalas}, {Konopacky}, {Larkin},
  {Maire}, {Marchis}, {Marley}, {Marois}, {Metchev}, {Millar-Blanchaer},
  {Oppenheimer}, {Palmer}, {Patience}, {Perrin}, {Poyneer}, {Pueyo}, {Rajan},
  {Rantakyr{\"o}}, {Savransky}, {Schneider}, {Sivaramakrishnan}, {Soummer},
  {Thomas}, {Wallace}, {Ward-Duong}, {Wiktorowicz}, \& {Wolff}}]{nielsen2016}
{Nielsen}, E.~L., {De Rosa}, R.~J., {Wang}, J., {et~al.} 2016, \aj, 152, 175

\bibitem[{{Norton} {et~al.}(2007){Norton}, {Wheatley}, {West}, {Haswell},
  {Street}, {Collier Cameron}, {Christian}, {Clarkson}, {Enoch}, {Gallaway},
  {Hellier}, {Horne}, {Irwin}, {Kane}, {Lister}, {Nicholas}, {Parley},
  {Pollacco}, {Ryans}, {Skillen}, \& {Wilson}}]{norton2007}
{Norton}, A.~J., {Wheatley}, P.~J., {West}, R.~G., {et~al.} 2007, \aap, 467,
  785

\bibitem[{{Oppenheimer} {et~al.}(1995){Oppenheimer}, {Kulkarni}, {Matthews}, \&
  {Nakajima}}]{oppenheimer1995}
{Oppenheimer}, B.~R., {Kulkarni}, S.~R., {Matthews}, K., \& {Nakajima}, T.
  1995, Science, 270, 1478

\bibitem[{{Park} {et~al.}(2014){Park}, {Jaffe}, {Yuk}, {Chun}, {Pak}, {Kim},
  {Pavel}, {Lee}, {Oh}, {Jeong}, {Sim}, {Lee}, {Nguyen Le}, {Strubhar},
  {Gully-Santiago}, {Oh}, {Cha}, {Moon}, {Park}, {Brooks}, {Ko}, {Han}, {Nah},
  {Hill}, {Lee}, {Barnes}, {Yu}, {Kaplan}, {Mace}, {Kim}, {Lee}, {Hwang}, \&
  {Park}}]{park2014}
{Park}, C., {Jaffe}, D.~T., {Yuk}, I.-S., {et~al.} 2014, in \procspie, Vol.
  9147, Ground-based and Airborne Instrumentation for Astronomy V, 91471D

\bibitem[{{P{\"o}hnl} \& {Paunzen}(2010)}]{pohnlpaunzen2010}
{P{\"o}hnl}, H., \& {Paunzen}, E. 2010, \aap, 514, A81

\bibitem[{{Rajan} {et~al.}(2017){Rajan}, {Rameau}, {De Rosa}, {Marley},
  {Graham}, {Macintosh}, {Marois}, {Morley}, {Patience}, {Pueyo}, {Saumon},
  {Ward-Duong}, {Ammons}, {Arriaga}, {Bailey}, {Barman}, {Bulger}, {Burrows},
  {Chilcote}, {Cotten}, {Czekala}, {Doyon}, {Duch{\^e}ne}, {Esposito},
  {Fitzgerald}, {Follette}, {Fortney}, {Goodsell}, {Greenbaum}, {Hibon},
  {Hung}, {Ingraham}, {Johnson-Groh}, {Kalas}, {Konopacky}, {Lafreni{\`e}re},
  {Larkin}, {Maire}, {Marchis}, {Metchev}, {Millar-Blanchaer}, {Morzinski},
  {Nielsen}, {Oppenheimer}, {Palmer}, {Patel}, {Perrin}, {Poyneer},
  {Rantakyr{\"o}}, {Ruffio}, {Savransky}, {Schneider}, {Sivaramakrishnan},
  {Song}, {Soummer}, {Thomas}, {Vasisht}, {Wallace}, {Wang}, {Wiktorowicz}, \&
  {Wolff}}]{rajan2017}
{Rajan}, A., {Rameau}, J., {De Rosa}, R.~J., {et~al.} 2017, \aj, 154, 10

\bibitem[{{Rayner} {et~al.}(2003){Rayner}, {Toomey}, {Onaka}, {Denault},
  {Stahlberger}, {Vacca}, {Cushing}, \& {Wang}}]{rayner2003}
{Rayner}, J.~T., {Toomey}, D.~W., {Onaka}, P.~M., {et~al.} 2003, \pasp, 115,
  362

\bibitem[{{Rebolo} {et~al.}(1995){Rebolo}, {Zapatero Osorio}, \&
  {Mart{\'\i}n}}]{rebolo1995}
{Rebolo}, R., {Zapatero Osorio}, M.~R., \& {Mart{\'\i}n}, E.~L. 1995, \nat,
  377, 129

\bibitem[{{Rizzuto} {et~al.}(2017){Rizzuto}, {Mann}, {Vanderburg}, {Kraus}, \&
  {Covey}}]{rizzuto2017}
{Rizzuto}, A.~C., {Mann}, A.~W., {Vanderburg}, A., {Kraus}, A.~L., \& {Covey},
  K.~R. 2017, \aj, 154, 224

\bibitem[{{Saumon} \& {Marley}(2008)}]{saumon&marley2008}
{Saumon}, D., \& {Marley}, M.~S. 2008, \apj, 689, 1327

\bibitem[{{Schlieder} {et~al.}(2010){Schlieder}, {L{\'e}pine}, \&
  {Simon}}]{schlieder2010}
{Schlieder}, J.~E., {L{\'e}pine}, S., \& {Simon}, M. 2010, \aj, 140, 119

\bibitem[{{Service} {et~al.}(2016){Service}, {Lu}, {Campbell}, {Sitarski},
  {Ghez}, \& {Anderson}}]{service2016}
{Service}, M., {Lu}, J.~R., {Campbell}, R., {et~al.} 2016, \pasp, 128, 095004

\bibitem[{{Shkolnik} {et~al.}(2017){Shkolnik}, {Allers}, {Kraus}, {Liu}, \&
  {Flagg}}]{shkolnik2017}
{Shkolnik}, E.~L., {Allers}, K.~N., {Kraus}, A.~L., {Liu}, M.~C., \& {Flagg},
  L. 2017, \aj, 154, 69

\bibitem[{{Shkolnik} {et~al.}(2012){Shkolnik}, {Anglada-Escud{\'e}}, {Liu},
  {Bowler}, {Weinberger}, {Boss}, {Reid}, \& {Tamura}}]{shkolnik2012}
{Shkolnik}, E.~L., {Anglada-Escud{\'e}}, G., {Liu}, M.~C., {et~al.} 2012, \apj,
  758, 56

\bibitem[{{Simon} {et~al.}(2019){Simon}, {Guilloteau}, {Beck}, {Chapillon}, {Di
  Folco}, {Dutrey}, {Feiden}, {Grosso}, {Pi{\'e}tu}, {Prato}, \&
  {Schaefer}}]{simon2019}
{Simon}, M., {Guilloteau}, S., {Beck}, T.~L., {et~al.} 2019, \apj, 884, 42

\bibitem[{{Simons} \& {Tokunaga}(2002)}]{tokunaga&simons2002}
{Simons}, D.~A., \& {Tokunaga}, A. 2002, \pasp, 114, 169

\bibitem[{{Snellen} \& {Brown}(2018)}]{snellenbrown2018}
{Snellen}, I.~A.~G., \& {Brown}, A.~G.~A. 2018, Nature Astronomy, 2, 883

\bibitem[{{Song} {et~al.}(2006){Song}, {Schneider}, {Zuckerman}, {Farihi},
  {Becklin}, {Bessell}, {Lowrance}, \& {Macintosh}}]{song2006}
{Song}, I., {Schneider}, G., {Zuckerman}, B., {et~al.} 2006, \apj, 652, 724

\bibitem[{{Tokunaga} {et~al.}(2002){Tokunaga}, {Simons}, \&
  {Vacca}}]{tokunga2002}
{Tokunaga}, A.~T., {Simons}, D.~A., \& {Vacca}, W.~D. 2002, \pasp, 114, 180

\bibitem[{{Tokunaga} \& {Vacca}(2007)}]{tokunaga2007}
{Tokunaga}, A.~T., \& {Vacca}, W.~D. 2007, in Astronomical Society of the
  Pacific Conference Series, Vol. 364, The Future of Photometric,
  Spectrophotometric and Polarimetric Standardization, ed. C.~{Sterken}, 409

\bibitem[{{Torres} {et~al.}(2006){Torres}, {Quast}, {da Silva}, {de La Reza},
  {Melo}, \& {Sterzik}}]{torres2006}
{Torres}, C.~A.~O., {Quast}, G.~R., {da Silva}, L., {et~al.} 2006, \aap, 460,
  695

\bibitem[{{Torres} {et~al.}(2008){Torres}, {Quast}, {Melo}, \&
  {Sterzik}}]{torres2008}
{Torres}, C.~A.~O., {Quast}, G.~R., {Melo}, C.~H.~F., \& {Sterzik}, M.~F. 2008,
  {Young Nearby Loose Associations}, ed. B.~{Reipurth}, 757

\bibitem[{{Vacca} {et~al.}(2003){Vacca}, {Cushing}, \& {Rayner}}]{vacca2003}
{Vacca}, W.~D., {Cushing}, M.~C., \& {Rayner}, J.~T. 2003, \pasp, 115, 389

\bibitem[{{van Dokkum}(2001)}]{dokkum2001}
{van Dokkum}, P.~G. 2001, Publications of the Astronomical Society of the
  Pacific, 113, 1420

\bibitem[{{Witte} {et~al.}(2011){Witte}, {Helling}, {Barman}, {Heidrich}, \&
  {Hauschildt}}]{witte2011}
{Witte}, S., {Helling}, C., {Barman}, T., {Heidrich}, N., \& {Hauschildt},
  P.~H. 2011, \aap, 529, A44

\bibitem[{{Yelda} {et~al.}(2010){Yelda}, {Lu}, {Ghez}, {Clarkson}, {Anderson},
  {Do}, \& {Matthews}}]{yelda2010}
{Yelda}, S., {Lu}, J.~R., {Ghez}, A.~M., {et~al.} 2010, \apj, 725, 331

\bibitem[{{Zacharias} {et~al.}(2012){Zacharias}, {Finch}, {Girard}, {Henden},
  {Bartlett}, {Monet}, \& {Zacharias}}]{zacharias2012}
{Zacharias}, N., {Finch}, C.~T., {Girard}, T.~M., {et~al.} 2012, VizieR Online
  Data Catalog, I/322A

\bibitem[{{Zuckerman} \& {Song}(2004)}]{zuckerman&song2004}
{Zuckerman}, B., \& {Song}, I. 2004, \araa, 42, 685

\bibitem[{{Zuckerman} {et~al.}(2001){Zuckerman}, {Song}, {Bessell}, \&
  {Webb}}]{zuckerman2001}
{Zuckerman}, B., {Song}, I., {Bessell}, M.~S., \& {Webb}, R.~A. 2001, \apjl,
  562, L87

\end{thebibliography}
\nocite{*}
\end{document}